 \documentclass[prb,twocolumn,showpacs,preprintnumbers]{revtex4-1}

\usepackage{graphicx,bm}
\usepackage{amsmath,amssymb}

\usepackage[dvips]{color}

\begin{document}
\title{
Kondo effects in a triangular triple quantum dot
 with lower symmetries 
}

\author{
A. Oguri$^1$, 
S. Amaha$^{2,3}$, 
Y. Nishikawa$^1$, 
T. Numata$^1$, 
M. Shimamoto$^1$, 
A. C. Hewson$^4$
and 
S. Tarucha$^5$
}

\affiliation{
$^{1}$Department of Physics, Osaka City University, Sumiyoshi-ku, 
Osaka 558-8585, Japan\\
$^{2}$Quantum Spin Information Project, ICORP-JST, Atsugi-shi, 
Kanagawa 243-0198, 
Japan\\
$^{3}$ Low Temperature Physics Laboratory, 
RIKEN, Wako-shi, Saitama 351-0198, Japan\\
$^{4}$ Department of Mathematics, Imperial College, 
180 Queen's Gate, London SW7 2BZ, UK \\
$^{5}$ Department of Applied Physics, School of Engineering,
University of Tokyo, Bunkyo-ku, Tokyo 133-8656, Japan
}


\date{\today}

\begin{abstract}
We study the low-energy properties 
and  characteristic Kondo energy scale 
of a triangular triple quantum dot, 
connected to two non-interacting leads,  
in a wide parameter range of a gate voltage and  
distortions which lower the symmetry of an equilateral structure, 
using the numerical renormalization group approach. 
For large Coulomb interactions, 
the ground states with different characters can be classified according to 
the plateaus of  
 $\Theta \equiv (\delta_\mathrm{e}-\delta_\mathrm{o})(2/\pi)$,
where  $\delta_\mathrm{e}$ and $\delta_\mathrm{o}$ are 
 the phase shifts for the even and odd partial waves.
At these plateaus of $\Theta$, both $\Theta$ and 
the occupation number 
$N_\mathrm{tot} \equiv (\delta_\mathrm{e}+\delta_\mathrm{o})(2/\pi)$ 
take  values close to integers,
and thus the ground states can be characterized by these two integers. 
The Kondo effect with a local moment with total spin $S=1$ 
due to a Nagaoka mechanism appears on the plateau, 
which can be identified by 
$\Theta \simeq 2.0$ and $N_\mathrm{tot}\simeq 4.0$.
For large distortions, however, 
the high-spin moment disappears through a singlet-triplet transition 
occurring within the four-electron region. 
It happens at a crossover to the adjacent plateaus for $\Theta \simeq 0.0$  
and  $\Theta \simeq 4.0$, and  the two-terminal conductance 
has a peak in the transient regions. 
For weak distortions,  
the SU(4) Kondo effect also takes place for $N_\mathrm{tot}\simeq 3.0$.
It appears as a sharp conductance valley between the  $S=1/2$ Kondo ridges 
on both sides. 
We also find that the characteristic energy scale $T^*$ 
reflect these varieties of the Kondo effect. 
Particularly, $T^*$ is sensitive to the 
distribution of the charge and spin in the triangular triple dot.

\end{abstract}



\maketitle

\section{Introduction}

The triangle is the simplest polygon,
and has a closed loop which plays an important 
role on various fascinating phenomena 
in the condensed matter physics.
The closed path in a metal and semiconductor   
allows the electrons to move around the loop, 
and causes a quantum-mechanical interference effects, 
such as an Aharanov-Bohm (AB) effect.\cite{AB,Webb}  
The closed path consisting of 
the odd-number of links also causes frustration,  
which leads to resonating valence bonds 
for some anti-ferromagnetic systems.\cite{RVB}

Furthermore, the interplay between the strong correlation and the interference 
effects caused by the triangular structure 
has also been one of the topics of the current interests 
in different fields of the condensed matter physics.
For instance, the single triangle 
is also a fundamental unit of the triangular and kagom\'{e} lattices.
In these systems the geometrical frustration affects significantly 
the magnetic properties 
and the behavior at the Mott-Hubbard 
metal-insulator transition.\cite{Koshibae1,Furukawa_Kawakami}
Another interesting example is the triangular trimer of 
Cr atoms placed upon an Au surface,\cite{Jamneala,Affleck3a,Zarand} 
and this system is expected to show 
a non-Fermi-liquid behavior due to the multi-channel 
Kondo effect.\cite{NozieresBlandin,CoxZawadowski}

Recently, 
the triangular triple quantum dot (TTQD) has been 
experimentally  realized and intensively studied using various 
 systems,  such as 
AlGaAs/GaAs heterostructures\cite{Vidan-Stopa,Canada,Amaha2,Amaha3,RoggeHaug}
and self-assembled InAs quantum dot.\cite{Amaha1} 
Theoretically, the TTQD has been shown to  
demonstrate various types of the Kondo effects, 
\cite{ONTN,Numata,Numata2,Zitko2,Logan3,Ulloa}
as well as the AB effect.\cite{KKA1,DelgadoShimKorkusinskiHawrylak} 
The closed path makes the TTQD different from 
a linear quantum-dot chain\cite{ao99,aoQuasi,OH,ONH,NO,KKA2,Zitko,Zitko_Bonca} 
and the other three-level systems.
\cite{Eto,Sakano,LeoFabrizio,Kita,Hecht_3,Paaske}
One of the most interesting points is 
that the appearance of a local moment with total spin $S=1$,   
at the filling where one additional electron is introduced 
into half-filling.\cite{ONTN,KorkusinskiGimenezHawrylakETAL}
This is caused by a Nagaoka ferromagnetic mechanism 
for the electrons moving around the triangular structure.
\cite{Nagaoka} 
The $S=1$ moment shows a Kondo behavior 
when the leads are coupled to 
the quantum dots.\cite{ONTN,Numata,Numata2}
Another interesting point is that the SU(4) Kondo effect 
takes place at half-filling, 
\cite{ONTN,Numata,Numata2,Logan3,Zitko2,Ulloa}
in the case where the ground state 
has a 4-fold degeneracy caused by 
the orbital and spin degrees of freedoms.
The TTQD has provided a new variety to the SU(4) Kondo effect, 
which had been studied for the double-dot systems.
\cite{Borda,Logan_2dotA,Mravlje_Ramsak_Rejec,Anders_Logan}

The number of leads connected to 
the TTQD also affects  significantly the Kondo behavior.
This is because whether or not the local moment 
can be screened depends on the relation between 
 the dimension of the Hilbert space for the local moment 
and the number of conducting channels.\cite{NozieresBlandin}
The low-temperature properties of the TTQD 
have been studied, so far, by several theoretical groups, 
for the configurations with one,\cite{Logan3}
two,\cite{ONTN,Numata,Numata2,Zitko2,Ulloa} and 
three leads.\cite{Ulloa}
These studies complement each other 
the wide parameter space of the TTQD.
\v{Z}itko {\it et al\/}\cite{Zitko2} 
  and Mitchell {\it et al\/}\cite{Logan3} 
studied the Kondo effect at half-filling in some situations,  
but the dependence on the electron filling was not examined.   
Vernek {\it et al\/} \cite{Ulloa} examined the gate voltage 
$\epsilon_d$ dependence in a wide range of the electron filling, 
but the parameters used were confined 
to a region of small interaction $U$ 
and a large dot-lead coupling $\Gamma$  
where the Kondo behavior is still rather smeared.

We have studied the Kondo behavior 
of the TTQD away from half-filling 
in the series of the works.\cite{ONTN,Numata,Numata2} 
Our research in the early stage\cite{ONTN} started 
with a theoretical observation of 
a two-stage Kondo screening 
of the $S=1$ Nagaoka high-spin at four-electron filling 
and  a sharp conductance dip caused by the 
SU(4) Kondo effect at half-filling (with three electrons),
appearing in the gate-voltage dependence. 
The precise features of these Kondo effects 
have been clarified further in the previous paper,\cite{Numata2}
for the parameter values which cover 
the weak and strong couplings 
with respect to  both  $U$ and $\Gamma$.
We have also examined the effects of 
the perturbations which break the full symmetry 
of the equilateral triangle,\cite{Numata2}
as the real TTQD systems have some deviations from the regular 
structure in most of cases.
Our results, obtained with the Wilson numerical 
renormalization group (NRG),\cite{KWW,KWW2}
have shown that the conductance dip 
typical of the SU(4) Kondo effect in the TTQD 
is quite sensitive to the perturbation,  
while the $S=1$ Kondo behavior is robust.\cite{Numata2} 

The distortions of the triangular structure discussed in the previous paper, 
however, were still relatively small, so that 
the overall features  of the Kondo effect in the TTQD  
have not yet been fully revealed and much remains to be explored.
Particularly, 
a singlet-triplet transition between a local singlet 
and the Nagaoka high-spin state 
occurs in the isolated TTQD cluster for large deformations,  
and this transition will 
affect the Kondo behavior at four-electron filling. 
Furthermore, the behavior of 
the conductance dip due to the SU(4) Kondo effect 
also needs to be clarified in more detail.

The purpose of the present work is to provide  
a comprehensive overview of the Kondo effect  
in the TTQD and to study the effects of large distortions. 
 Specifically, we examine two different types of distortion: 
 ($i$) an irregular inter-dot coupling, 
 and ($ii$) an inhomogeneity 
 in the level position of the quantum dots.
We calculate the phase shifts, 
 $\delta_\mathrm{e}$ and $\delta_\mathrm{o}$,  
for the even and odd partial waves  
of the renormalized quasi-particles, 
in a wide parameter region  
of the gate voltage $\epsilon_d$ and the distortions.
These two phase shifts determine the ground 
state properties of the TTQD connected to two leads.

In the parameter space for large $U$ and small $\Gamma$, 
we find plateau with the integer values 
of the phase 
difference $\Theta \equiv (\delta_\mathrm{e}-\delta_\mathrm{o})(2/\pi)$, 
and at each plateau 
the occupation number given by the Friedel sum rule
$N_\mathrm{tot} 
\equiv (\delta_\mathrm{e}+\delta_\mathrm{o})(2/\pi)$   
also approaches to an integer. 
These plateaus, therefore,  can be classified with 
the two integer set  $(N_\mathrm{tot},\,\Theta)$, 
and 
each plateau corresponds to the ground state 
realized  
in each parameter region.  
For instance, the plateau for the $S=1$ Kondo region    
can be labelled as $(N_\mathrm{tot},\,\Theta) \simeq (4.0,\,2.0 )$. 
The singlet-triplet transition 
emerges as a steep rise in $\Theta$ to the adjacent plateaus 
with the label $(N_\mathrm{tot},\,\Theta) \simeq(4.0,\, 0.0)$ 
and $(4.0,\,4.0)$, 
situated in the regions of a large distortion.
Therefore, the two-terminal conductance $g_\mathrm{s}$ shows 
a peak of the unitary-limit 
value $g_\mathrm{s}=2e^2/h$ in the middle of the rise.
We also find that the SU(4) Kondo behavior appears  
in the parameter space along the contour line 
for $\Theta=2.0$, which traverses the middle of the steep rise in $\Theta$ 
between the plateaus with $\Theta \simeq 1.0$ and $\Theta \simeq 3.0$,   
for $N_\mathrm{tot} \simeq 3.0$.

\begin{figure}[t]

\begin{minipage}{0.68\linewidth}
 \includegraphics[width=1\linewidth]{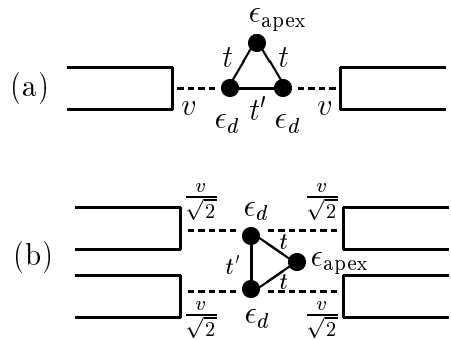}
\end{minipage}

\caption{Triangular triple quantum dot 
in (a) series and (b) parallel configurations.
The dot which has no direct 
connection to the leads is referred to 
as the {\it apex\/} site, 
and labeled as $i=2$ in the text.
}

\label{fig:system}
\end{figure}

We also estimate the characteristic energy scale $T^*$ of the 
Kondo screening in the wide parameter region, 
from the flow of the low-lying excitation energies in the NRG.
The energy scale  $T^*$ depends strongly on the local charge 
distribution in the TTQD. 
The screening is protracted significantly in the case where 
the partial component of the local moment becomes 
large at the apex site, which is located away from 
the leads as shown in Fig.\ \ref{fig:system} (a).
This is because that 
the tunneling processes of the conduction electrons 
 from the leads to the apex site tend to be suppressed 
in the intermediate states on the other two sites.

The paper is organized as follows.
We describe the model and the formulation 
in Sec.\ \ref{sec:formulation}. 
Some characteristics of the TTQD, 
seen already in the non-interacting case of $U=0$, 
are summarized in Sec.\ \ref{eq:U=0_results}. 
Then, the molecular limit $\Gamma \to 0$ for finite $U$ is 
considered in Sec.\ \ref{sec:molecule_limit} 
in order to see the basic features of the local 
charge and spin states of the TTQD. 
The NRG results for the ground-state properties 
are shown in Sec.\ \ref{sec:results_I}. 
The results for the characteristic energy scale $T^*$ are 
presented in Sec.\ \ref{sec:TK}. 
A summary is given in Sec.\ \ref{sec:summary}.

\section{Formulation}
\label{sec:formulation}

 \subsection{TTQD connected to two non-interacting leads}
 \label{subsec:model}

We consider a three-site Hubbard model on 
a triangular cluster as a model for the TTQD. 
The cluster is connected  to two non-interacting leads 
on the left ($L$) and right ($R$) 
as illustrated in Fig.\ \ref{fig:system} (a). 
The Hamiltonian is given by 
\begin{align}
&\mathcal{H} =  
\mathcal{H}_\mathrm{dot}^0 +
\mathcal{H}_\mathrm{dot}^U 
+  \mathcal{H}_\mathrm{mix}  +  
\mathcal{H}_\mathrm{lead}\;, 
\label{eq:H}
\\
& 
\mathcal{H}_\mathrm{dot}^0  = \,    
 - \sum_{<ij>}^{N_D}\sum_{\sigma} t_{ij} 
 \left(
 d^{\dagger}_{i\sigma}d^{\phantom{\dagger}}_{j\sigma}  
+ d^{\dagger}_{j\sigma}d^{\phantom{\dagger}}_{i\sigma}  
\right) 
 \nonumber
 \\
& \qquad \quad \ 
+ 
\sum_{i=1}^{N_D}\sum_{\sigma}  
\epsilon_{d,i} \,
 d^{\dagger}_{i\sigma}d^{\phantom{\dagger}}_{i\sigma} , 
\label{eq:Hdot^0}
\\
& 
\mathcal{H}_\mathrm{dot}^U  =     
  U\sum_{i=1}^{N_D} 
n_{d,i\uparrow}\,n_{d,i\downarrow} , 
\qquad \ \ 
n_{d,i\sigma} \equiv d^{\dagger}_{i\sigma}d^{}_{i\sigma},
\label{eq:HC^U}\\
&
\mathcal{H}_\mathrm{mix} =   
 \sum_{\sigma}   
 \left(  \,
v_L^{}\, d^{\dagger}_{1,\sigma}
 C^{\phantom{\dagger}}_{L \sigma}
   +   
v_R^{}\, 
d^{\dagger}_{N_D, \sigma} C^{\phantom{\dagger}}_{R\sigma}
+\, \mathrm{H.c.}
\, \right)   ,
 \label{eq:Hmix}
\\
& \mathcal{H}_\mathrm{lead} =  
\sum_{\nu=L,R} 
 \sum_{k\sigma} 
  \epsilon_{k}^{\phantom{0}}\,
         c^{\dagger}_{k \nu \sigma} 
         c^{\phantom{\dagger}}_{k \nu \sigma}
\,.
\label{eq:H_lead}
\end{align}
Here, 
$d^{\dagger}_{i\sigma}$ creates an electron with spin $\sigma$ 
at the $i$-th dot,  $\epsilon_{d,i}$ the onsite potential, 
 $U$ the Coulomb interaction, 
and the number of the dots is given by $N_D\equiv 3$.
The hopping matrix element $t_{ij}$ between the dots is chosen 
to be real and positive ($t_{ij}\geq 0$). 
The dots labelled by $i=1$ and $i=3$ are 
directly coupled, respectively, 
to the left and right leads 
via the tunnelling matrix element $v_{L/R}^{}$.
The coupling causes the level broadening 
of $\Gamma_{L/R}^{} \equiv \pi  \rho \, v_{L/R}^2$, 
with $\rho$ the density of states for the conduction band 
described by $\epsilon_k$, and we will take $\Gamma$ to be 
a constant assuming a wide flat band. 
The conduction electrons are described by the operators 
 $c^{\dagger}_{k \nu \sigma}$ and
 $C_{\nu \sigma}^{\phantom{\dagger}} 
\equiv \sum_k c_{k \nu \sigma}^{\phantom{\dagger}}/\sqrt{N}$. 
In the present work, 
we consider the case that the system has an inversion symmetry 
choosing 
 $\Gamma_L^{} = \Gamma_R^{}$ ($\equiv \Gamma$),  
namely $v_L^{} = v_R^{}$ ($\equiv v$), 
$\epsilon_{d,1}=\epsilon_{d,3}$ ($\equiv \epsilon_d$), 
$t_{12}=t_{23}$ ($\equiv t$) and $t_{13}$ ($\equiv t'$). 
We shall refer to the dot which has no 
direct connection to the leads as the {\em apex} site, 
and will use a notation  $\epsilon_{d,2} \equiv \epsilon_\mathrm{apex}$ 
for $i=2$.  
 We also choose the Fermi energy $E_F$ 
as the origin of the energy $E_F=0$.

\subsection{Phase shift, conductance and local charge}
\label{subsec:cond_phase_shift}

Charge transfer between the dots and leads 
makes the low-energy states of the whole system  
a local Fermi liquid, 
which can be described by renormalized quasi-particles. 
Specifically, in the inversion symmetry case 
the two phase shifts, $\delta_\mathrm{e}$ and $\delta_\mathrm{o}$,
for the {\em even\/} and {\em odd\/} partial waves 
become the essential parameters which characterize the ground state 
 [see also Appendix \ref{sec:app_green}].
At zero temperature $T=0$, 
the series conductance $g_\mathrm{s}$ for the current 
flowing through the two-channel configuration shown 
in Fig.\ \ref{fig:system} (a), 
and the total number of electrons $N_\mathrm{tot}$ in the dots 
can be expressed in terms of these phase shifts,\cite{Numata2,Izumida2} 
\begin{align}
 g_\mathrm{s} \,= & \   
g_0^{}\,
\sin^{2} \left(\delta_\mathrm{e} -\delta_\mathrm{o} \right) \;,
\label{eq:gs}
\\
 N_\mathrm{tot}
\equiv &  \  \sum_{i=1}^{N_D}\sum_{\sigma} \,
\langle d_{i\sigma}^{\dagger}d_{i\sigma}^{\phantom{\dagger}}\rangle
\, = \, \frac{2}{\pi}\left(\delta_\mathrm{e}+\delta_\mathrm{o}\right) \:,
\label{eq:N_tot}
\end{align}
where  $g_0^{} \equiv 2e^2/h$. 
Note that the sum $N_\mathrm{tot}$ and the difference 
$\Theta \equiv 
(\delta_\mathrm{e} -\delta_\mathrm{o})(2/\pi)$ between the two phase shifts 
link directly to the ground-state properties in the series configuration. 
Specifically, $\Theta$ 
becomes a more natural measure for classifying the parameter space 
than $g_\mathrm{s}$ for the quantum dots consisting of 
more than three local orbitals $N_D \geq 3$. 
This is because the phase difference can be greater than $\pi$, 
for instance, 
it takes a value in the range  
 $0\leq \delta_\mathrm{e}-\delta_\mathrm{o} \leq 2\pi$ 
for $N_D=3$.

The parallel conductance $g_\mathrm{p}$ for the current flowing 
along the horizontal direction in the four-terminal geometry, 
shown in Fig.\ \ref{fig:system} (b),    
can also be deduced from these two  phase 
shifts $\delta_\mathrm{e}$ and $\delta_\mathrm{o}$ 
defined with respect to the series configuration,
\begin{align}
 &
 g_\mathrm{p} \,= \, 
g_0^{}\,
  \left(\sin^{2}\delta_\mathrm{e}  + 
  \sin^{2}\delta_\mathrm{o}\right)\;.
       \label{eq:gp}
 \end{align}
The even and odd channels contribute to 
the parallel conductance separately 
with no cross terms which would represent  interference effects.
Note that in the case where the series conductance 
reaches the unitary-limit value $g_\mathrm{s}^{} = 2e^2/h$, 
namely 
at  $\delta_\mathrm{e}-\delta_\mathrm{o}=(n+1/2)\pi$ for $n=0,\,\pm 1,
\,\pm 2,\ldots$, 
the parallel conductance also takes 
the same value $g_\mathrm{p}= 2e^2/h$ which 
is the half of its maximum possible value $4e^2/h$.

The phase shifts for the interacting case 
can be expressed in terms of the renormalized hopping matrix 
element $\widetilde{t}_{ij}$ for the 
quasi-particles,\cite{Numata2,aoFermi} 
\begin{align}
 -\widetilde{t}_{ij} 
 \,\equiv & \ -t_{ij} + \epsilon_{d,i}\, \delta_{ij} 
 + \mathrm{Re}\, \Sigma_{ij}^+(0) \;. 
\end{align}
Here, $\Sigma_{ij}^+(\omega)$ is the self energy 
due to the Coulomb interaction $\mathcal{H}_\mathrm{dot}^U$,  
defined in Appendix \ref{sec:app_green}.
For the TTQD, the renormalized matrix elements expressed in the form 
\begin{align}
 \{-\widetilde{t}_{ij} \} = 
\left[
\begin{matrix}
 \widetilde{\epsilon}_d & -\widetilde{t} & -\widetilde{t}' \cr
 -\widetilde{t} & \widetilde{\epsilon}_\mathrm{apex} & -\widetilde{t}  \cr
 -\widetilde{t}' & -\widetilde{t} &  \widetilde{\epsilon}_d\cr
\end{matrix}
\right] \;.
\end{align}
The explicit form of the phase shifts can be obtained by 
solving the scattering problem of the renormalized quasi-particles,
or equivalently from the Dyson equation given 
in given in \eqref{eq:Dyson}, as   
\begin{align}
\!
\cot \delta_\mathrm{e}  
= \frac{\widetilde{\epsilon}_d \, 
-\widetilde{t}' -2\widetilde{t}^2/\widetilde{\epsilon}_\mathrm{apex}}{\Gamma }
, \quad   
\cot \delta_\mathrm{o} = \frac{\widetilde{\epsilon}_d+\widetilde{t}'}{\Gamma}.
\label{eq:phase_shift_renormalized}
\end{align}
Specifically, the zero points of $g_\mathrm{s}$ 
can be determined by the condition between the renormalized parameters 
\begin{align}
\widetilde{\epsilon}_\mathrm{apex}\,\widetilde{t}'
+\widetilde{t}^2 = 0 \;,
\end{align}
which follows from 
the relation $\cot \delta_\mathrm{e} = \cot \delta_\mathrm{o}$. 
Similarly, $g_\mathrm{s}$  takes the unitary-limit value   
$g_\mathrm{s}=2e^2/h$ 
in the case of $\cot \delta_\mathrm{e} = -1/\cot \delta_\mathrm{o}$,
which corresponds to the condition
\begin{align}
\left(
\widetilde{\epsilon}_d 
-\widetilde{t}' -2\widetilde{t}^2/\widetilde{\epsilon}_\mathrm{apex}
\right)
\left(
\widetilde{\epsilon}_d +\widetilde{t}' 
\right) +\Gamma^2=0 \;.
\end{align}

\section{Effects of distortions in the non-interacting case}
\label{eq:U=0_results}

We first of all discuss the level structure of 
an isolated cluster  for $\Gamma=0$ 
in order to trace out the particular characteristics of the TTQD. 
We then calculate the conductances through the dots 
for $U=0$ to see how they reflect the level structure. 
These examples provide us with  essential information   
for understanding  the variety of forms  which we will encounter   
in the wider parameter space.

\subsection{Level structure of the TTQD}

The one-particle energy levels for the non-interacting TTQD cluster 
 which is described by $\mathcal{H}_\mathrm{dot}^0$  are given by
\begin{align}
E_{\mathrm{e},\pm}^{(1)} =& \  
\frac{ \epsilon_\mathrm{apex}+\epsilon_d-t'}{2} 
\pm \sqrt{
\left(\frac{\epsilon_\mathrm{apex}-\epsilon_d+t'}{2}\right)^2 
+2t^2
},   
\label{eq:U0_eigen_cluster1}
\\
E_\mathrm{o}^{(1)} =& \ \epsilon_d + t' \;.
\label{eq:U0_eigen_cluster2}
\end{align}
Here, 
$E_{\mathrm{e},\pm}^{(1)}$ and 
$E_{\mathrm{o}}^{(1)}$ are, 
respectively, 
the energy for the eigenstates with the even and odd parities 
[see also Appendix \ref{sec:even_odd}].
Among the three eigenstates, the one with the energy 
$E_\mathrm{e,-}^{(1)}$ is the lowest 
for $t>0$ and $t'\geq 0$. 
 The excited states become degenerate,
 $E_\mathrm{e,+}^{(1)}=E_\mathrm{o}^{(1)}$,
for an equilateral triangle  
with $\epsilon_\mathrm{apex} = \epsilon_d$ and $t'=t$.
The degeneracy is lifted as the symmetry is lowered  
by a site diagonal distortion  
$\Delta \epsilon \equiv \epsilon_\mathrm{apex}-\epsilon_d$ 
and also by an off-diagonal distortion $\Delta t \equiv t'-t$.
The first order correction is given by 
$E_{\mathrm{e},+}^{(1)} - E_\mathrm{o}^{(1)} 
\,\simeq \, ({2}/{3}) \,\Delta \epsilon - ({4}/{3}) \,\Delta t.$ 
Thus, for $\Delta \epsilon -2 \Delta t >0\,$ ($<0$),  
the energy of the even excited state becomes larger (smaller) than 
that of the odd one. 
It reflects the fact that $\epsilon_\mathrm{apex}$ 
belongs to the even part of the basis, and  
the odd energy $E_\mathrm{o}^{(1)}$ increases with $t'$

\begin{figure}[t]
\begin{minipage}[t]{0.47\linewidth}
\includegraphics[width=1\linewidth]{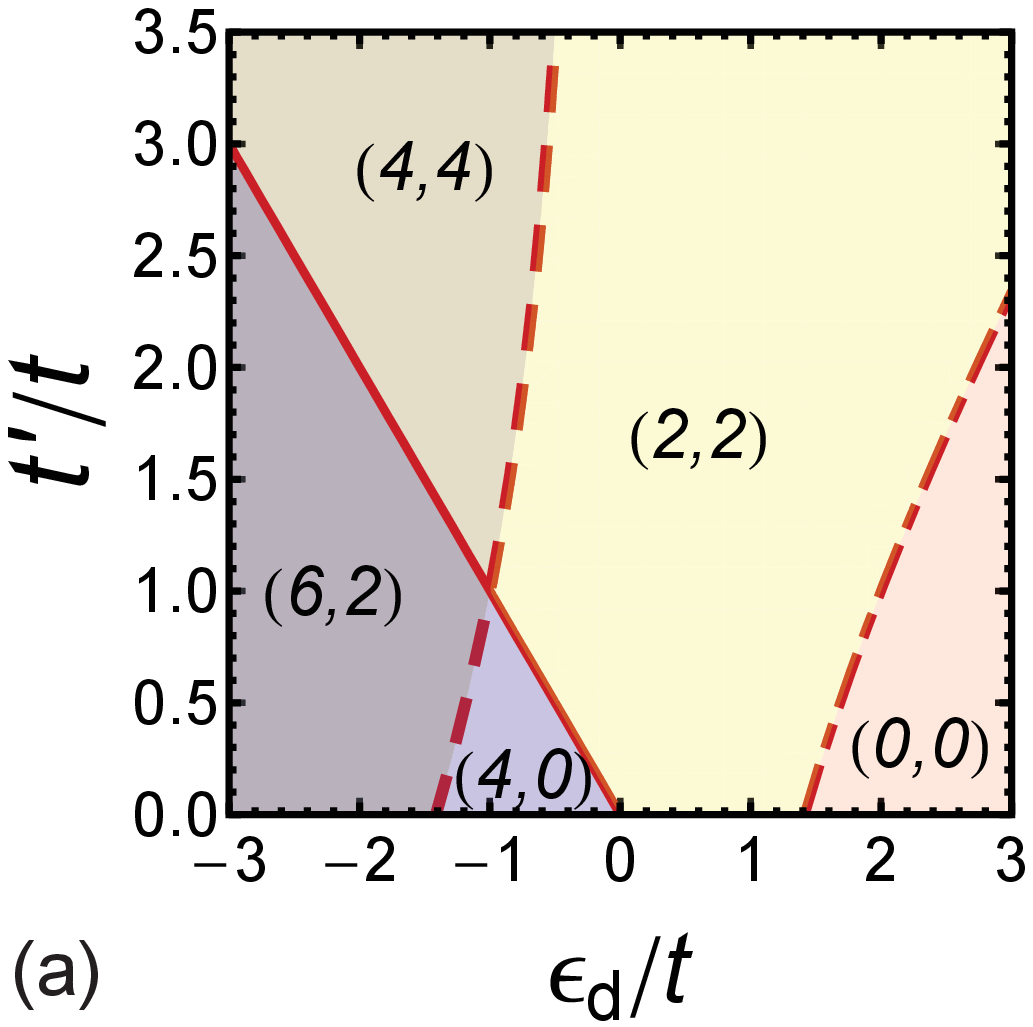}
\end{minipage}
\rule{0.18cm}{0cm}
\begin{minipage}[t]{0.47\linewidth}
\includegraphics[width=1\linewidth]{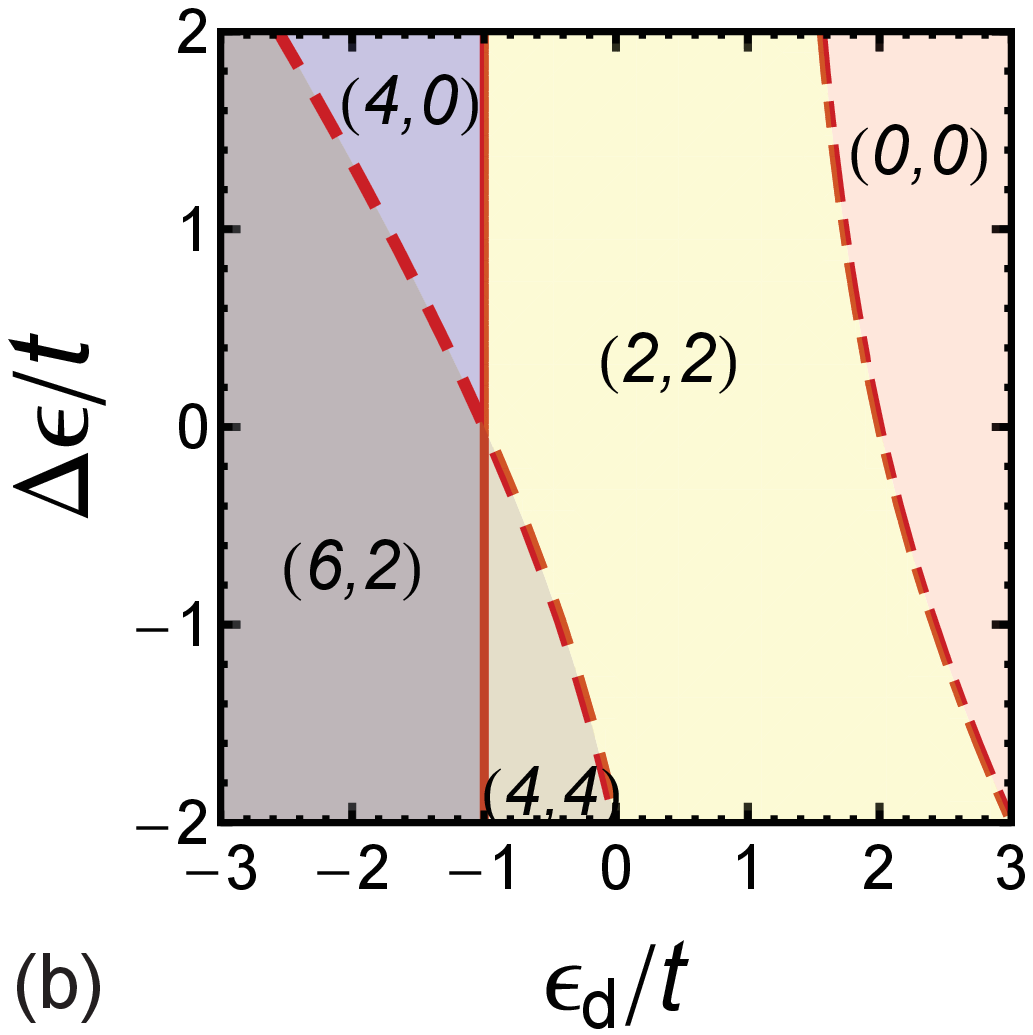}
\end{minipage}
\caption{(Color online) 
Ground-state phase diagram of the isolated TTQD 
for $\Gamma \to 0$ and $U = 0$ is plotted in  
(a) an $\epsilon_d$  vs $t'$ plane 
for $\epsilon_\mathrm{apex}=\epsilon_d$, and  
in (b) an $\epsilon_d$  vs $\Delta \epsilon$ plane 
for $t'=t$.
Here, $\Delta \epsilon \equiv \epsilon_\mathrm{apex}- \epsilon_d$. 
The phase boundaries are given by the contours for 
the energy levels, crossing the Fermi level: 
 $E_{\mathrm{o}}^{(1)}=0$ (solid line), 
 $E_{\mathrm{e},+}^{(1)}=0$ (dashed line), and 
 $E_{\mathrm{e},-}^{(1)}=0$ (dot-dashed line). 
The label ($N_\mathrm{tot}$, $\Theta$) is assigned 
for each region divided by these lines,
and it represents the occupation number of the TTQD $N_\mathrm{tot}$,
 and $\Theta \equiv (\delta_\mathrm{e} -\delta_\mathrm{o})(2/\pi)$, 
which only for $U=0$ coincides with 
the difference in the occupation number 
of the even-parity levels and that of the odd-parity level.
}
 \label{fig:ground_state_isolated_u0}
\end{figure}

The number of the electrons which enter  
the TTQD is determined by the relative position of 
these levels with respect to the Fermi energy $E_F$ ($=0$).
Figure \ref{fig:ground_state_isolated_u0}  
shows the phase diagrams of the ground state 
of the TTQD cluster for $U = 0$ and $\Gamma \to 0$.
The boundaries are determined by 
the condition that the one-particle energy level 
crosses the Fermi energy:  
$E_{\mathrm{o}}^{(1)}=0$ (solid line), 
 $E_{\mathrm{e},+}^{(1)}=0$ (dashed line), and 
 $E_{\mathrm{e},-}^{(1)}=0$ (dot-dashed line).
The  left panel (a) is plotted  as a function of  $\epsilon_d$ and $t'$  
keeping the site diagonal part uniform $\epsilon_\mathrm{apex}=\epsilon_d$.
Similarly,  
the right panel (b) is plotted as a function 
of $\epsilon_d$ and $\Delta \epsilon$  
keeping the inter-dot couplings uniform $t'=t$.  
Therefore, in the case of off-diagonal distortion,
the cluster deforms from the regular triangle to 
a linear chain for $0 \leq t'<t$, 
and then for $t'>t$  
the coupling between the apex site  
and the other two becomes relatively weak 
as $t'$ increases. 
The diagonal distortion 
$\epsilon_\mathrm{apex} \neq \epsilon_d$ affects 
directly the charge density in the apex site,  
and in Fig.\ \ref{fig:ground_state_isolated_u0} (b) 
the contour for the odd level becomes a vertical line 
because $E_{\mathrm{o}}^{(1)}$ 
does not depend on $\epsilon_\mathrm{apex}$.

The occupation number  $N_\mathrm{tot}$, 
varies discontinuously 
as an energy level crosses the Fermi energy.
For finite $\Gamma$,  
it can be deduced from the Friedel sum rule 
 $(\delta_\mathrm{e} + \delta_\mathrm{o})(2/\pi)$. 
As shown in  Fig.\ \ref{fig:ground_state_isolated_u0},
it takes the values 
$N_\mathrm{tot} =0,\,2,\,4$, 
and $6$, depending on the region that is separated by the boundaries.  
The difference in the two phase shifts 
$\Theta \equiv (\delta_\mathrm{e} -\delta_\mathrm{o})(2/\pi)$ 
coincides, for $U=0$, with 
$N_\mathrm{even} -N_\mathrm{odd}$,   
where $N_\mathrm{even}$ and $N_\mathrm{odd}$
are the occupation number for    
the even and odd levels, respectively. 
In the limit of $\Gamma \to 0$, it can take the values 
of $\Theta =0,\,2$ and $4$ in the case of the TTQD, 
as shown in Fig.\ \ref{fig:ground_state_isolated_u0}.
For interacting electrons $U\neq 0$, 
however, there is no such general correspondence 
between $\Theta$ and the charge difference in the even and odd subspaces,  
while the Friedel sum rule remains valid.  
This is because the Coulomb interaction $\mathcal{H}_\mathrm{dot}^U$ 
breaks the charge conservation in each subspace, 
as seen explicitly in Eq.\ \eqref{eq:even_odd}.

We also see in Fig.\ \ref{fig:ground_state_isolated_u0} 
that the odd-parity level $E_{\mathrm{o}}^{(1)}$ (solid line) and 
the excited even-parity level $E_{\mathrm{e},+}^{(1)}$ (dashed line) 
cross each other at $t=t'$ and  $\epsilon_\mathrm{apex}=\epsilon_d$,
where the system has the full symmetry of the equilateral triangle.  
The crossing divides the region of four-electron occupation 
into two different spin-singlet regions, 
which can be classified according to the values 
of $\Theta$. In the region with $\Theta=0$ 
the highest occupied orbital is  
 the odd-parity $b_1$ orbital, 
while in the opposite side with $\Theta=4$ 
the the even-parity orbital with energy 
$E_{\mathrm{e},+}^{(1)}$ becomes the highest occupied orbital. 
Note that the lowest even-parity orbital with energy 
$E_{\mathrm{e},-}^{(1)}$ has already been 
occupied by two electrons in this area of the parameter space.

\begin{figure}[t]
 \leavevmode
 \begin{minipage}[t]{0.5\linewidth}
  \includegraphics[width=1\linewidth]{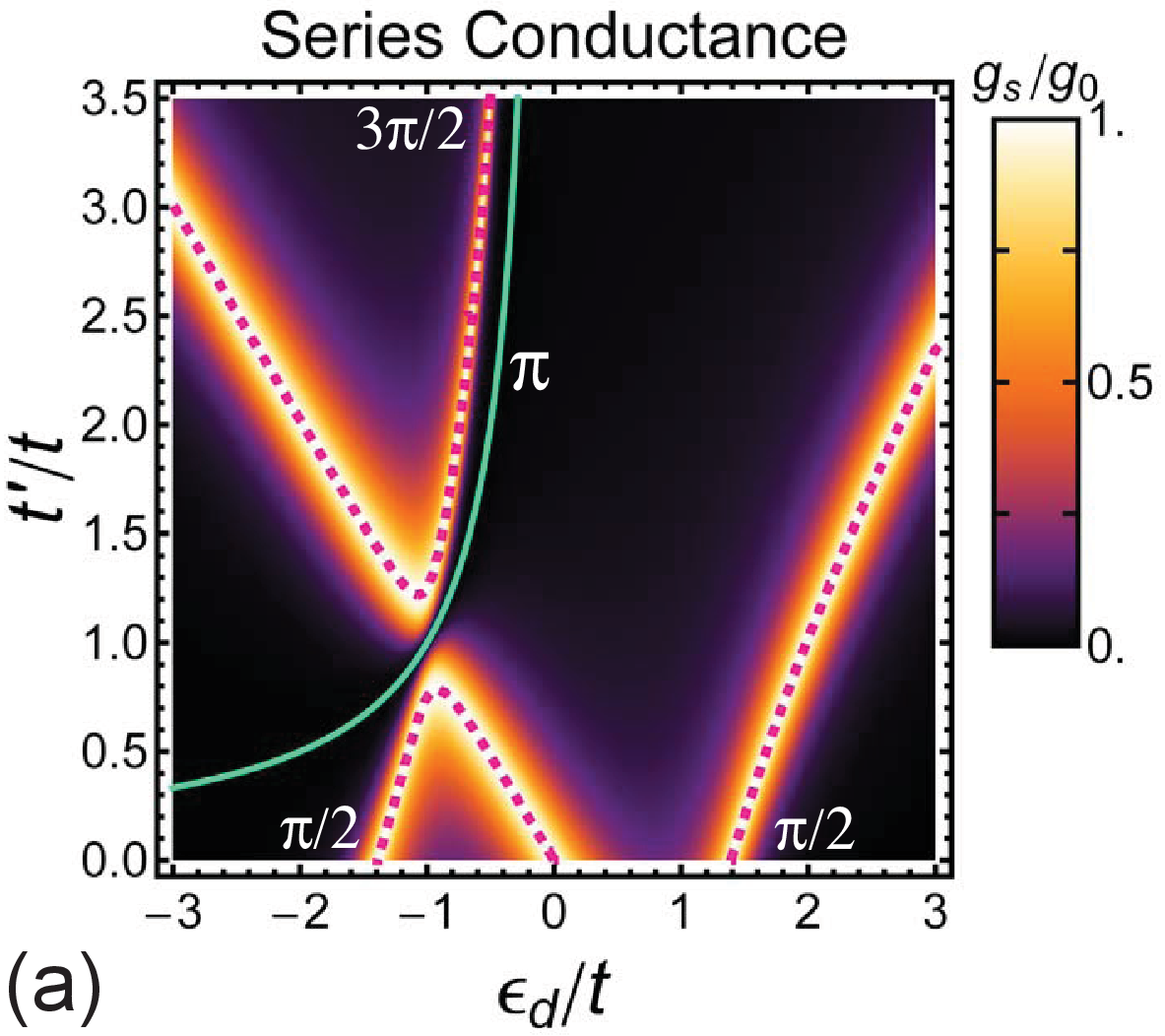}
 \end{minipage}
 \begin{minipage}[t]{0.485\linewidth}
  \includegraphics[width=1\linewidth]{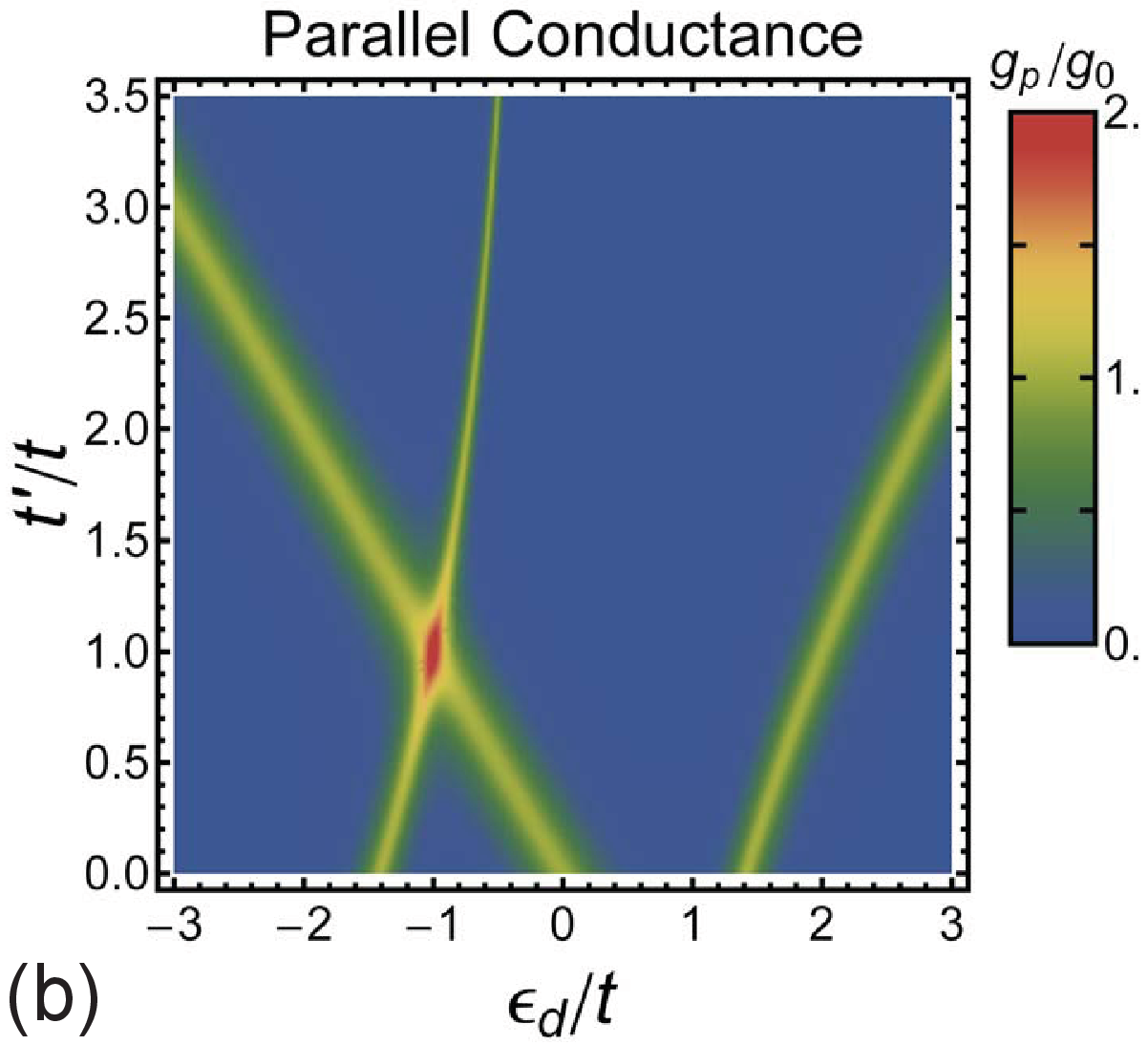}
 \end{minipage}
\caption{(Color online) 
Series (a) and parallel (b) conductances  
in the noninteracting case  $U = 0.0$ 
as functions 
of $\epsilon_d/t$ and $t'/t$, for 
$\Gamma/t =0.25$ and
$\epsilon_\mathrm{apex} = \epsilon_d$.
The values of the conductances are scaled 
by $g_0 \equiv 2e^2/h$.
In (a) the solid line denotes the contour for 
$\delta_\mathrm{e}- \delta_\mathrm{o} = \pi$,
and the dashed lines are the contours 
for the values of $\pi/2$ and $3\pi/2$. 
}
 \label{fig:conductance_ttqd_u0}
\end{figure}


\subsection{Conductance for $U=0$}

We next consider the noninteracting TTQD 
which are connected to the leads in a 
series or parallel configurations, as shown in Fig.\ \ref{fig:system}. 
In this case, the conduction electrons from the leads are  
 scattered at the TTQD.
The phase shifts $\delta_\mathrm{e}$ and  $\delta_\mathrm{o}$, 
caused by the scattering, are given by 
Eq.\ \eqref{eq:phase_shift_renormalized},
replacing the renormalized parameters there  
by the bare ones $t$, $t'$, and $\epsilon_\mathrm{apex}$. 
The conductance can be deduced from these phase shifts 
through Eqs.\ \eqref{eq:gs} and \eqref{eq:gp},
or equivalently from the Green's function using 
Eqs.\ \eqref{eq:cond_s} and 
\eqref{eq:def_g_p_Green} given in Appendix \ref{sec:app_green}.

\begin{figure}[t]
 \leavevmode
\begin{minipage}[t]{0.5\linewidth}
  \includegraphics[width=1\linewidth]{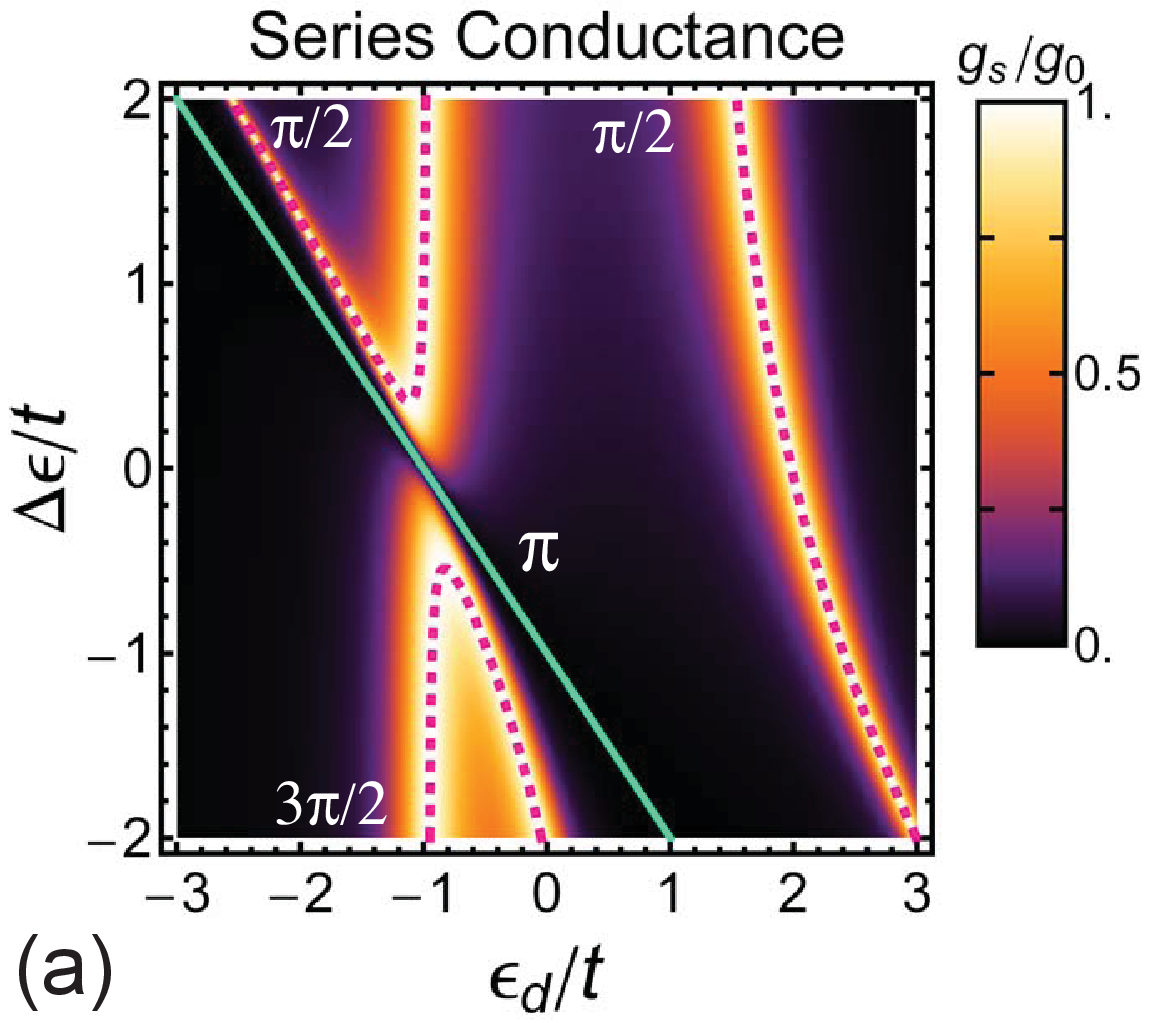}
 \end{minipage}
 \begin{minipage}[t]{0.48\linewidth}
  \includegraphics[width=1\linewidth]{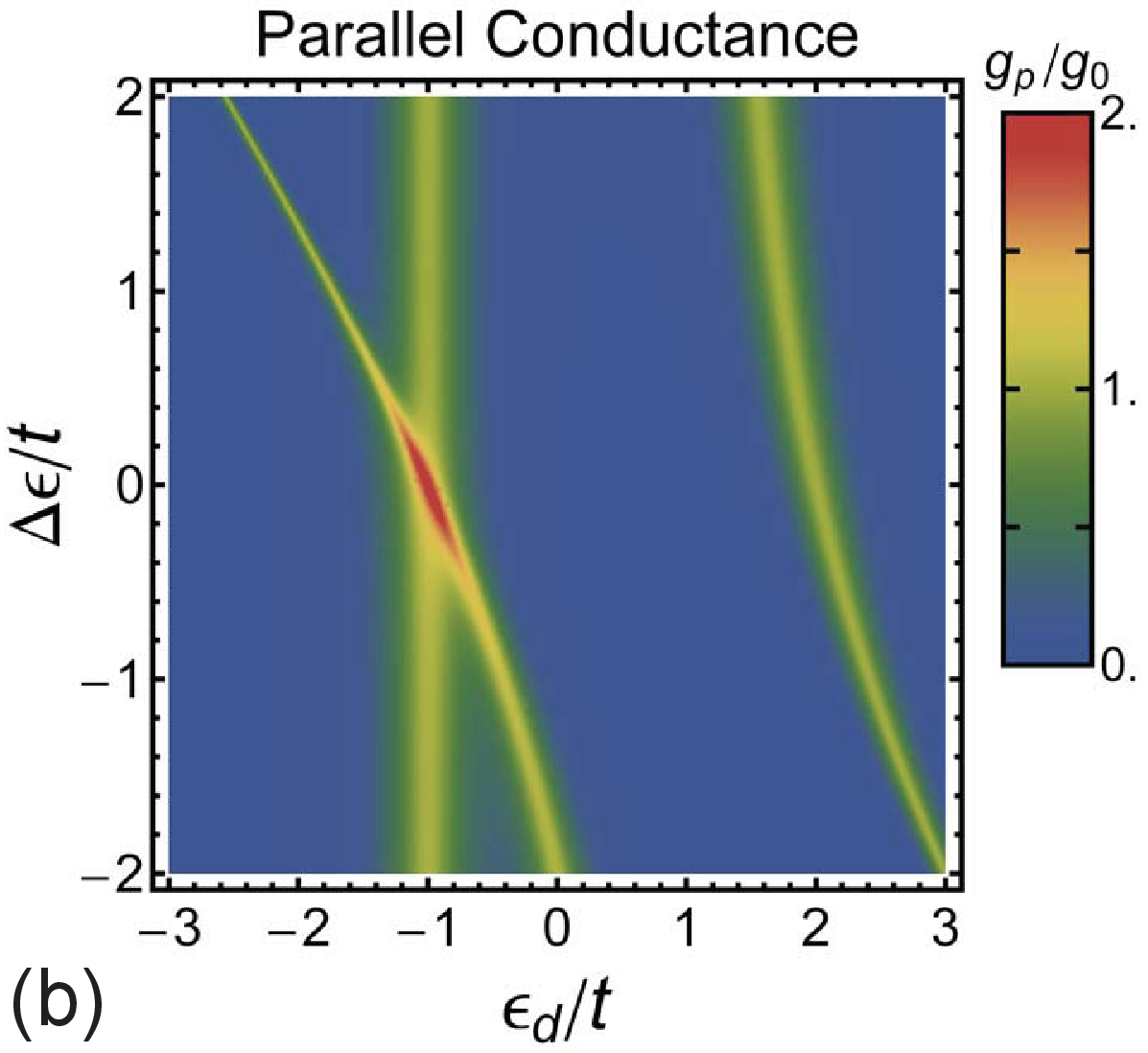}
 \end{minipage}
 \caption{(Color online) 
Series (a) and parallel (b) conductances  
in the noninteracting case  $U = 0.0$ 
as functions of $\epsilon_d/t$ and $\Delta \epsilon/t$,  
for $\Gamma/t =0.25$ and  $t'=t$. 
Here, $\Delta \epsilon \equiv \epsilon_\mathrm{apex}- \epsilon_d$. 
The values of the conductances are scaled 
by $g_0 \equiv 2e^2/h$.
In (a) the solid line denotes the contour for 
$\delta_\mathrm{e}- \delta_\mathrm{o} = \pi$,
and the dashed lines are the contours 
for the values of $\pi/2$ and $3\pi/2$. 
}
 \label{fig:conductance_deltaE_u0}
\end{figure}

The  series and parallel conductances 
in the  non-interacting case $U=0$ are plotted  
in Fig.\  \ref{fig:conductance_ttqd_u0} 
as functions of $\epsilon_d/t$ and $t'/t$ 
for $\Gamma/t =0.25$ keeping the onsite potential 
for the three dots to be the same  $\Delta \epsilon=0$.
Similarly, in Fig.\ \ref{fig:conductance_deltaE_u0}
the conductances are plotted as  
functions of $\epsilon_d/t$ and $\Delta \epsilon/t$,
taking the inter-dot hopping matrix elements to be uniform $t'=t$.  
The figures \ref{fig:conductance_ttqd_u0} 
and \ref{fig:conductance_deltaE_u0}  
can be compared, respectively, to the phase diagrams given in 
Fig.\ \ref{fig:ground_state_isolated_u0} (a) and (b).

Both $g_\mathrm{s}$ and $g_\mathrm{p}$ 
are enhanced as the resonance levels 
which correspond to the one-particle energies defined in 
Eq.\ \eqref{eq:U0_eigen_cluster1} and \eqref{eq:U0_eigen_cluster2} 
cross the Fermi level.   
The series and parallel conductances  
 show a similar behavior in most of the parameter regions.  
We can see, however, that they show 
a quite different behavior at the point $t=t'$ and $\Delta \epsilon=0$,
where the series conductance vanishes   
while the parallel conductance takes the maximum possible value 
$g_\mathrm{p}=4e^2/h$ for two conducting channels.
At this point, the two one-particle levels 
 $E_{\mathrm{e},+}^{(1)}$ and $E_{\mathrm{o}}^{(1)}$ cross the 
Fermi level simultaneously, and 
the phase shifts take the value   
$\delta_\mathrm{e}=3\pi/2$ and $\delta_\mathrm{o}=\pi/2$.

The solid line in Fig.\ \ref{fig:conductance_ttqd_u0} (a) 
and Fig.\ \ref{fig:conductance_deltaE_u0}  (a),     
denotes the contour of the difference in the two phase shifts 
for the value  
$\delta_\mathrm{e}-\delta_\mathrm{o}=\pi$.
Thus,  this contour corresponds to 
a zero line of the series conductance,
and it means that  destructive interference 
is most pronounced along this line. 
Similarly, the dashed lines 
in Fig.\ \ref{fig:conductance_ttqd_u0} (a)
and Fig.\ \ref{fig:conductance_deltaE_u0} (a) 
are the contours for  
$\delta_\mathrm{e}-\delta_\mathrm{o}=\pi/2$ and $\,3\pi/2$, 
on which the two conductances show  peaks 
of the same height  $g_\mathrm{s}=2e^2/h$ and  $g_\mathrm{p}=2e^2/h$.

We can also see in Figs.\ \ref{fig:conductance_ttqd_u0} 
and \ref{fig:conductance_deltaE_u0}  
that some conductance peaks are sharp  
and the others are relatively wide.
Particularly, the resonance peak for    
the excited even-parity level  
$E_{\mathrm{e},+}^{(1)}$, 
which corresponds to the dashed line in 
 Fig.\ \ref{fig:ground_state_isolated_u0} (a) and (b), 
is much sharper than the other peaks. This is because 
the eigenstate for $E_{\mathrm{e},+}^{(1)}$ 
has a large spectral weight at the apex site  
which has no direct couplings to the leads, 
and thus the hybridization with conduction band is suppressed. 
This feature can also be seen in 
the explicit expression for 
the spectral weight for the noninteracting TTQD 
is given in Eq.\ 
\eqref{eq:peak_even_excited}
 in Appendix \ref{sec:even_odd}.
Conversely, the resonance width becomes large for 
the local states, the spectral weight of which is mainly 
on the other two dots coupled directly to the leads.

\section{
Ground state in the molecular limit: $\Gamma=0$ and $U\neq 0$}
\label{sec:molecule_limit}

The model can be solved also for finite Coulomb interaction $U$ 
in the {\it molecular} limit $\Gamma \to 0$.\cite{Numata2,KorkusinskiGimenezHawrylakETAL}
In this case the TTQD is disconnected from the leads, 
and  described by the Hamiltonian,
\begin{align}
\mathcal{H}_\mathrm{dot} \equiv 
\mathcal{H}_\mathrm{dot}^0 + \mathcal{H}_\mathrm{dot}^U \;.
\end{align}
The eigenstates of $\mathcal{H}_\mathrm{dot}$ determine 
the high-energy properties of the system, 
particularly the properties of the local excitations near the quantum dots.
It gives us a knowledge how 
the parameter space could be classified;
such a classification relates to the fixed points 
of the renormalization group.  
In this section we examine the effects of 
the distortions 
on the local electronic states in the interacting case.

 \begin{figure}[t]
  \leavevmode
 \begin{minipage}[t]{0.475\linewidth}
  \includegraphics[width=1\linewidth]{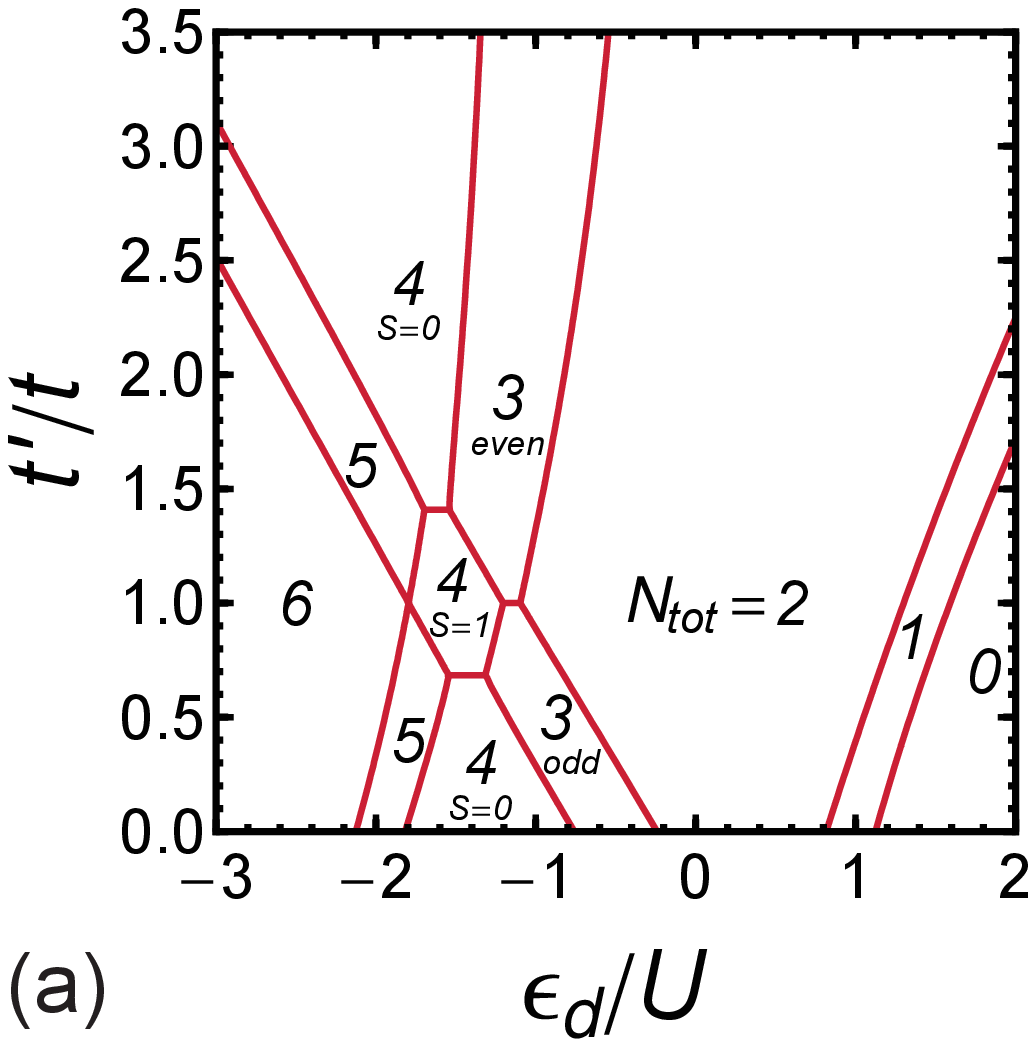}
 \end{minipage}
 \begin{minipage}[t]{0.49\linewidth}
  \includegraphics[width=1\linewidth]{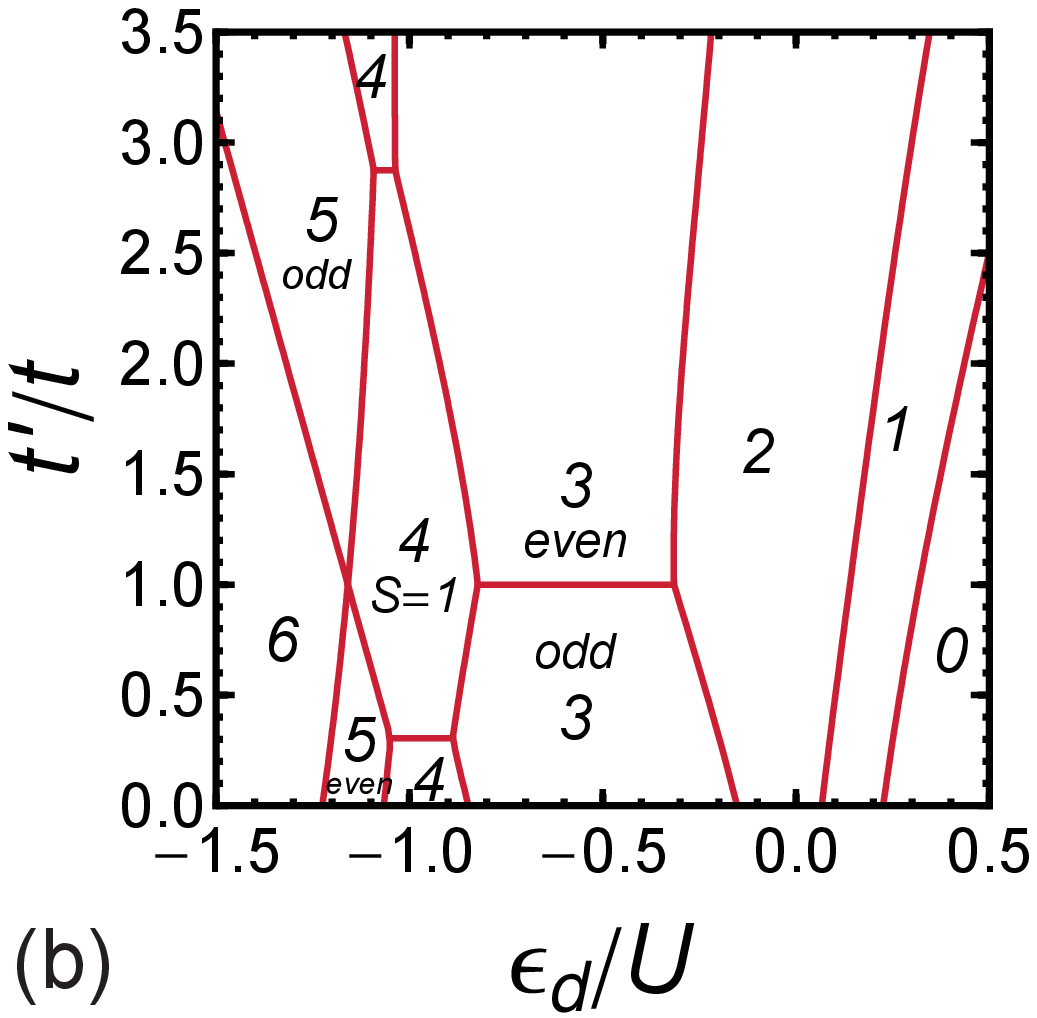}
 \end{minipage}
  \caption{(Color online) 
 Ground-state phase diagram of the TTQD for $\Gamma=0$ 
is plotted 
in a $\epsilon_d/U$ vs $t'/t$ plane for $\Delta \epsilon =0$.
The Coulomb interaction is chosen to be 
 (a) $U/(2\pi t) = 0.2$, and (b)  $U/(2\pi t) = 1.0$.
The occupation number $N_\mathrm{tot}$ 
is shown for each region. 
The total spin $S$  is $1/2$ for odd $N_\mathrm{tot}$,
and $S=0$ for even $N_\mathrm{tot}$, 
except it becomes $S=1$ 
in the middle of three $N_\mathrm{tot}=4$ regions. 
Along the horizontal line in the $N_\mathrm{tot}=3$ region, 
the ground state has an SU(4) symmetry 
due to the parity and spin degeneracies. 
The eigenstate has an even (odd) parity  
above (below) this horizontal line.
}
  \label{fig:ground_state_isolated_u1_ts}
 \end{figure}

Figures  
\ref{fig:ground_state_isolated_u1_ts} and 
\ref{fig:ground_state_isolated_u1_de} 
show the phase diagram of the ground state 
of the isolated TTQD for $\Gamma \to 0$.
The Coulomb interaction is chosen to be 
 (a) $U/(2\pi t) =0.2$,  and (b) $U/(2\pi t) =1.0$.  
These figures can be compared with
the phase diagrams in the non-interacting case 
shown in Fig.\ \ref{fig:ground_state_isolated_u0}. 
We can see that 
the eigenstate with an odd-number of electrons     
($N_\mathrm{tot}=1,\,3,$ and $5$)   
and total spin $S=1/2$ becomes a ground state 
due to the Coulomb interaction.
The odd-number electron regions 
emerge between the even-number electron regions 
in the parameter space, and become wider as $U$ increases.
We can also see that three horizontal border lines  
appear in the phase diagrams for $U>0$, 
and along each line  a level crossing takes place  
between the two different eigenstates 
with the same occupation number.

The ground state   
for $N_\mathrm{tot}=3$ is separated by one of these horizontal lines 
at $t'=t$ in  Fig.\ \ref{fig:ground_state_isolated_u1_ts}, 
and similarly by the one at $\Delta \epsilon = 0$ 
in Fig.\ \ref{fig:ground_state_isolated_u1_de}. 
The ground state 
has an SU(4) symmetry along this  border  
due to the orbital degeneracy 
caused by the symmetry of the equilateral triangle 
and the spin degeneracy.  
Away from this horizontal line,   
the distortions 
lower the equilateral symmetry,  
and lift the orbital degeneracy.
An even-parity (odd-parity) state becomes the ground state 
for $t'>t$ ($t'<t$)  
in Fig.\ \ref{fig:ground_state_isolated_u1_ts}. 
Correspondingly, in Fig.\ \ref{fig:ground_state_isolated_u1_de} 
the ground state is an even-parity (odd-parity) state
for  $\epsilon_\mathrm{apex} < \epsilon_d$ 
 ($\epsilon_\mathrm{apex} > \epsilon_d$).
Note that there are some similarities between  
the phase diagrams in Fig.\ \ref{fig:ground_state_isolated_u1_ts} 
and  Fig.\ \ref{fig:ground_state_isolated_u1_de}:
the features seen for $t'>t$  ($t'<t$) 
are similar qualitatively (graphically) to  
those for $\Delta \epsilon <0$ ($\Delta \epsilon >0$).
This is because the two types of the distortion, 
$t'/t$  and $\epsilon_\mathrm{apex}-\epsilon_d$,  
lift the degeneracy in an opposite way, 
as mentioned in the above 
with Eqs.\ \eqref{eq:U0_eigen_cluster1} and \eqref{eq:U0_eigen_cluster2}.

The Coulomb interaction also causes the high spin $S=1$ 
ground state 
seen in Figs.\ \ref{fig:ground_state_isolated_u1_ts} and 
\ref{fig:ground_state_isolated_u1_de} 
in the middle of the $N_\mathrm{tot}=4$ regions      
where the TTQD has one extra electron introduced 
into the  half-filled cluster.  
The $S=1$ region evolves in the parameter space 
from the level crossing point for $U=0$,
seen in Fig.\ \ref{fig:ground_state_isolated_u0} 
at the point of
$t'=t$  and $\epsilon_\mathrm{apex}=\epsilon_d$.  
 The degeneracy at this level crossing point is 
lifted for infinitesimal $U$,
and 
the high-spin state evolves continuously, 
as $U$ increases, to the Nagaoka ferromagnetic state 
which is  usually defined  in the large $U$ limit. 
For large distortions,
however, the transition to a singlet ground state takes place 
on the horizontal lines, running on the top and bottom 
of the $S=1$ region 
in Figs.\ \ref{fig:ground_state_isolated_u1_ts} and 
\ref{fig:ground_state_isolated_u1_de}.

The isolated TTQD which is not connected to the leads 
has a local moment of $S=1/2$  for odd-number fillings, 
and a high-spin $S=1$ for $N_\mathrm{tot}=4$, 
as mentioned in the above.  
In the case where two leads are coupled to the cluster, 
however, the local moment is screened eventually 
at low energies by the conduction electrons tunneling 
from the leads, and the ground state 
of the whole system becomes a spin singlet.
 We show the results of the ground-state properties 
of the TTQD connected to the leads in the next section, 
and then discuss also the characteristic energy scale of 
the Kondo screening in Sec.\ \ref{sec:TK}.

\section{
NRG results for ground-state properties of the interacting TTQD 
} 
\label{sec:results_I}

 \begin{figure}[t]
  \leavevmode
  \begin{minipage}[t]{0.475\linewidth}
   \includegraphics[width=1\linewidth]{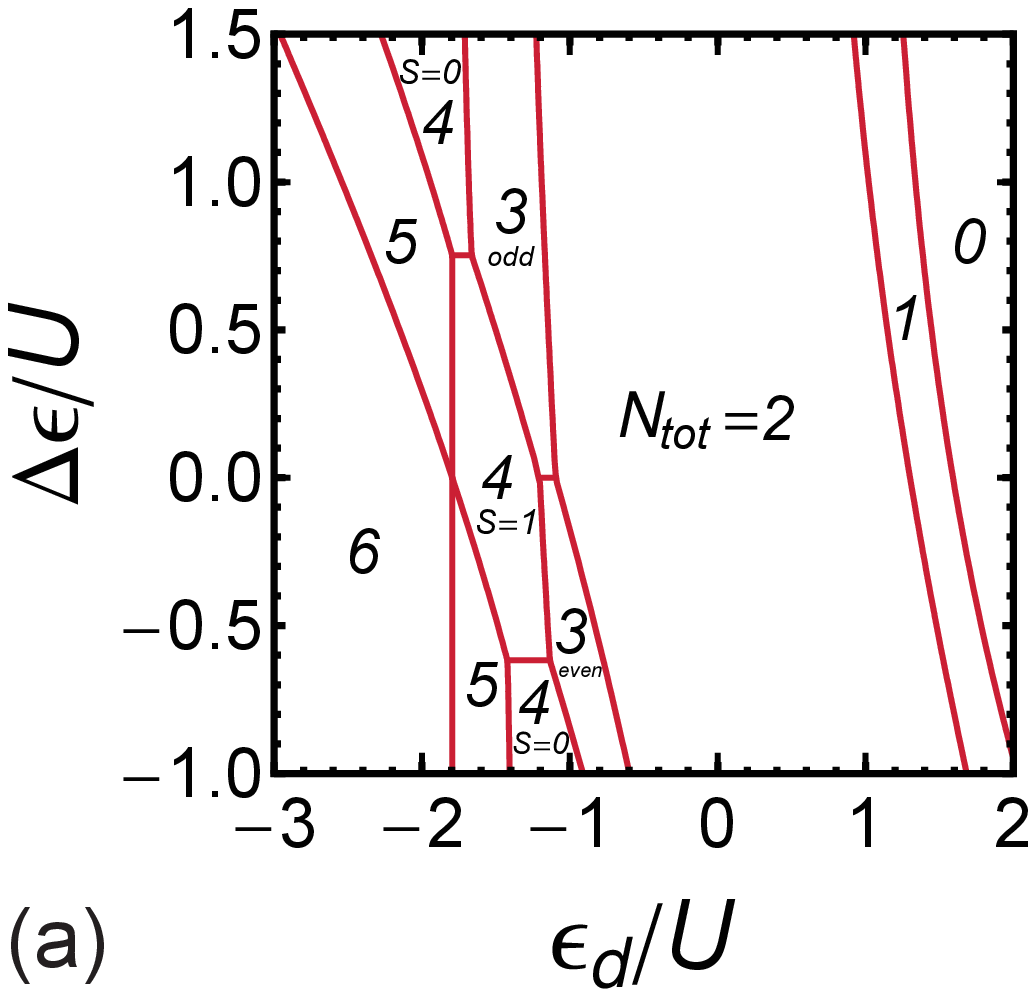}
  \end{minipage}
 \begin{minipage}[t]{0.49\linewidth}
   \includegraphics[width=1\linewidth]{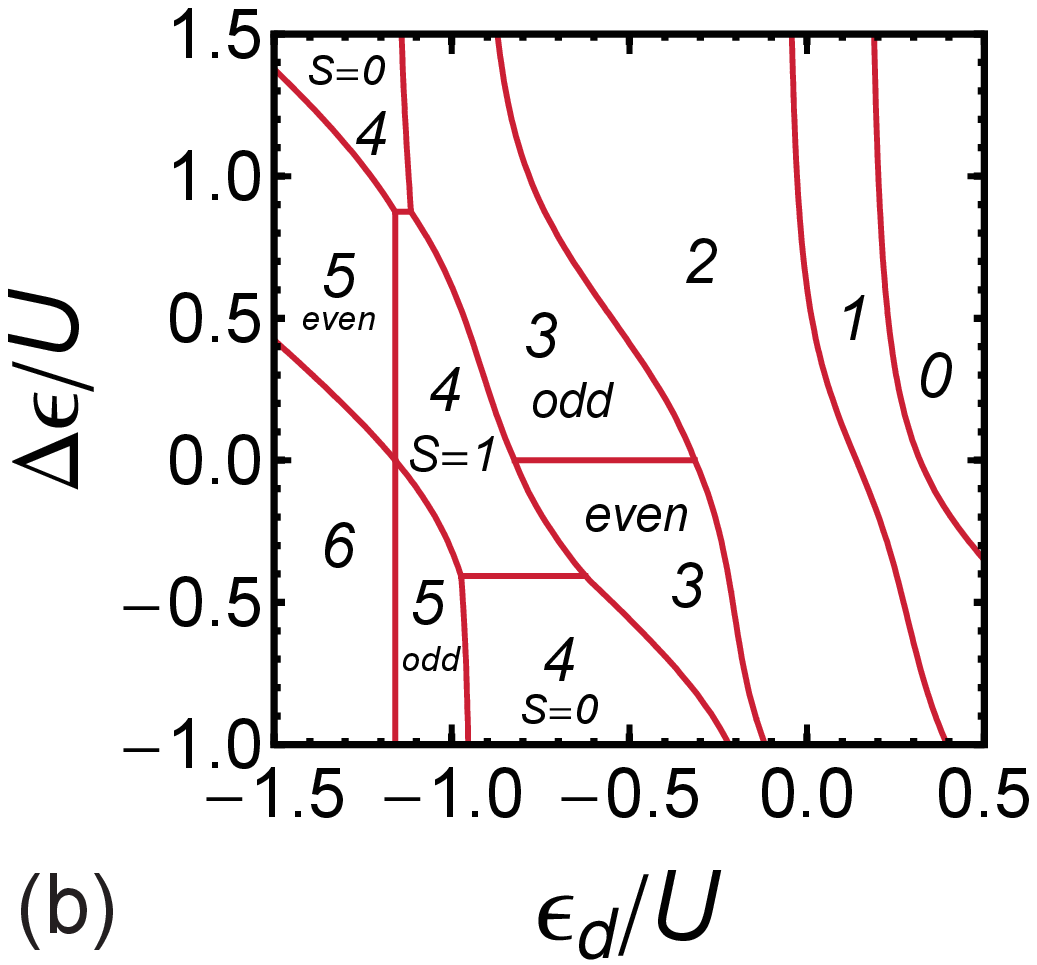}
  \end{minipage}
  \caption{(Color online)  
 Ground-state phase diagram of the TTQD for $\Gamma=0$ 
is plotted 
in a $\epsilon_d/U$  vs $\Delta \epsilon/U$ plane 
for $t' = t$.
The Coulomb interaction is chosen to be 
 (a) $U/(2\pi t) = 0.2$, and (b) $U/(2\pi t) = 1.0$.
The occupation number $N_\mathrm{tot}$ 
is shown for each region. 
The total spin $S$  is $1/2$ for odd $N_\mathrm{tot}$,
and $S=0$ for even $N_\mathrm{tot}$, 
except it becomes $S=1$ 
in the middle of three $N_\mathrm{tot}=4$ regions. 
Along the horizontal line in the $N_\mathrm{tot}=3$ region, 
the ground state has an SU(4) symmetry 
due to the parity and spin degeneracies.
The eigenstate has an odd (even) parity  
above (below) this horizontal line.
}
  \label{fig:ground_state_isolated_u1_de}
 \end{figure}

We now consider an interacting TTQD coupled 
to two leads  
via tunneling matrix elements  
$\mathcal{H}_\mathrm{mix}$ defined in 
Eq.\  \eqref{eq:Hmix}.  
In this case the phase shifts 
 $\delta_\mathrm{e}$ and  $\delta_\mathrm{o}$ 
play an central role on the low-energy properties. 
The effects of the Coulomb interaction enter 
through these two phase shifts, 
which can be expressed in terms of the 
renormalized parameters for the 
quasi-particles of the local Fermi liquid,
as described in  Eq.\ \eqref{eq:phase_shift_renormalized}.

 We have calculated the many-body phase shifts 
using the NRG method,\cite{Numata2,hewsonEPJ}
and have deduced the conductance and the occupation number of the TTQD 
at zero temperature from the phase shifts,\cite{Numata2} 
using Eqs.\  \eqref{eq:gs}--\eqref{eq:gp}.
In our calculations, 
the ratio of the inter-dot hopping matrix element $t$ 
and the half width of the conduction band $D$, 
 defined in Appendix \ref{sec:NRG_approach},
is chosen to be $t/D =0.1$.
The iterative diagonalization has been carried out 
by using the even-odd basis, 
described in Appendix \ref{sec:even_odd}.\cite{Numata2}
For constructing the Hilbert space in each NRG step, 
instead of adding two orbitals from even and odd orbitals simultaneously, 
we add one orbital from the even part first and retain  
3600 low-energy states after carrying out the diagonalization 
of the Hamiltonian. Then, we add the other orbital from the odd part, 
and again keep the lowest 3600 eigenstates after the diagonalization. 
The discretization parameter is chosen to be $\Lambda=6.0$, 
which has been confirmed to reproduce the noninteracting results 
with a sufficient accuracy.\cite{OH,ONH,NO}
In the following, we set the strength of the 
 Coulomb interaction to be $U/(2 \pi t) =1.0$,
which is adequate for observing typical results  
caused by $U$, as seen in Figs.\
  \ref{fig:ground_state_isolated_u1_ts} and 
  \ref{fig:ground_state_isolated_u1_de}.
We have carried out some calculations 
changing $U$ and $\Gamma$ for the equilateral triangle  
in the previous work.\cite{Numata2} 
Our results have clarified how the value of $U$ affects 
the width of the Kondo ridges and the Kondo energy scale. 
Furthermore, a large $\Gamma$ smears 
the electronic structure of the TTQD origin.
Through these observations, we have confirmed that 
 the characteristic feature of 
the Kondo behavior can be seen clearly 
for typical a parameter set of  $U/(2 \pi t) =1.0$ and  
$\Gamma/t = 0.12,\, 0.25$.

\subsection{Off-diagonal distortions: $\,t' \neq t$}
\label{subsec:off-diagonal}

We discuss in this subsection 
the transport properties in the presence of the 
off-diagonal distortion $t' \neq t$ keeping 
the site-diagonal potential uniform $\Delta \epsilon =0$. 
The effects of the diagonal 
distortion $\epsilon_\mathrm{apex} \neq \epsilon_d$ 
are examined in the next subsection \ref{subsec:diagonal}.

\subsubsection{Local charge: 
$\,N_\mathrm{tot} =
\frac{\displaystyle\protect\mathstrut 2}{\displaystyle\protect\mathstrut \pi}
(\delta_\mathrm{e}+\delta_\mathrm{o})\,$ 
for $\,t\neq t'$
}

\begin{figure}[t]
 \leavevmode
 \begin{minipage}[t]{0.5\linewidth}
 \includegraphics[width=1\linewidth]{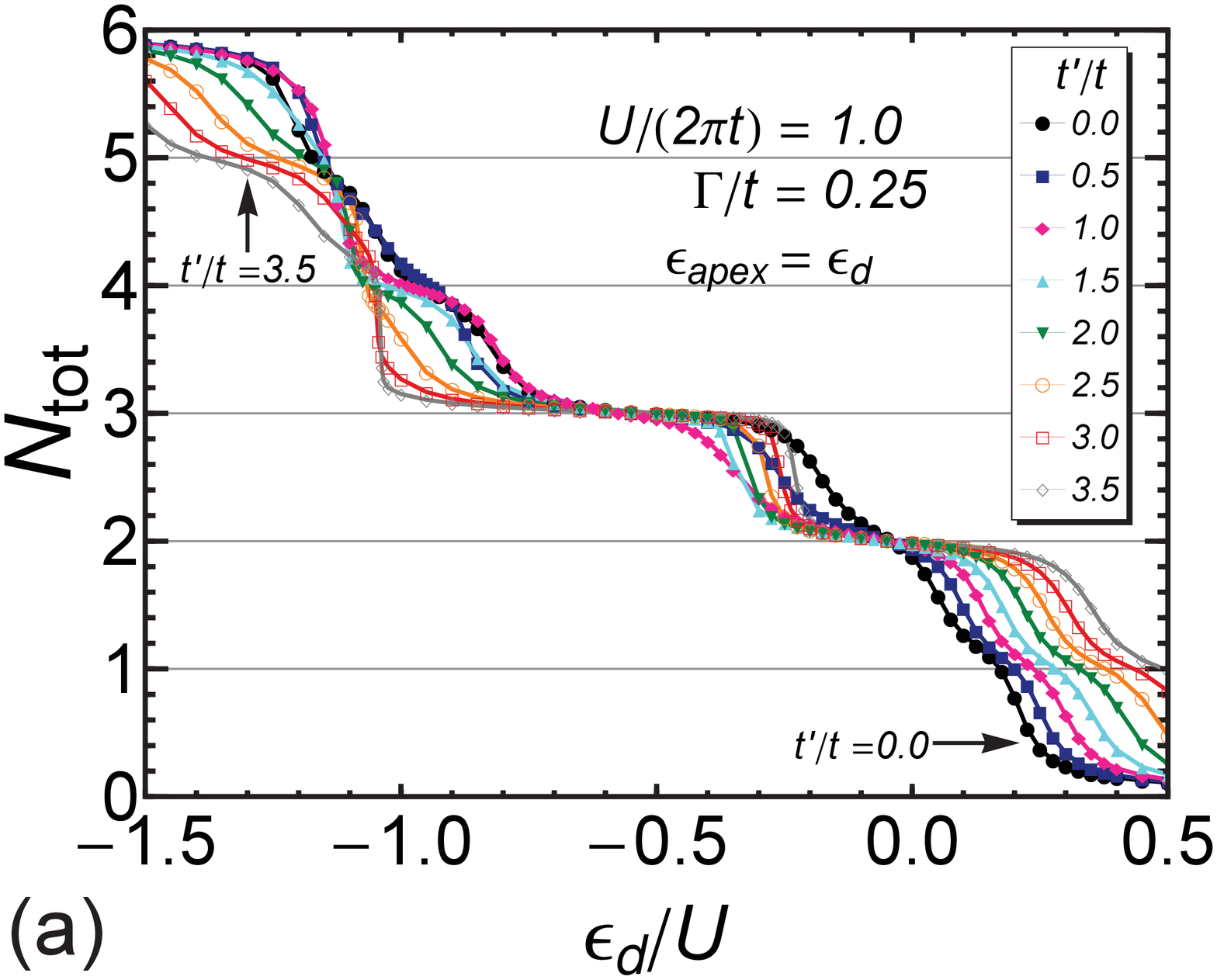}
 \end{minipage}
\rule{0.01\linewidth}{0cm}
 \begin{minipage}[t]{0.465\linewidth}
 \includegraphics[width=1\linewidth,clip,trim = 0.1cm 0cm 0cm 0cm]{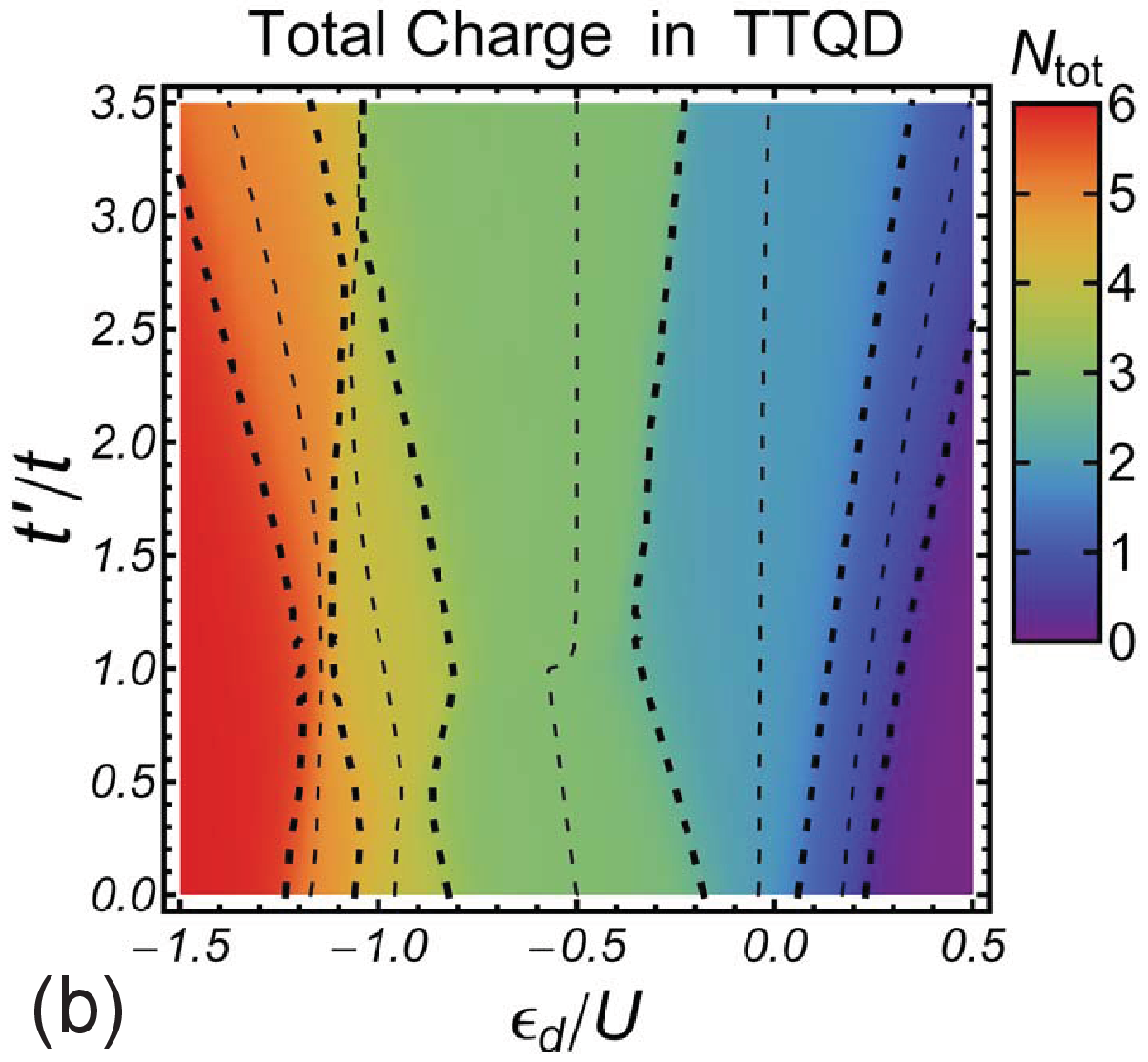}
 \end{minipage}
 \caption{(Color online) 
NRG results for the occupation number 
$N_\mathrm{tot}$ for $U/(2\pi t) = 1.0$,  
 $\Gamma/t=0.25$, and $\Delta \epsilon =0$.
The left panel (a) shows $N_\mathrm{tot}$  
as a function of $\epsilon_d/U$  for  
$t'/t=0.0,\, 0.5,\, 1.0, \ldots,$ and $3.5$. 
The right panel (b) shows $N_\mathrm{tot}$ in the 
 $\epsilon_d/U$ vs $t'/t$ plane. 
The dotted lines  in (b) are the contours for    
 $N_\mathrm{tot} = 0.5,\,1.0,\,1.5,\,\ldots,$ and $5.5$ 
(in steps of $0.5$ from the right to the left).
}
 \label{fig:nd_u1_ts}
\end{figure}

Figure  \ref{fig:nd_u1_ts} shows the NRG results of 
the occupation number $N_\mathrm{tot}$ 
 for $\Gamma/t=0.25$, and $U/(2\pi t) =1.0$.
In (a),  the results are plotted 
as a function of $\epsilon_d/U$ for 
several of values of $t'/t$ 
($=0.0,\,0.5,\,\ldots,$ and $3.5$, in steps of $0.5$).
We can see clearly that the plateaus emerge near 
integer values of $N_\mathrm{tot}$ due to the Coulomb interaction, 
especially the one for $N_\mathrm{tot} \simeq 3.0$ 
becomes almost flat for large $t'/t \gtrsim 3.0$. 
These results show that  
the coupling  strength $\Gamma =0.25 t$ 
is small enough to distinguish the different charge 
states for $U=2 \pi t$. 
The plateau for the five-electron filling 
emerges  due to the distortion 
and it becomes wider as $t'/t$ deviates from $1.0$.
These features are consistent with 
that for the  isolated TTQD discussed in the above.

We have carried out the calculations more densely 
for a number of points in the parameter space 
than those presented in Fig.\  \ref{fig:nd_u1_ts} (a),  
and the results are plotted in the $\epsilon_d/U$ vs $t'/t$ plane 
in Fig.\  \ref{fig:nd_u1_ts} (b). 
The dotted lines are the contours for    
 $N_\mathrm{tot} = 0.5,\,1.0,\,1.5,\,\ldots,$ and $5.5$ 
(in steps of $0.5$ from the right to the left).
Note that this figure can be compared with
Fig.\ \ref{fig:ground_state_isolated_u1_ts} (b)
 where $N_\mathrm{tot}$ for $\Gamma \to 0$ is shown.
We can see in Fig.\  \ref{fig:nd_u1_ts} (b)
that the contours of $N_\mathrm{tot}$ for half integers  
($0.5$, $1.5$, \ldots, and $5.5$), 
which are shown with the thicker dotted lines, 
follow almost faithfully the phase boundaries between the 
different charge states for $\Gamma \to 0$ 
shown 
 in Fig.\ \ref{fig:ground_state_isolated_u1_ts} (b).
The electron filling $N_\mathrm{tot}$ changes rapidly near 
these contours for the half integers. 
This can be seen explicitly in Fig.\  \ref{fig:nd_u1_ts} (a), 
and it reflects the fact that 
$\Gamma$ is much smaller than the inter dot matrix 
elements $t$ and the Coulomb interaction $U$ in the present case. 
From these observations, we see that  
the charge distribution in the plateaus regions 
is almost completely determined by the high energy states, 
and it can be approximated by 
the one in the limit of $\Gamma \to 0$. 
The low-lying energy states are required, however, 
to describe correctly the transport properties and 
the conduction-electron screening of the local moment 
of the TTQD.

\begin{figure}[t]
 \leavevmode
 \begin{minipage}[t]{0.49\linewidth}
  \includegraphics[width=1\linewidth]{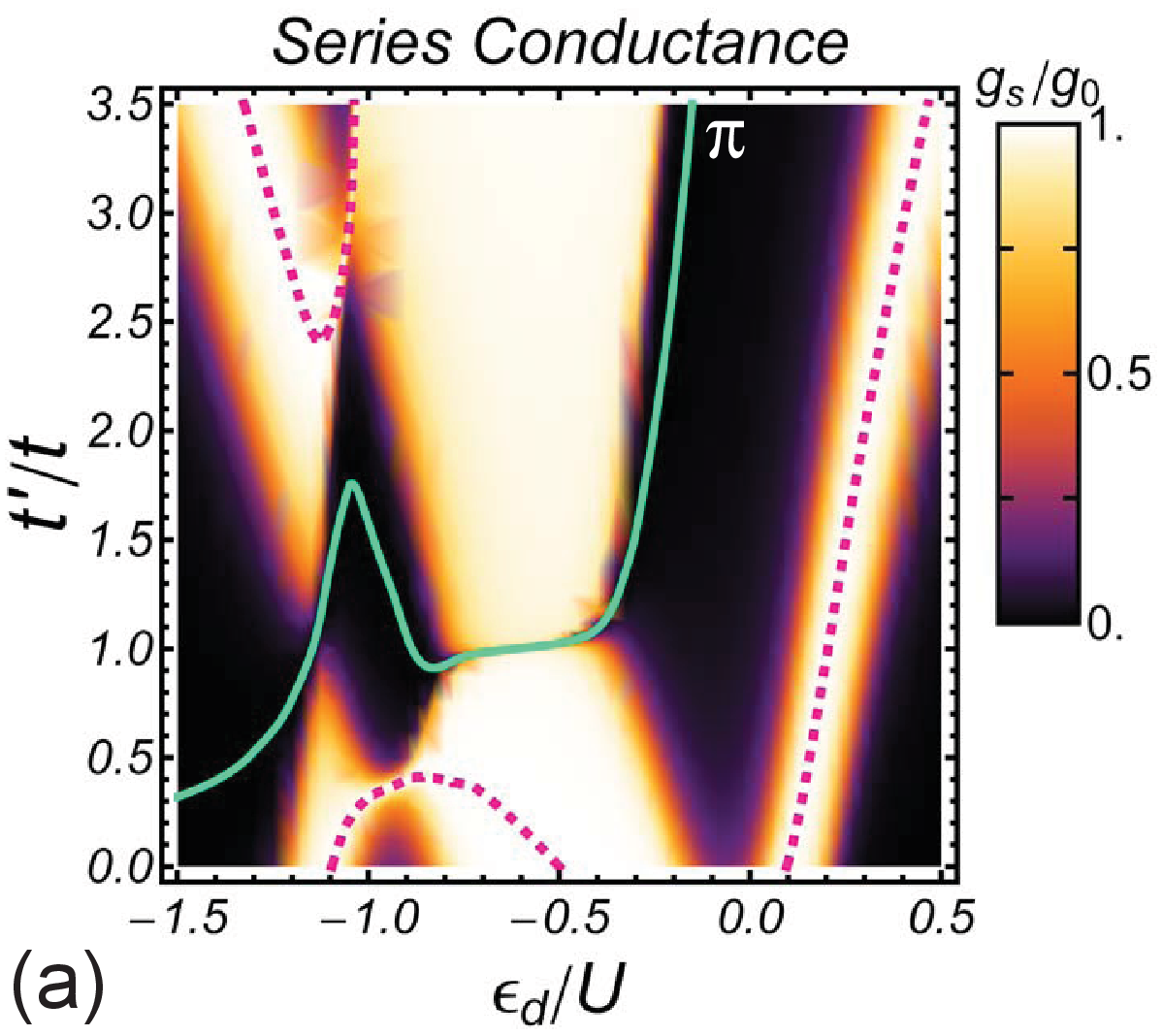}
 \end{minipage}
 \begin{minipage}[t]{0.48\linewidth}
  \includegraphics[width=1\linewidth]{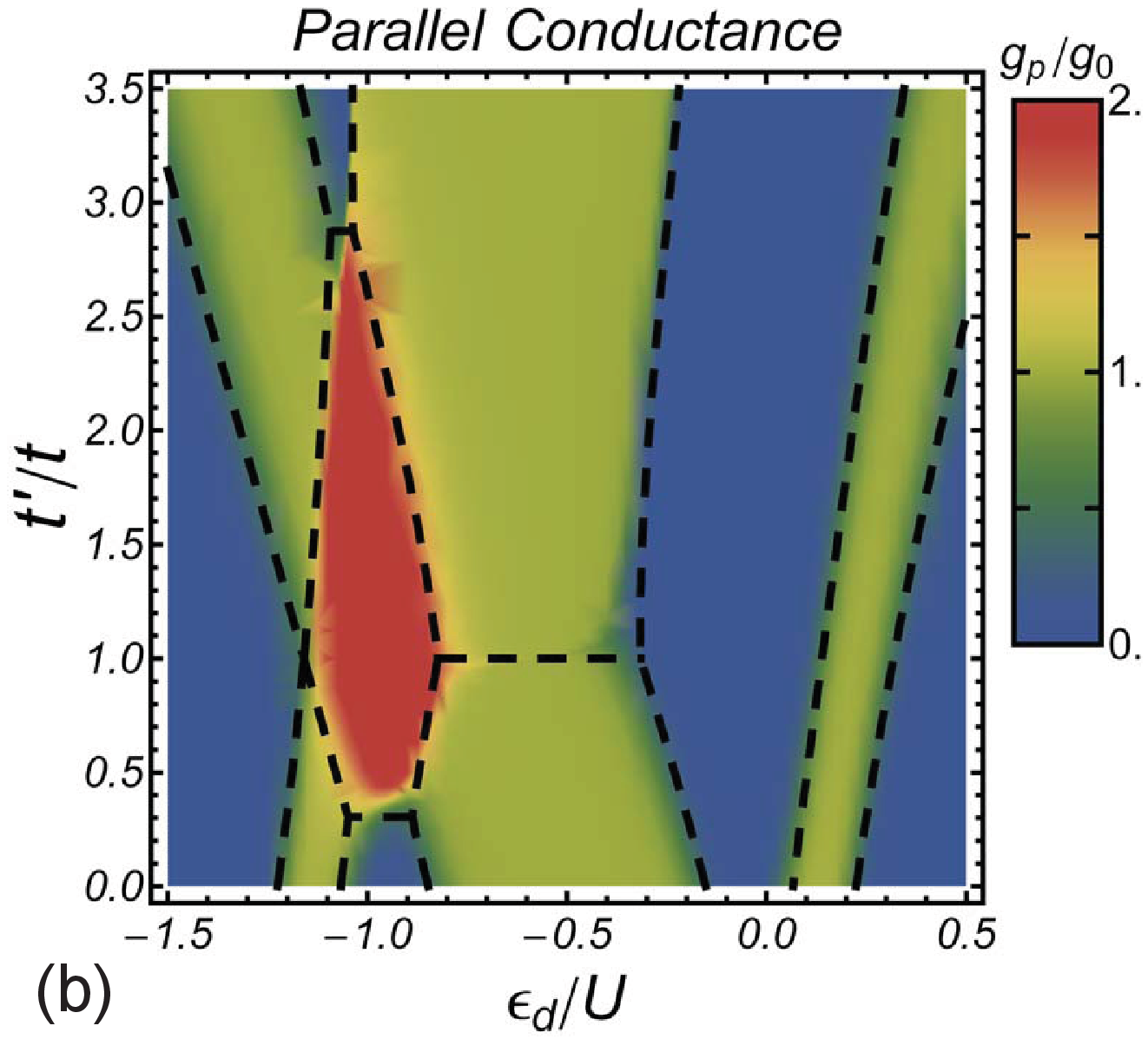}
 \end{minipage}
\\
  \begin{minipage}[t]{0.485\linewidth}
   \includegraphics[width=1\linewidth]{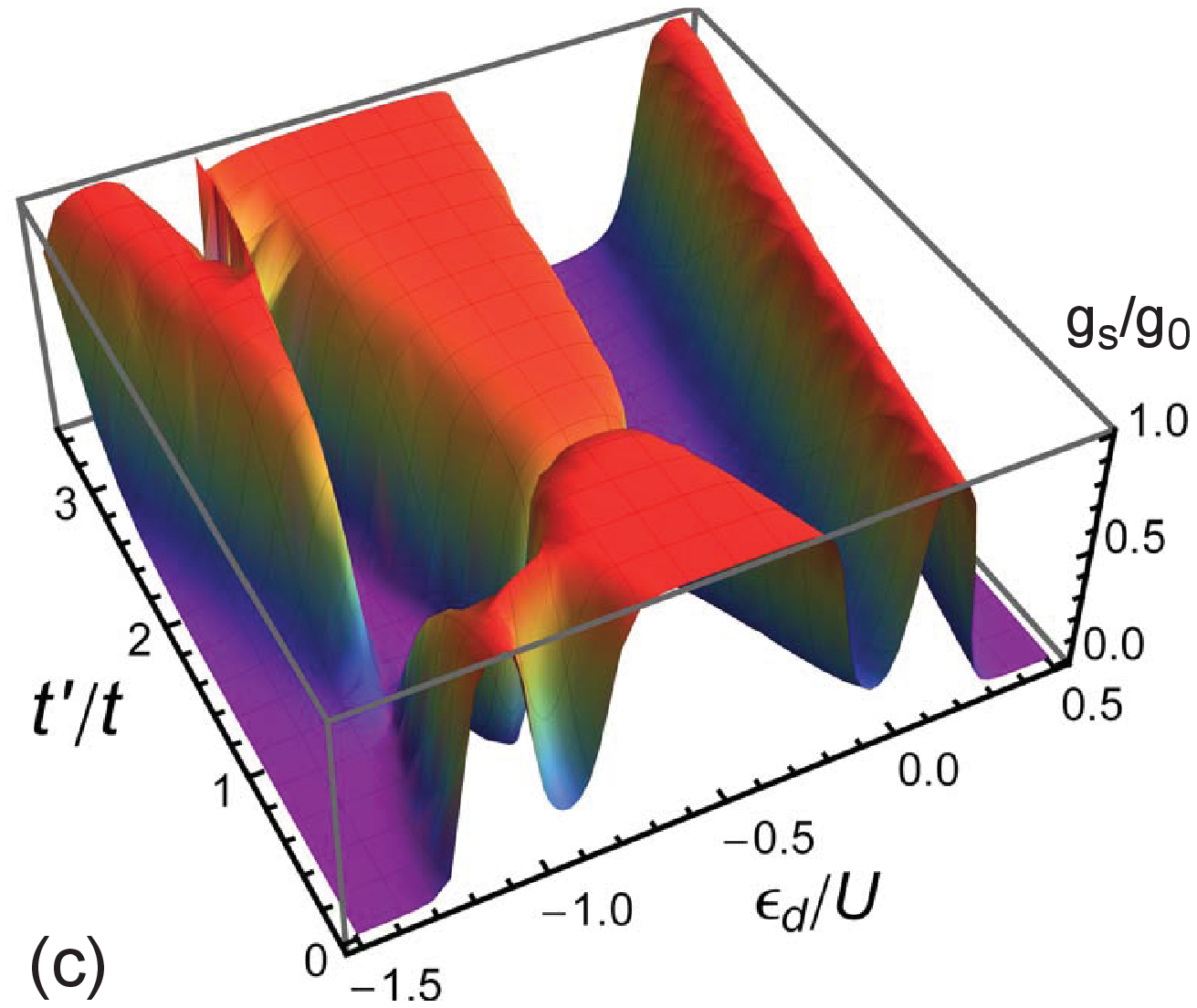}
\end{minipage}
\begin{minipage}[t]{0.482\linewidth}
   \includegraphics[width=1\linewidth]{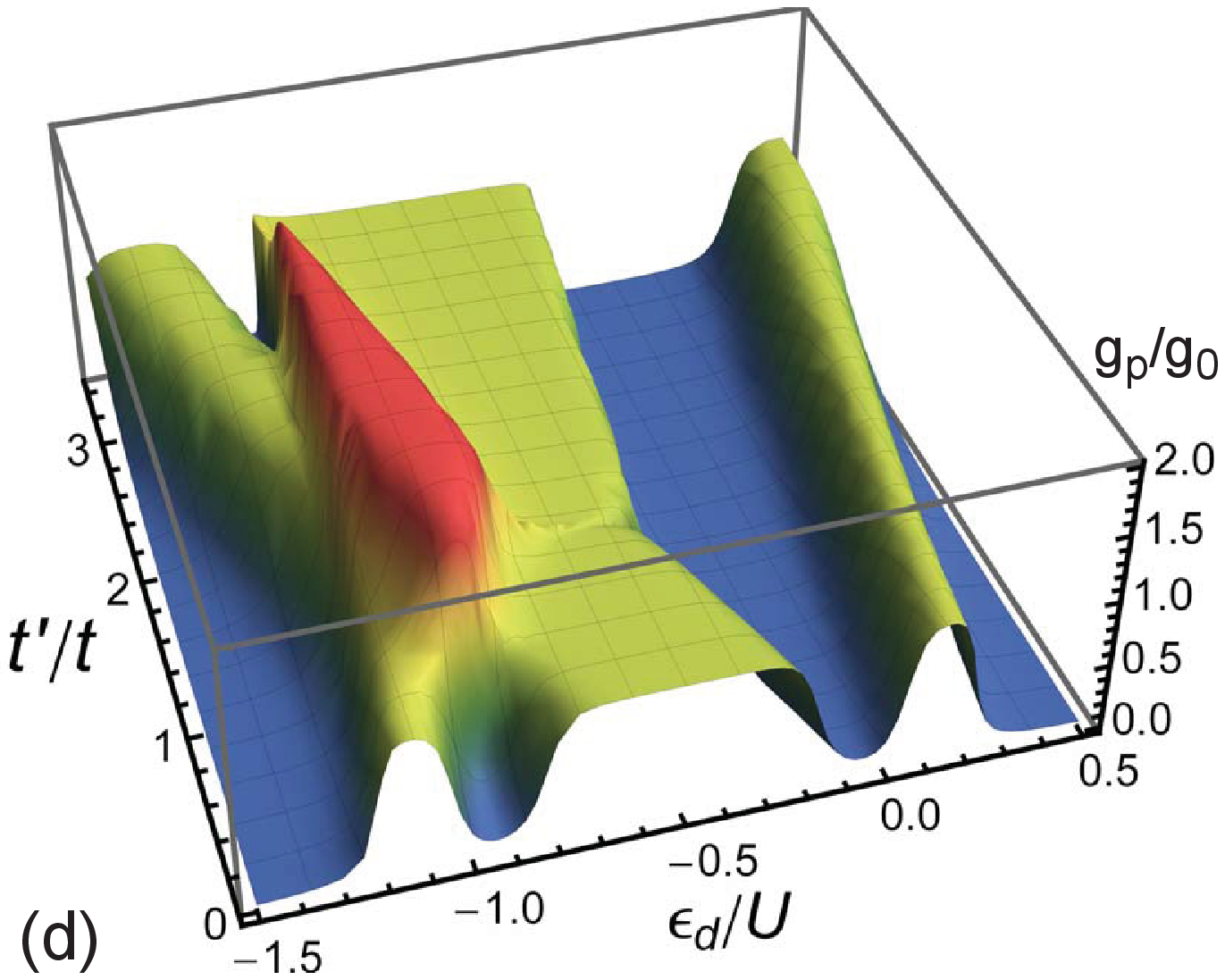}
 \end{minipage}
 \caption{(Color online)  
Series (a) and parallel (b) 
conductances for $U/(2\pi t) = 1.0$, $\Gamma/t=0.25$ 
and $\Delta \epsilon =0$ are 
plotted as functions of $\epsilon_d/U$ and $t'/t$.
In (a), the solid line is a contour 
for the phase-shift difference
 $\delta_\mathrm{e} - \delta_\mathrm{o}=\pi$, 
and the dotted lines are also the contours for 
 $\delta_\mathrm{e} - \delta_\mathrm{o}=\pi/2$ and $3\pi/2$. 
In (b), the dashed lines denote the phase boundaries 
in the isolated limit $\Gamma \to 0$, corresponding 
to the ones shown in Fig.\ \ref{fig:ground_state_isolated_u1_ts} (b).  
Lower panels: (c) and (d) are 
the surface plots of the series and parallel conductances 
corresponding to (a) and (b), respectively.
}
 \label{fig:conductance_u1}
\end{figure}

\subsubsection{
Series and Parallel Conductances 
 for $\,t\neq t'$
}

The NRG results for the  
conductances at zero temperature 
are shown in the $\epsilon_d/U$ vs $t'/t$ plane 
in Fig.\  \ref{fig:conductance_u1} 
for $U/(2\pi t) = 1.0$, $\Gamma/t=0.25$, 
and $\Delta \epsilon =0$. 
In order to show  more clearly the overall features, 
we have provided two types of the plots 
seen from different points in the parameter space 
for each of the conductances. 
The series conductance $g_\mathrm{s}$ is plotted in
 (a) and (c). Similarly,  
the parallel conductance $g_\mathrm{p}$ is shown  
in (b) and (d). 
For comparison, the phase boundary for $\Gamma \to 0$ 
 given in Fig.\ \ref{fig:ground_state_isolated_u1_ts} (b) 
is also superposed onto Fig.\  \ref{fig:conductance_u1} (b) 
with the dashed lines.
We see that the feature of the conductances 
reflects the occupation number 
in each of the regions in the parameter space. 
Note that the ground state becomes a spin singlet 
in the whole region of the parameter space 
due to the screening by the conduction electron.

We can also see in Figs.\  \ref{fig:conductance_u1} (a) and (c) 
that typical Kondo ridges for the series conductance  
with $g_\mathrm{s} \simeq 2e^2/h$ emerge for odd-number fillings 
 $N_\mathrm{tot} \simeq 1.0,\,3.0$, and $5.0$.  
Furthermore, 
both $g_\mathrm{s}$ and $g_\mathrm{p}$ almost vanish 
for even-number fillings 
$N_\mathrm{tot} \simeq 0.0,\, 2.0$, $4.0$ and $6.0$ 
 except for the $S=1$ Kondo region.
Particularly, 
the behavior at small fillings $N_\mathrm{tot} \lesssim 2.0$,  
for $\epsilon_d/U \gtrsim -1.0$,  
can be explained simply by  
the $S=1/2$ Kondo effect due to the 
lowest single molecular orbital of $E_{\mathrm{e},-}^{(1)}$.
Therefore, the characteristic features of the TTQD 
appear in the region of $\epsilon_d \lesssim -0.2 U$, 
where the two excited levels  
$E_{\mathrm{o}}^{(1)}$ and $E_{\mathrm{e},+}^{(1)}$ 
are partially filled.

The solid line in Fig.\  \ref{fig:conductance_u1} (a)   
denotes the contour for the difference in the two phase shifts 
corresponding to the value  $\delta_\mathrm{e} - \delta_\mathrm{o}=\pi$.
Along this line, the series conductance becomes exactly zero.
Specifically, in the three-electron region, 
for $-0.8 \lesssim \epsilon_d/U \lesssim -0.3$, 
this contour runs near the horizontal line 
for $t'/t = 1.0$ where the triangle has the equilateral symmetry.
The contour line tilts slightly from 
the horizontal line 
 because the coupling to the two leads breaks 
the equilateral symmetry  already  at $t'=t$.
This contour for $\delta_\mathrm{e} - \delta_\mathrm{o}=\pi$ 
appears in Fig.\  \ref{fig:conductance_u1} (c)  
as a very sharp valley of the series conductance.
The Kondo ridges on each side of this valley have a different parity. 
Just at the bottom of the valley, 
the low-lying quasi-particle states for the even and odd channels 
become degenerate, 
and the low-energy properties 
can be described by the SU(4) Fermi-liquid theory.\cite{Numata2}
Furthermore, along this valley   
the two phase shifts are almost constant with the values, 
 $\delta_\mathrm{o}\simeq \pi/4$ and 
$\delta_\mathrm{e}\simeq 5\pi/4$,   
since the Coulomb interaction keeps  
the sum of the two to be $\delta_\mathrm{e} +\delta_\mathrm{o} \simeq 3 \pi/2$ 
in the three-electron region 
through the Friedel sum rule. 
Therefore, the parallel conductance does not change 
so much near this valley of the series conductance, 
keeping the value of  $g_\mathrm{p} \simeq 2e^2/h$.

\begin{figure}[t]
 \leavevmode
 \begin{minipage}[t]{0.48\linewidth}
 \includegraphics[width=1.0\linewidth]{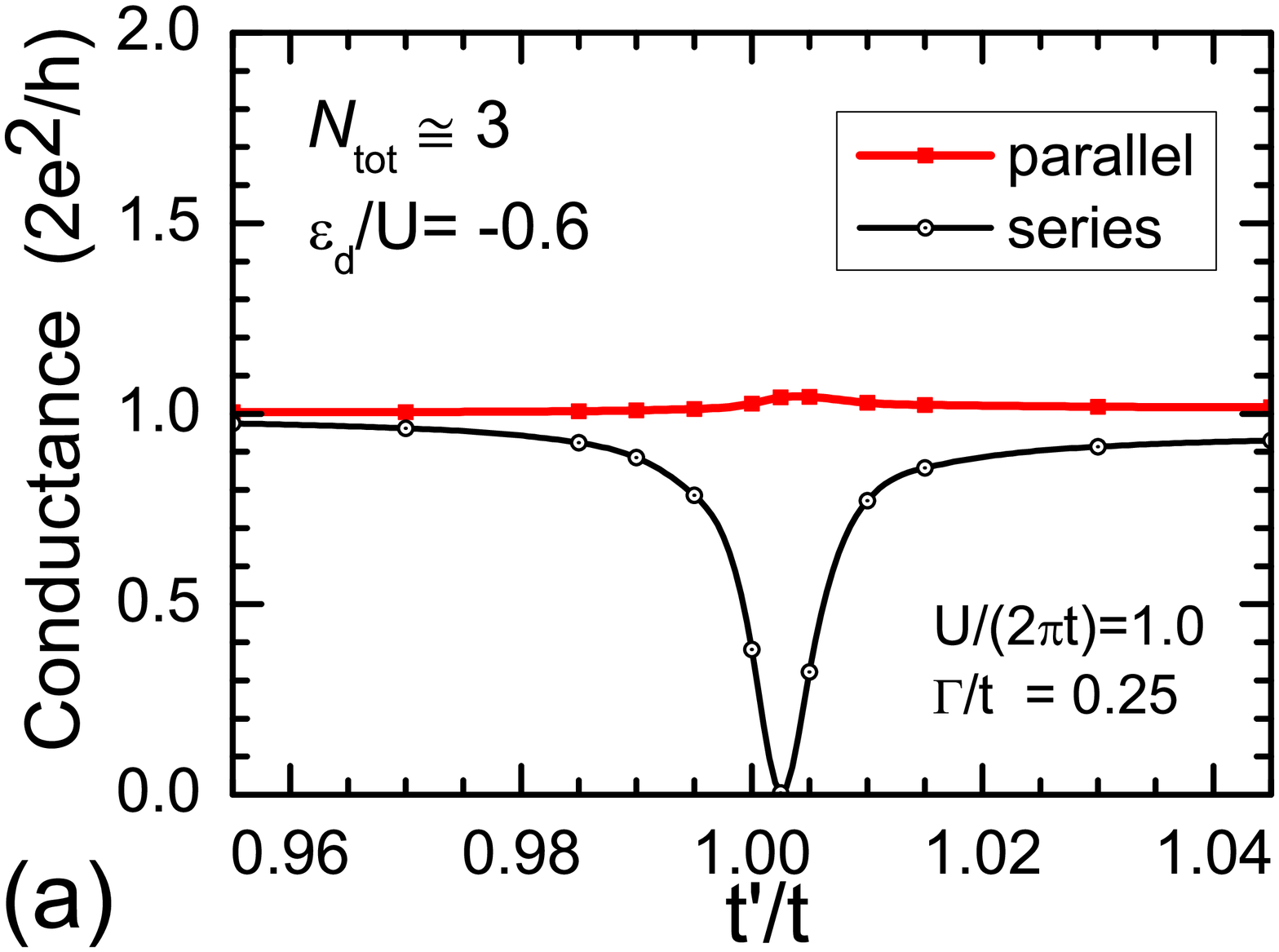}
 \end{minipage}
\rule{0.02\linewidth}{0cm}
 \begin{minipage}[t]{0.47\linewidth}
\includegraphics[width=0.99\linewidth]{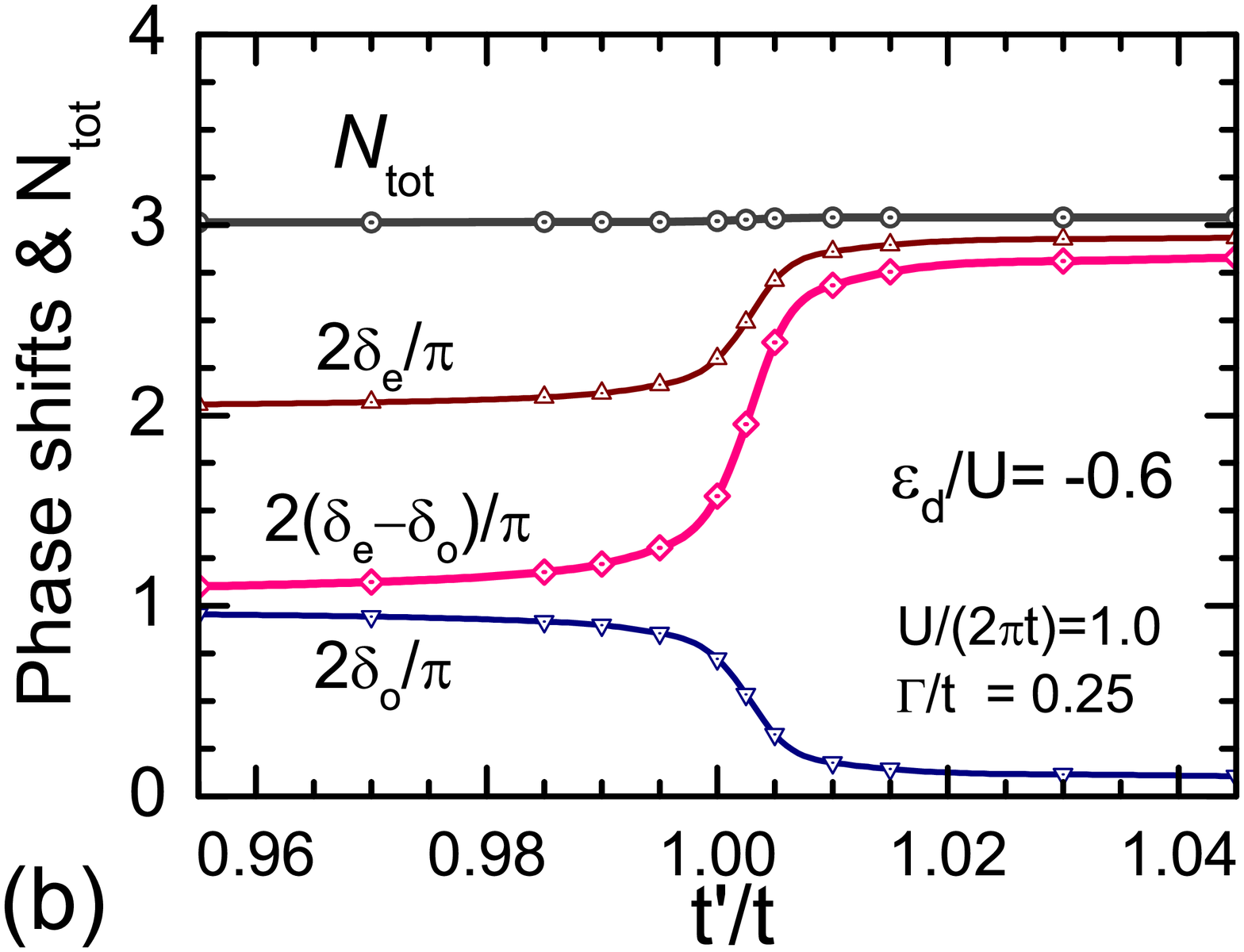}
 \end{minipage}
 \caption{(Color online)  
Ground-state properties at $\epsilon_d = -0.6U$,    
(a) conductances and 
(b) phase shifts $(\delta_\mathrm{e} \pm \delta_\mathrm{o})(2/\pi)$, 
are plotted in a narrow region of $t'$ near $t'/t=1.0$ 
for 
 $U/(2\pi t) = 1.0$, $\Gamma/t=0.25$, 
and $\epsilon_\mathrm{apex}= \epsilon_d$.     
In this parameter region, 
the occupation number is almost 
constant $N_\mathrm{tot} \simeq 3.0$.
}
 \label{fig:cond_su4_vs_ts}
\end{figure}

In order to see the sharp SU(4) Kondo valley in more detail,
the conductances and the phase shifts 
at $\epsilon_d = -0.6U$ are plotted  
in Fig.\ \ref{fig:cond_su4_vs_ts}
as functions of $t'/t$.
Particularly, the two lines in Fig.\ \ref{fig:cond_su4_vs_ts} (a) 
correspond 
to a cross section of the surface plots given 
in Fig.\  \ref{fig:conductance_u1} (c)  and (d)  
in the middle of the three-electron region along the vertical direction. 
We can see in Fig.\ \ref{fig:cond_su4_vs_ts} (b) 
that the phase-shift difference $\delta_\mathrm{e}-\delta_\mathrm{o}$ 
increases  with $t'/t$ 
showing a kink, the value of which varies from $\pi/2$ to $3 \pi/2$ 
as $t'/t$ increases, and taking the value of $\pi$ at $t'/t \simeq 1.0$ 
in the middle of the transient region. 
This kink determines the structure of the series conductance valley  
 seen in Fig.\ \ref{fig:cond_su4_vs_ts} (a).
Therefore, the slope of the phase difference 
 $\delta_\mathrm{e}-\delta_\mathrm{o}$ 
in the middle of the kink determines the width of the valley.
Note that it is quite general to the local Fermi-liquid systems  
that the derivative of the phase shift with 
respect to the parameters, such as the frequency and 
the external fields, plays an important role 
on the renormalization of some correlation functions.

The $S=1$ Kondo behavior can 
be seen for the four-electron filling 
in the diamond-shape region 
in Figs.\  \ref{fig:conductance_u1} (a) and (b). 
The series conductance 
almost vanishes $g_\mathrm{s}\simeq 0.0$ in this region, 
while the parallel conductance 
is enhanced $g_\mathrm{p}\simeq 4e^2/h$ 
despite an even-number electron filling.
This contrast between $g_\mathrm{s}$ and $g_\mathrm{p}$ can be 
seen clearly, particularly 
in Figs.\  \ref{fig:conductance_u1} (c) and (d). 
We can also see in Fig.\  \ref{fig:conductance_u1} (a)
that the contour for 
$\delta_\mathrm{e} - \delta_\mathrm{o}=\pi$   
is winding in the center of the diamond region 
near $\epsilon_d/U \simeq -1.0$. 
Such a bend is not seen in the noninteracting case,
for which the contour varies monotonically as shown  
in Fig.\ \ref{fig:conductance_ttqd_u0} (a). 
The contour lines of the phase shifts evolve, however, 
continuously from the non-interacting form. 
This is because the ground state 
of the whole system evolves adiabatically from a 
singlet described by a single Slater determinant 
to a correlated singlet described by
the local Fermi-liquid theory for finite $\Gamma$. 
Note that the $S=1$ moment
is screened at low temperature 
by the conduction electrons from the two leads 
via a two-stage screening processes.\cite{ONTN,Numata,Numata2}

\begin{figure}[t]
 \leavevmode
 \begin{minipage}[t]{0.48\linewidth}
 \includegraphics[width=1.0\linewidth]{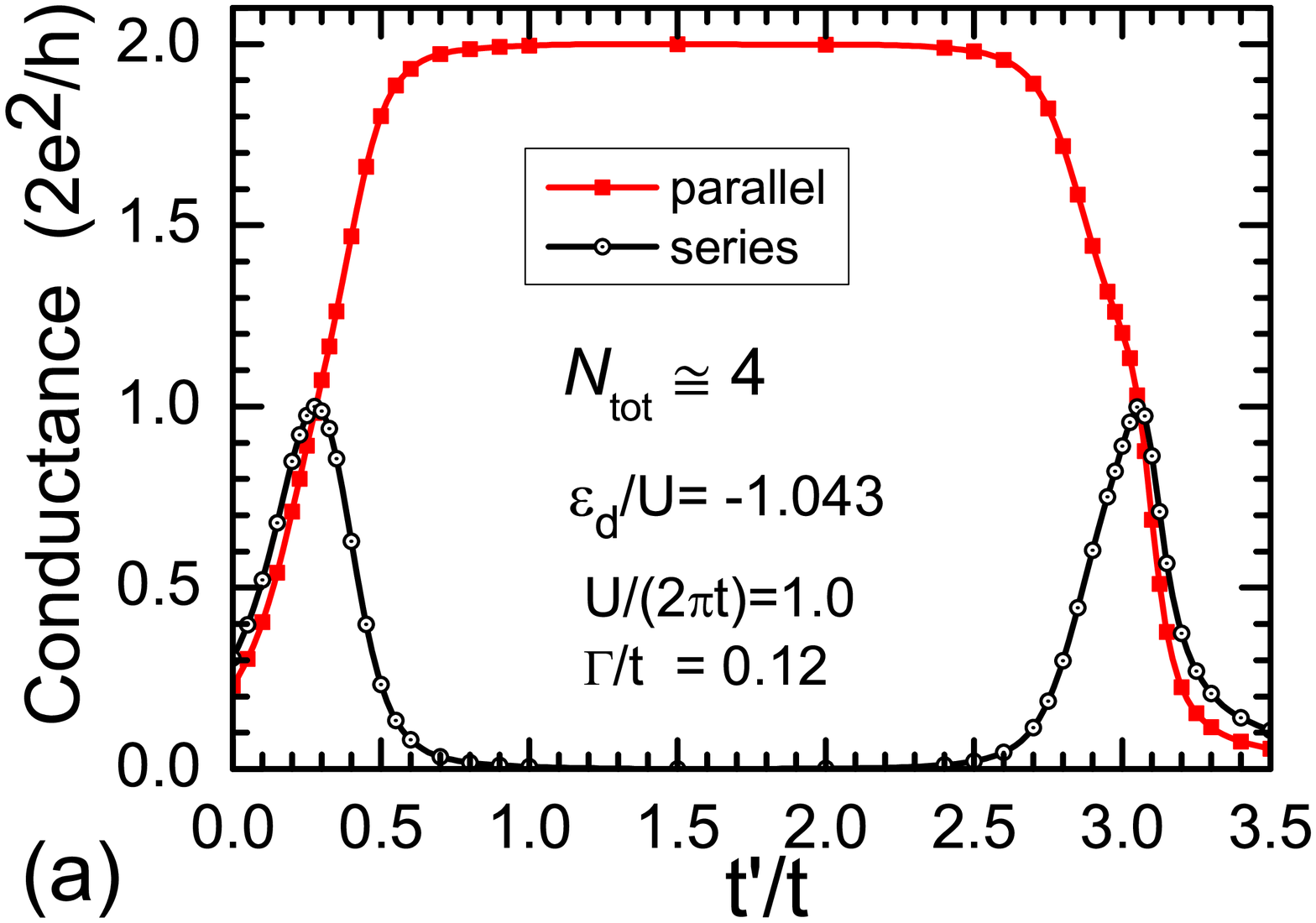}
 \end{minipage}
\rule{0.02\linewidth}{0cm}
 \begin{minipage}[t]{0.47\linewidth}
\includegraphics[width=1\linewidth]{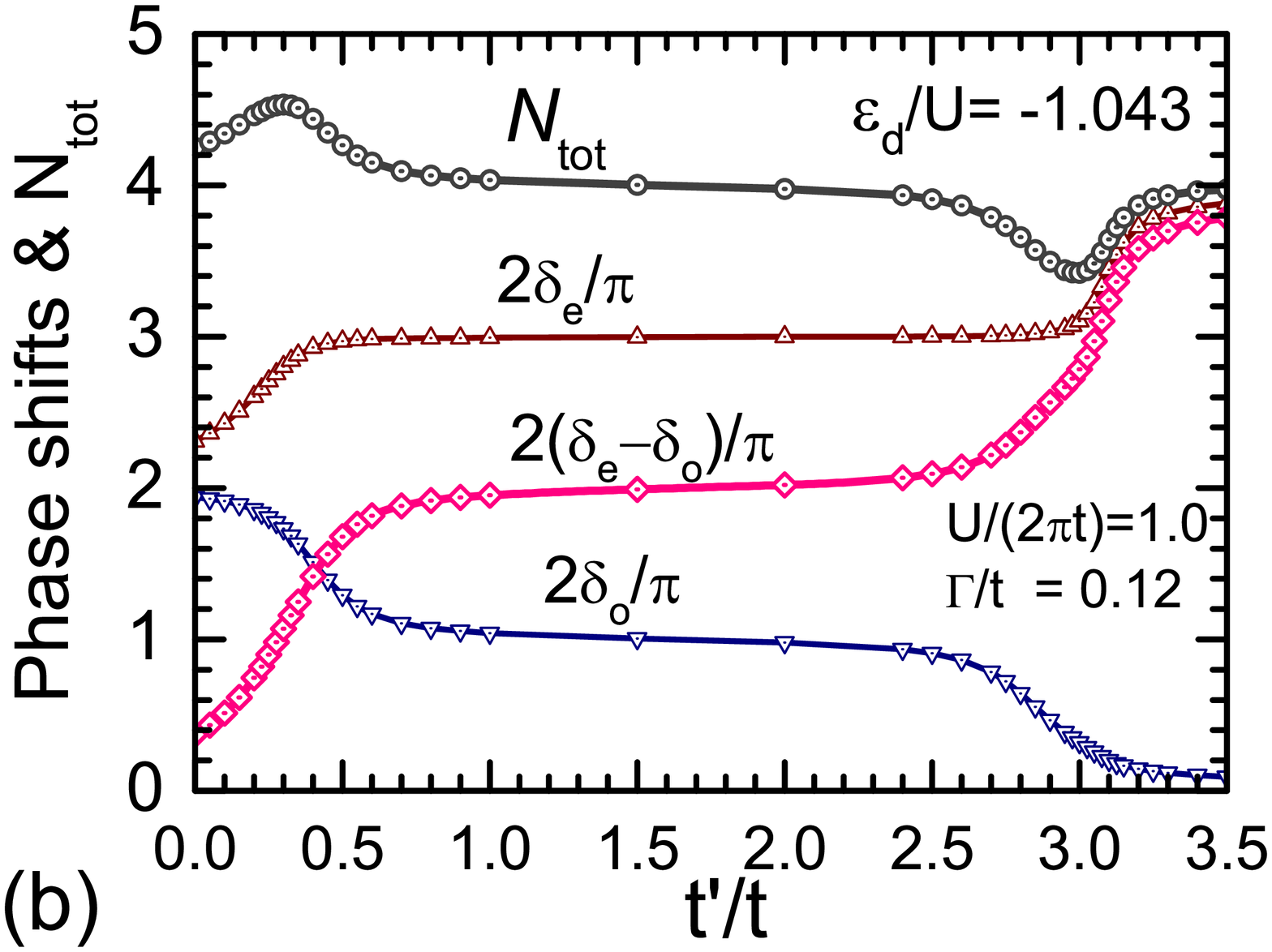}
 \end{minipage}
 \caption{(Color online) 
Ground-state properties at $\epsilon_d = -1.043U$,
(a) conductances and 
(b) phase shifts $(\delta_\mathrm{e} \pm \delta_\mathrm{o})(2/\pi)$, 
are plotted  as functions of $t'/t$ for 
 $U/(2\pi t) = 1.0$, $\Gamma/t=0.12$, 
and $\epsilon_\mathrm{apex}= \epsilon_d$.     
In this parameter region, 
the occupation number is almost 
constant $N_\mathrm{tot} \simeq 4.0$.
Note that 
in the limit of $\Gamma \to 0$ 
the Nagaoka state is the ground state for the 
isolated cluster for $0.30 <t'/t<2.88$. 
}
 \label{fig:cond_s0_s1_s0_vs_ts}
\end{figure}

The dotted lines in Fig.\ \ref{fig:conductance_u1} (a)     
express the contours 
for $\delta_\mathrm{e} - \delta_\mathrm{o}=\pi/2$ 
(below the solid line) and $3\pi/2$ (above the solid line), 
on which the series conductance reaches 
the unitary-limit value $g_\mathrm{s} = 2e^2/h$.  
One of the dotted lines on the right, 
 at $\epsilon_d/U \gtrsim 0.1$, 
follows simply the Kondo ridge 
caused by the lowest orbital $E_{\mathrm{e},-}^{(1)}$. 
The other two lines pass on 
the top and bottom of the diamond of the $S=1$ Kondo region. 
These two contours can be compared  
to the phase boundaries for the singlet-triplet transition, 
seen in a narrow range of $\epsilon_d$ 
at $t'/t \simeq 0.30$ and $2.88$
in Fig.\ \ref{fig:ground_state_isolated_u1_ts} (b).
In order to clarify the precise feature 
of the corresponding crossover between the $S=1$ Kondo 
and non-Kondo singlet states,
 the conductance and phase shifts
are plotted in Fig.\ \ref{fig:cond_s0_s1_s0_vs_ts} 
as functions of $t'/t$, 
 choosing the level position $\epsilon_d$ to be 
in the middle of the four-electron region at 
$\epsilon_d \simeq -1.04U$. 
At each of the crossover points, near $t'/t \simeq 0.3$ and $3.0$, 
the series conductance has a peak.   
The feature of these conductance peaks reflects the kink in the phase 
difference  $\delta_\mathrm{e} - \delta_\mathrm{o}$,
the value of which varies from $0.2\pi$ to $1.0\pi$ near $t'/t \simeq 0.3$, 
and  from $1.0\pi$ to $3.0\pi$  near $t'/t \simeq 3.0$. 
Therefore, 
the slope of these kink determines the width of the conductance of peak.
Furthermore at the crossover region, 
the electron occupation fluctuates  
slightly from $4.0$ as $N_\mathrm{tot} -4.0 \simeq \pm 0.5$.
In the $S=1$ Kondo-singlet region situating between 
the two peaks of $g_\mathrm{s}$,
the phase shifts are almost locked  
at $\delta_\mathrm{e} \simeq 3\pi/2$ and 
$\delta_\mathrm{o} \simeq \pi/2$, and thus 
the parallel conductance takes the value $g_\mathrm{p} = 4e^2/h$.
In one of the non-Kondo regions for $t'/t \gtrsim 3.2$,
the phase shifts approach to $\delta_\mathrm{e} \simeq 2\pi$ and 
$\delta_\mathrm{o} \simeq 0$.
The phase shifts take the values of $\delta_\mathrm{e} \simeq 1.2 \pi$, 
and  $\delta_\mathrm{o} \simeq \pi$ in the limit of $t' \to 0$, 
in the other  non-Kondo region for 
 $t'/t \lesssim 0.2$.
Note that for Fig.\ \ref{fig:cond_s0_s1_s0_vs_ts},
the coupling between the TTQD and the leads 
has been chosen to be 
$\Gamma/t=0.12$, which is  
smaller than that ($\Gamma/t=0.25
$) for the previous figures, 
in order to see clearly the typical features 
of the narrow crossover regions.

\subsubsection{Phase-shift difference:
$\,\Theta =
\frac{\displaystyle\protect\mathstrut 2}{\displaystyle\protect\mathstrut \pi}
(\delta_\mathrm{e}-\delta_\mathrm{o})\,$
for $\,t\neq t'$
}

\begin{figure}[t]
 \leavevmode
 \begin{minipage}[t]{0.51\linewidth}
  \includegraphics[width=1.0\linewidth]{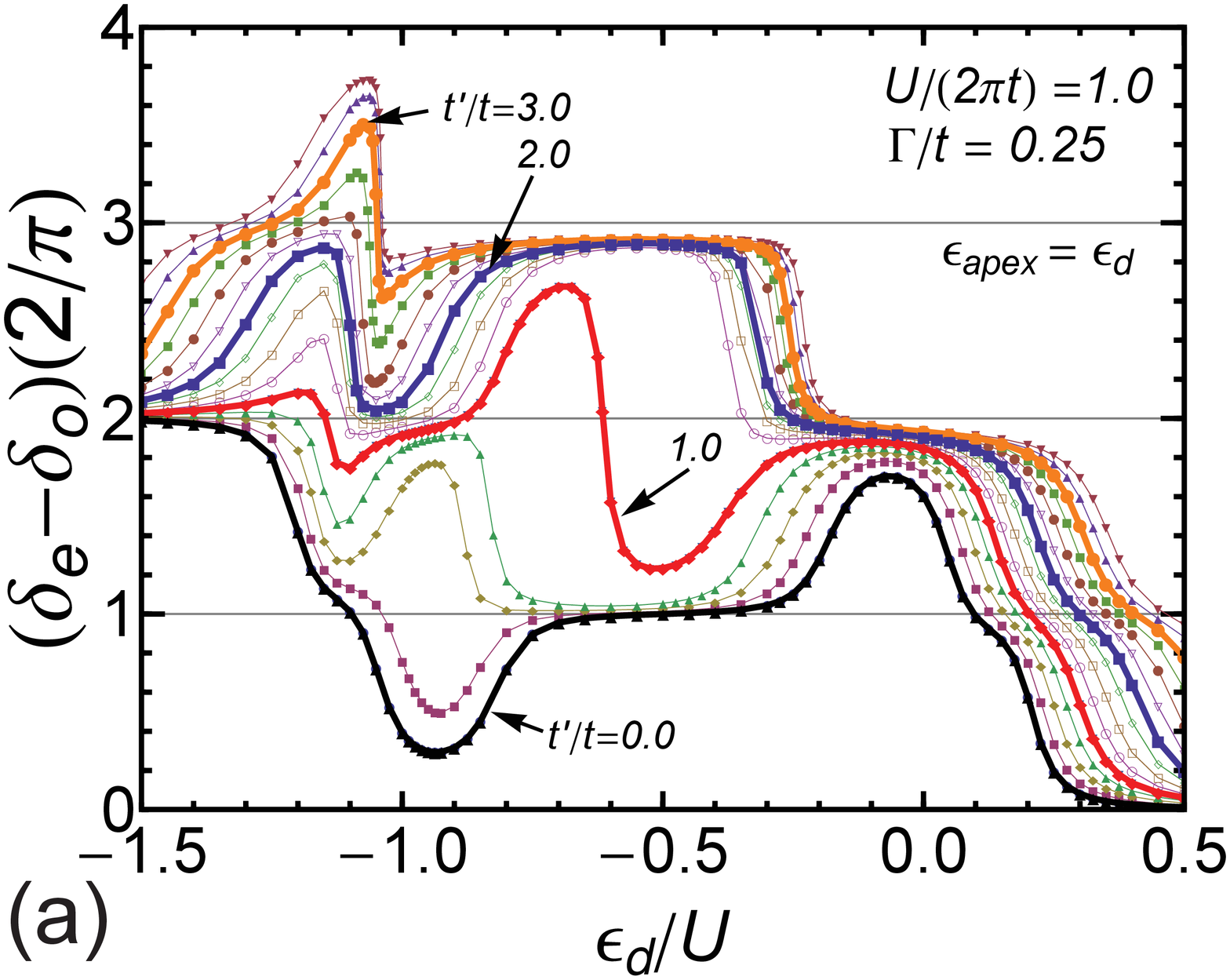}
 \end{minipage}
 \begin{minipage}[t]{0.45\linewidth}
 \includegraphics[width=1.0\linewidth,clip,trim = 0.1cm 0cm 0cm 0cm]{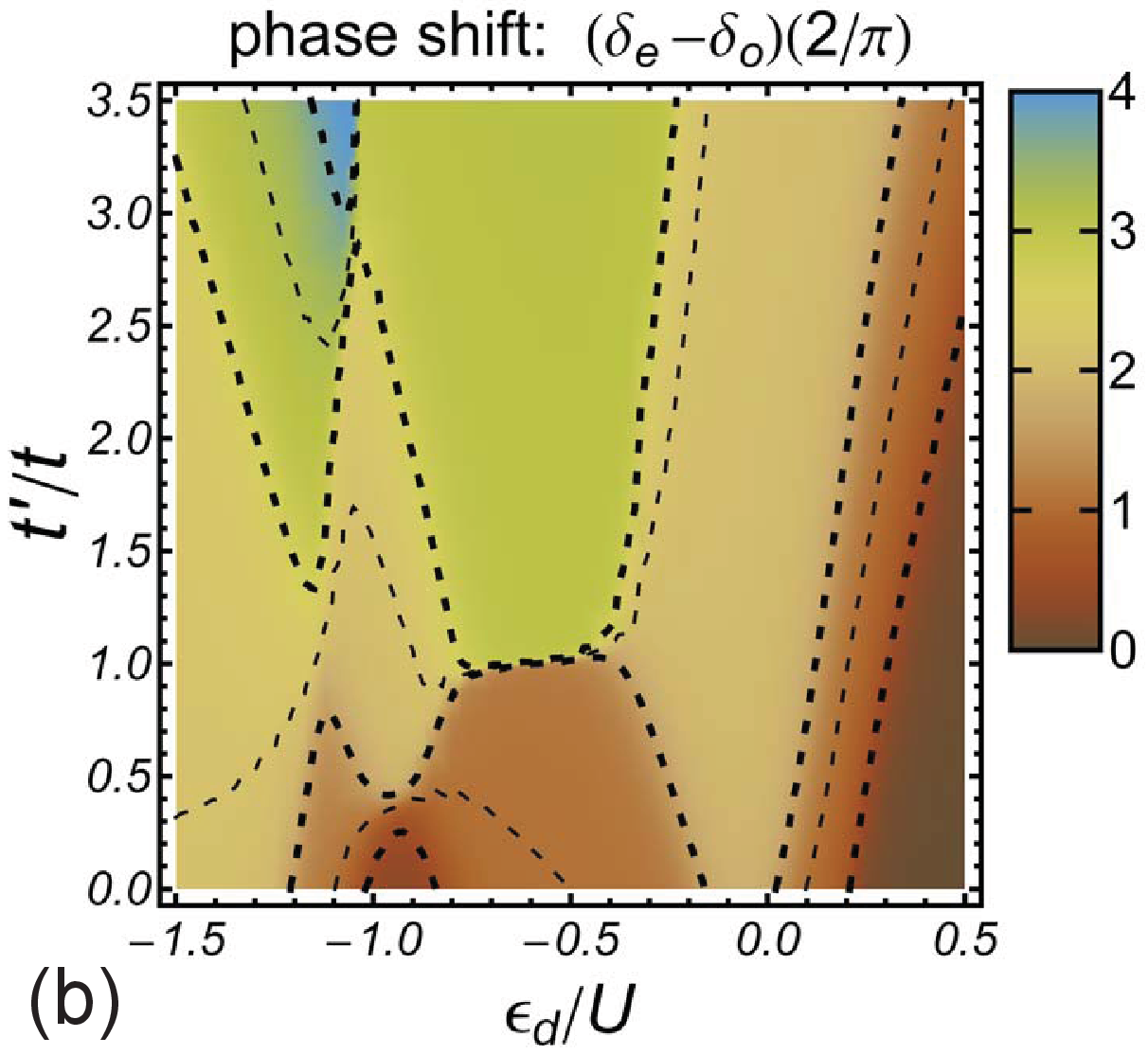}
 \end{minipage}
 \caption{(Color online) 
Difference between the even and odd phase shifts 
$\Theta \equiv (\delta_\mathrm{e} -\delta_\mathrm{o})(2/\pi)$ 
for $U/(2\pi t) =1.0$, $\Gamma/t =0.25$ 
and $\Delta \epsilon =0$. 
The left panel (a)  shows
$\Theta$  
as a function of $\epsilon_d/U$  for 
the values of 
$t'/t=0.0,\, 0.25,\, 0.5, \ldots,$ and $3.5$ 
(in steps of $0.25$ from the bottom to the top).  
The right panel (b) shows $\Theta$  
 in  
the $\epsilon_d/U$ vs $t'/t$ plane. 
The dotted lines  in (b) are the contours for    
$(\delta_\mathrm{e} -\delta_\mathrm{o})(2/\pi) 
= 0.5,\, 1.0,\, 1.5,\, \ldots,$ and $3.5$ 
(in steps of $0.5$ from the bottom to the top).   
}
\label{fig:u1_dedo_ts}
\end{figure}

The difference between 
the two phase shifts $\delta_\mathrm{e}-\delta_\mathrm{o}$ 
is a fundamental parameter that 
contains the essential information of the interference effects 
between the even and odd conducting channels. 
It affects the series conductance, 
while each channel contributes independently 
to the parallel conductance, 
through the expressions given in 
eqs.\ \eqref{eq:gs} and \eqref{eq:gp}. 
Specifically,  
peaks and dips of the series conductance 
correspond directly to the kinks of the phase-shift difference 
 $\delta_\mathrm{e}-\delta_\mathrm{o}$, 
as seen in Figs.\ \ref{fig:cond_su4_vs_ts} 
and \ref{fig:cond_s0_s1_s0_vs_ts}. 
It is also much easier for a numerical purpose 
to trace the kink structure of $\delta_\mathrm{e}-\delta_\mathrm{o}$ 
than to find directly the dips and peaks of $g_\mathrm{s}$.  

We also provide the NRG results 
for $\Theta \equiv (\delta_\mathrm{e}-\delta_\mathrm{o})(2/\pi)$  
in  Fig.\ \ref{fig:u1_dedo_ts} 
in order to clarify its behavior in the wide parameter space. 
In the left panel (a), 
$\Theta$ is  plotted as a function of $\epsilon_d/U$ 
for the values of $t'/t$ varying 
from $0.0$ to $3.5$ in steps of $0.25$.
Furthermore,  Fig.\ \ref{fig:u1_dedo_ts} (b) shows the results  
obtained in the $\epsilon_d/U$ vs $t'/t$ plane: 
the dotted lines are the contours 
for the values of  
$\Theta$ 
varying from $0.5$ (bottom and right) to $3.5$ (top) in steps of $0.5$.
We can see that there are several plateaus, or shelves,  
in these figures  near the integer values of  
$\Theta \simeq 1.0,\, 2.0,$ and $3.0$, 
on which $g_\mathrm{s}$ becomes almost transparent or zero. 
Furthermore, the occupation number also approaches to 
an integer value on each of these plateaus, 
and thus  they can be classified according to  
a set of the two integers  $(N_\mathrm{tot},\,\Theta)$.   
For instance, 
in Fig.\ \ref{fig:u1_dedo_ts} (a),
we can see a wide plateau which can labelled by 
 $(N_\mathrm{tot},\,\Theta) \simeq (3.0,\,3.0)$ 
for $-1.0 \lesssim \epsilon_d/U \lesssim -0.3$ and $t'/t \gtrsim 1.0$. 
The height of the plateau, however, is still 
somewhat smaller than the exact integer $3.0$. 
Such a deviation of 
the plateau height from an integer value decreases 
 as $\Gamma$ decreases. 
This has been confirmed  explicitly 
for the equilateral triangle 
in the previous work [see Fig.\ 6 of Ref.\ \onlinecite{Numata2}].
Furthermore, we can see 
 another example for smaller $\Gamma$ 
in the next section [see Fig.\ \ref{fig:u1_dedo_de}].

The feature of $\Theta$ in 
the parameter space can also be compared 
to the phase diagram for the isolated TTQD.
Particularly, the contours for
$\Theta = 0.5,\, 1.5,\,2.5,\,$ and $3.5$, which are shown with 
the thicker dotted lines in  Fig.\ \ref{fig:u1_dedo_ts} (b),
divide the parameter space in a similar way that 
the phase boundaries did in Fig.\ \ref{fig:ground_state_isolated_u1_ts} (b).
The contour lines, however, do not cross each other 
while the border lines for $\Gamma=0$ are crossing at some points. 
We can see in Fig.\ \ref{fig:u1_dedo_ts} 
that the SU(4) Kondo effect is manifest  
in the parameter space as a sheer {\it cliff\/} 
at $-0.8 \lesssim \epsilon_d/U \lesssim -0.3$ 
near $t'/t \simeq 1.0$. 
It also corresponds to the kink that we have  
seen in Fig.\ \ref{fig:cond_su4_vs_ts} (b).
Between the bottom and top 
of the cliff the value of $\Theta$ varies from $1.0$ to $2.9$,
respectively. The slope of the cliff determines the width 
of the SU(4) valley which corresponds 
to the contour line for $\Theta = 2.0$, running in the middle of the cliff.
The Kondo ridges of $g_\mathrm{s}$ on both sides 
of the valley can be classified according to 
the plateau value of $\delta_\mathrm{e}-\delta_\mathrm{o} =\pi/2$ 
or $3\pi/2$, as the phase difference 
varies by $\pi$ across the valley. 

The $S=1$ Kondo region can also be seen as 
a diamond-shape plateau in Fig.\ \ref{fig:u1_dedo_ts} (b), 
appearing at $\epsilon_d \simeq -1.0U$ and $0.3 \lesssim t'/t \lesssim 2.9$.
This plateau is characterized by the two parameters, 
$\Theta \simeq 2.0$ and $N_\mathrm{tot} \simeq 4.0$.
Thus the phase shifts are almost fixed at the value of 
 $\delta_\mathrm{e} \simeq 3\pi/2$ and 
$\delta_\mathrm{o} \simeq \pi/2$ in this region.

 \subsection{Diagonal distortions:  $\epsilon_\mathrm{apex} \neq \epsilon_d$}
\label{subsec:diagonal}

We next examine the effects of the diagonal 
distortion $\epsilon_\mathrm{apex} \neq \epsilon_d$,  
keeping the inter-dot hopping matrix elements  uniform $t' = t$
and taking the Coulomb interaction to be $U = 2\pi t$.
In this subsection  
we choose the coupling between the leads and the TTQD  
such that $\Gamma/t=0.12$,  
which is approximately a half of the one used 
for Figs.\  \ref{fig:nd_u1_ts}, \ref{fig:conductance_u1} 
and \ref{fig:u1_dedo_ts} in the off-diagonal case.

\subsubsection{Local charge: $\,N_\mathrm{tot}
= \frac{\displaystyle\protect\mathstrut 2}{\displaystyle\protect\mathstrut \pi}
(\delta_\mathrm{e}+\delta_\mathrm{o})\,$
for $\,\Delta \epsilon \neq 0$
}
\label{subsubsec:N_diag}

\begin{figure}[t]
 \leavevmode
 \begin{minipage}[t]{0.5\linewidth}
\includegraphics[width=1\linewidth]{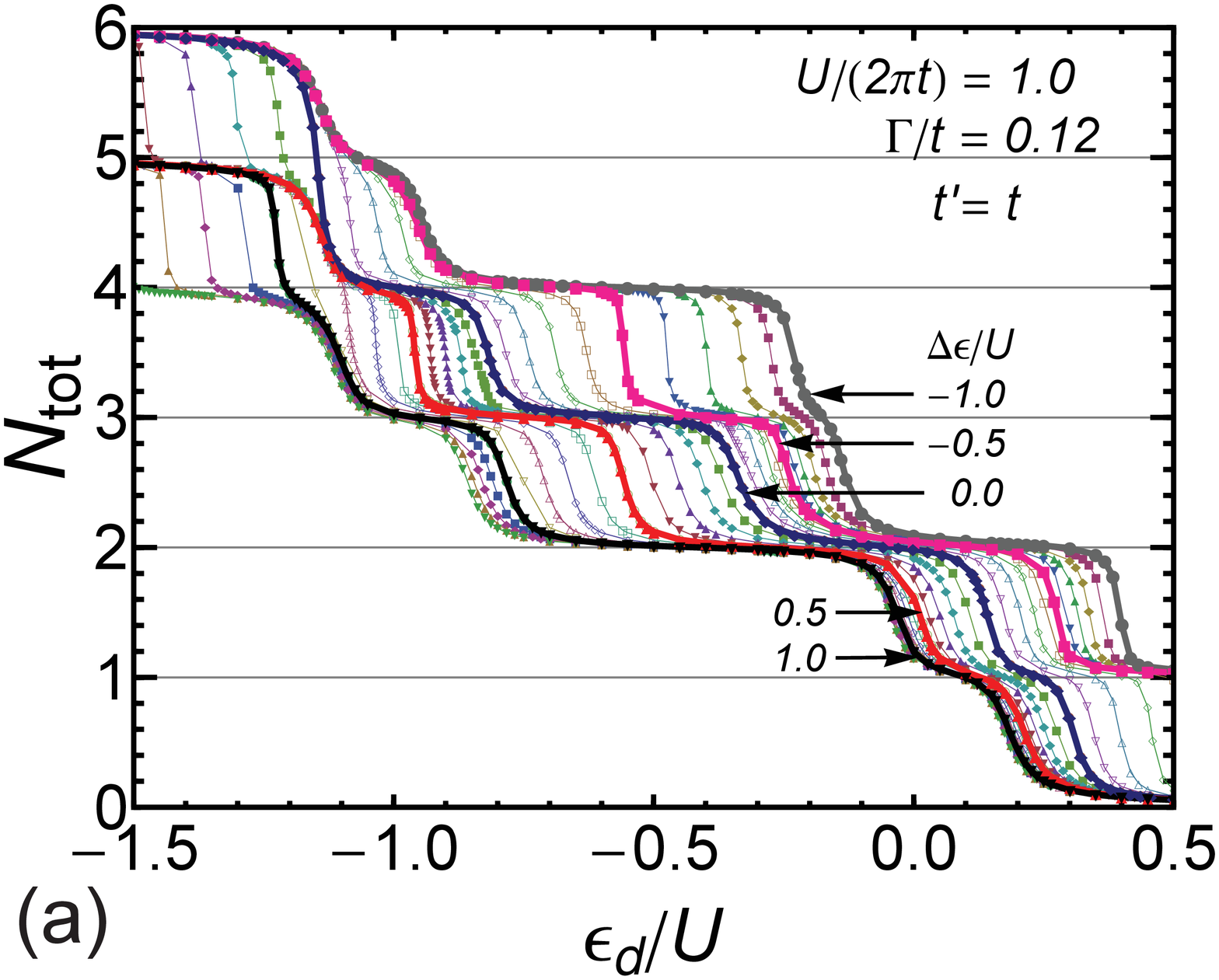}
 \end{minipage}
\rule{0.01\linewidth}{0cm} 
 \begin{minipage}[t]{0.465\linewidth}
\includegraphics[width=1\linewidth,clip,trim = 0.1cm 0cm 0cm 0cm]{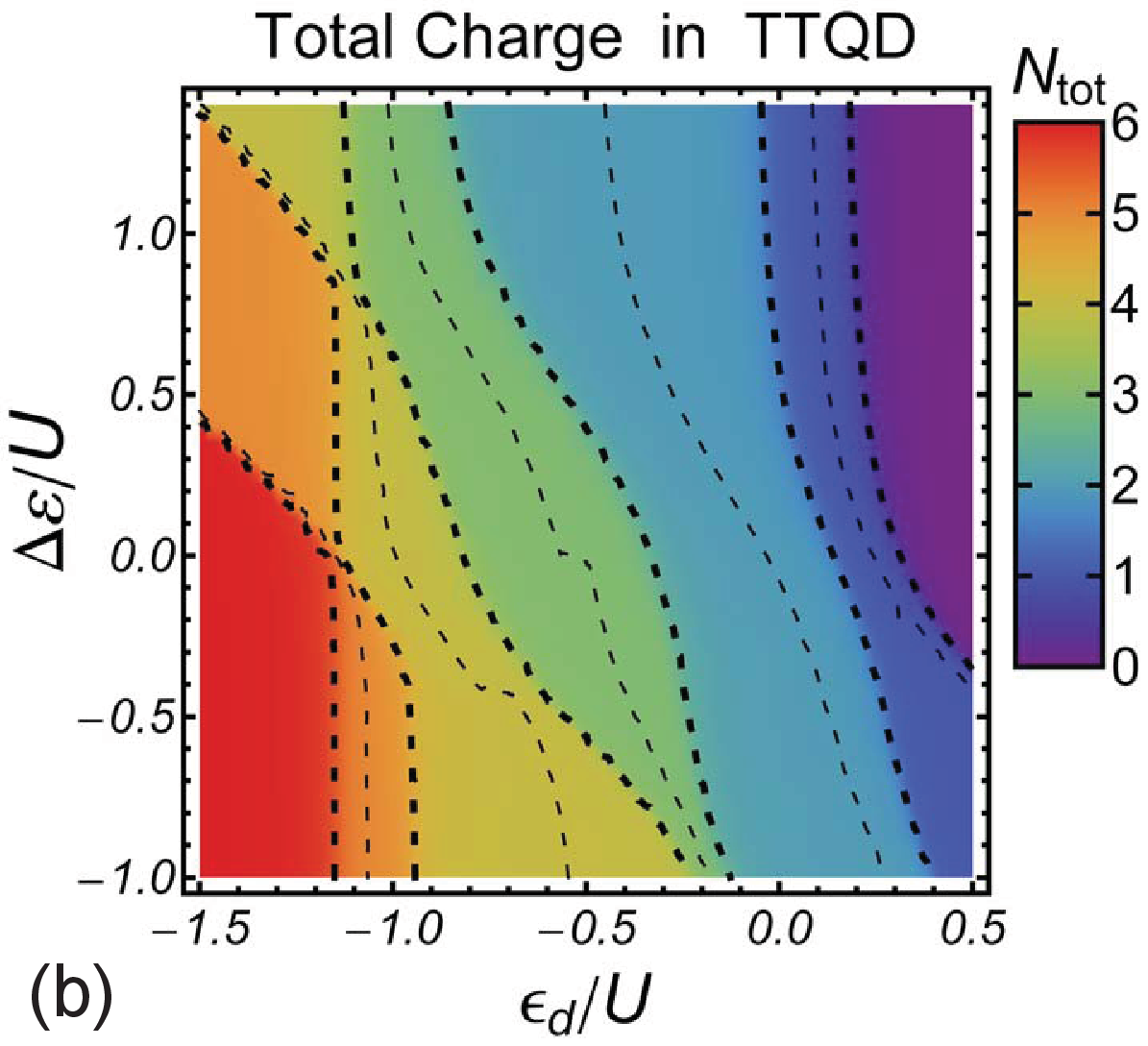}
 \end{minipage}
 \caption{(Color online) 
NRG results of the occupation number 
 $N_\mathrm{tot}$ for $U/(2\pi t) = 1.0$,
for $\Gamma/t=0.12$ and $t' = t$.
The left panel (a) shows 
$N_\mathrm{tot}$   
as a function of $\epsilon_d/U$  for  values of 
for the values of 
$\Delta \epsilon/U=-1.0,\,-0.9,\,-0.8,\, \ldots,$ and $1.4$ 
(in steps of $0.1$ from the top to the bottom).  
The right panel (b) shows $N_\mathrm{tot}$  
in the $\epsilon_d/U$ vs $\Delta \epsilon/U$ plane. 
The dotted lines in (b) are the contours for 
 $N_\mathrm{tot} = 0.5,\,1.0,\,1.5,\,\ldots,$ and $5.5$ 
(in steps of $0.5$ from the right to the left).
}
 \label{fig:nd_u1_de}
\end{figure}


Figure  \ref{fig:nd_u1_de} (a)
shows the occupation number $N_\mathrm{tot}$ 
in the TTQD as a function of $\epsilon_d/U$ 
for the values of  
$\Delta \epsilon/U=-1.0,\,-0.9,\,-0.8,\, \ldots,$ and $1.4$ 
(in steps of $0.1$ from the top to the bottom).  
We can see clearly that the plateaus appear near integer values 
of  $N_\mathrm{tot}$. 
In the present case the coupling strength $\Gamma$ 
is much smaller than $U$ and $t$, 
so that the different charge states can be distinguished clearly.
In other words, 
the crossover between two adjacent charge states becomes sharp, 
and thus the border can be determined reasonably 
by the middle point where $N_\mathrm{tot}$ takes a half-integer value.

We have also carried out the calculations for 
a number of parameter sets, 
more than the ones which are 
shown explicitly in Fig.\ \ref{fig:nd_u1_de} (a),  
and have plotted the results in Fig.\  \ref{fig:nd_u1_de} (b) 
in the $\epsilon_d/U$ vs $\Delta \epsilon/U$ plane.
In this figure the dotted lines denote 
the contours for 
 $N_\mathrm{tot} = 0.5,\,1.0,\,1.5,\,\ldots,$ and $5.5$ 
(in steps of $0.5$ from the right to the left).
Particularly, the thick dotted lines 
are the contours for  
the half-integer values; $0.5$, $1.5$, $2.5$, $3.5$, $4.5$, and $5.5$. 
We can see that these thick dotted lines 
almost follow  the phase boundaries 
between the different charge states in the isolated TTQD, 
shown in Fig.\ \ref{fig:ground_state_isolated_u1_de} (b). 
 The local charge $N_\mathrm{tot}$ changes 
rapidly near these thick contours, 
and has a plateau of an integer value between 
the thick dotted lines, 
as can be seen explicitly in Fig.\  \ref{fig:nd_u1_de} (a).
Therefore the charge in the plateau regions is 
determined at high energy scale, 
and the sum of the phase shifts 
$(2/\pi)(\delta_\mathrm{e}+\delta_\mathrm{o})$ in the plateaus 
can be approximated reasonably by the value of $N_\mathrm{tot}$ 
in the $\Gamma \to 0$ limit.
However, the transport properties at zero temperature  
are determined by each of the two phase shifts or the difference 
between them, which are determined essentially 
by the low-lying energy states 
of the whole system including the leads.

\subsubsection{
Series and Parallel Conductances 
for $\,\Delta \epsilon \neq 0$
}

\begin{figure}[t]
  \leavevmode
  \begin{minipage}[t]{0.485\linewidth}
   \includegraphics[width=1\linewidth]{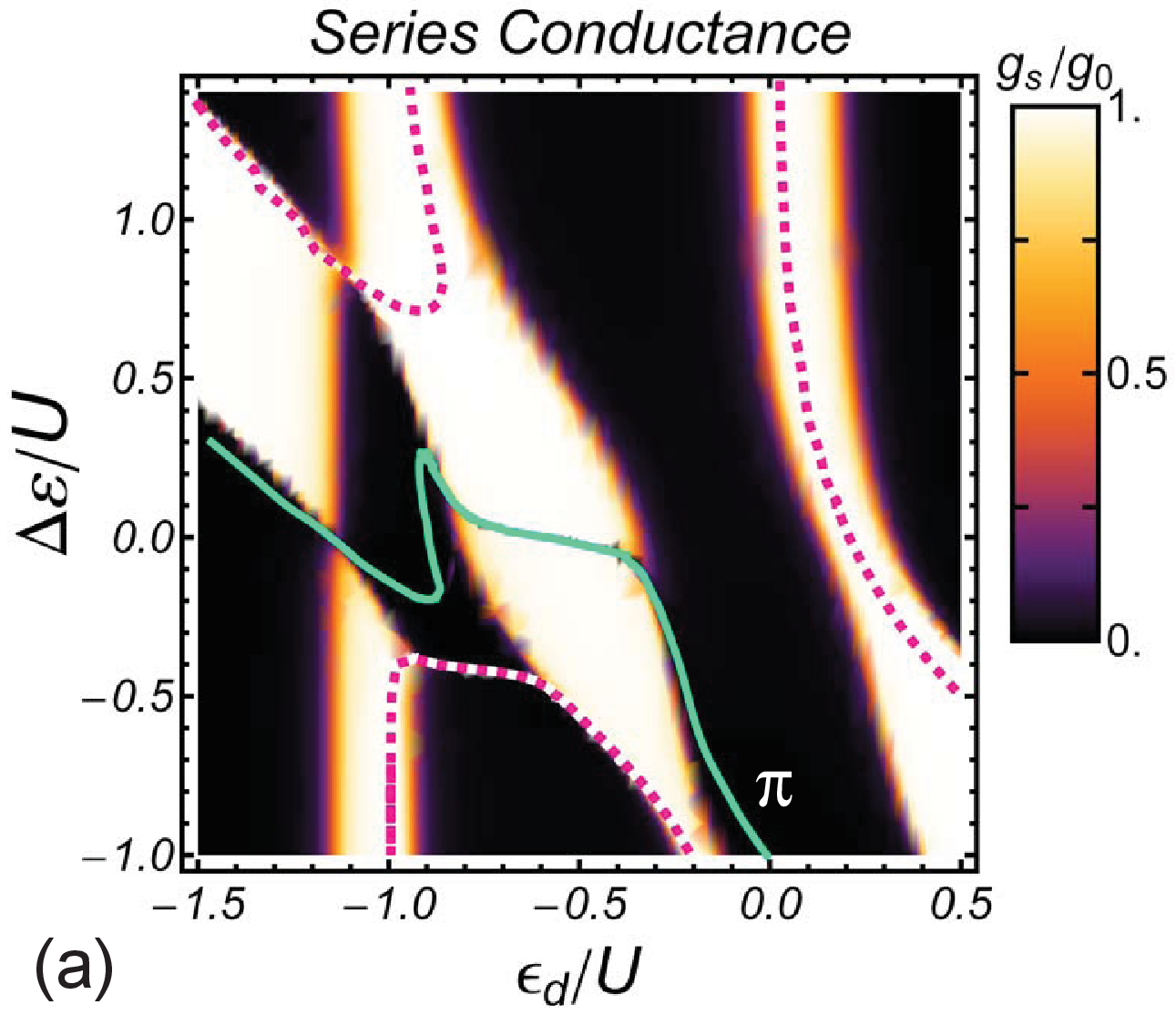}
  \end{minipage}
  \begin{minipage}[t]{0.48\linewidth}
 \includegraphics[width=1\linewidth]{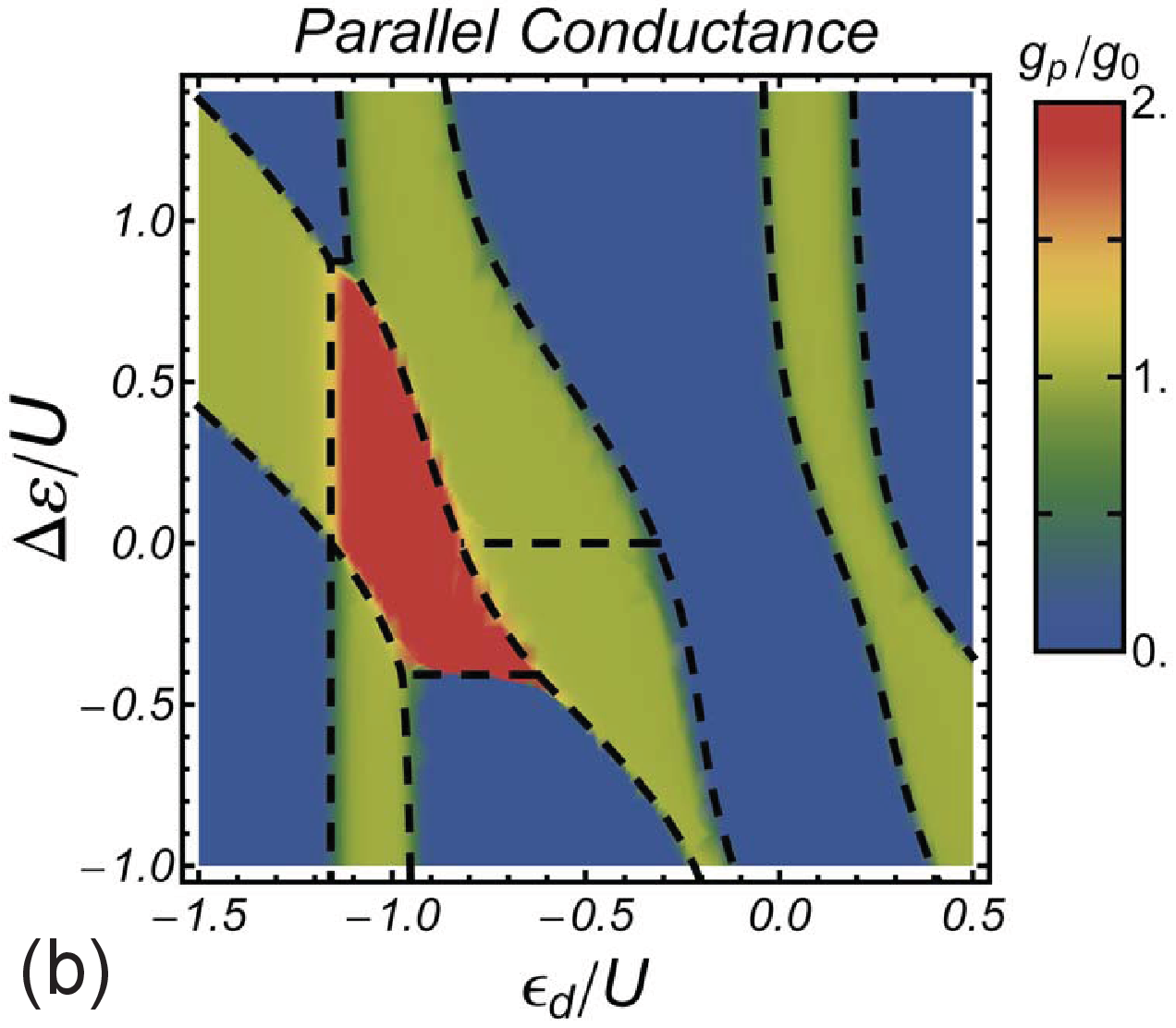}
  \end{minipage}
  \caption{(Color online) 
 Series (a) and parallel (b) 
conductances for $U/(2\pi t) = 1.0$, $\Gamma/t=0.12$ 
and $t=t'$ are 
 plotted as functions of $\epsilon_d/U$ and $\Delta \epsilon/U$.
In (a), the solid line is a contour 
for the phase-shift difference $\delta_\mathrm{e} - \delta_\mathrm{o}=\pi$, 
and the dotted lines are also the contours 
for the values of $\pi/2$ and $3\pi/2$. 
In (b), the dashed lines denote the phase boundaries 
for the isolated TTQD, 
corresponding to Fig.\ \ref{fig:ground_state_isolated_u1_de} (b).  
}
  \label{fig:conductance_de_u1}
 \end{figure}

The series (a) and parallel (b) conductances  
are plotted in Fig.\  \ref{fig:conductance_de_u1}
in the parameter space of $\epsilon_d/U$ and $\Delta \epsilon/U$.  
For comparison,  the phase diagram for $\Gamma \to 0$ 
given in Fig.\ \ref{fig:ground_state_isolated_u1_de} (b) 
is superposed onto Fig.\  \ref{fig:conductance_de_u1} (b)  
with the dashed lines. 
We can see that 
the behavior of the conductances in this parameter space 
also reflects the feature of 
the phase diagram for the isolated TTQD.
In the regions of the odd-number electron filling
the both conductances  $g_\mathrm{s}$ and $g_\mathrm{p}$ 
have the Kondo plateaus with the height of  $2e^2/h$.   
Furthermore, the $S=1$ Kondo effect takes place 
in a trapezoidal region 
near $\Delta \epsilon \simeq 0.0$ and $\epsilon_d/U \simeq -1.0$.
In this region, the series and parallel conductances show a clear contrast,
namely  $g_\mathrm{s} \simeq 0$ while $g_\mathrm{p}\simeq 4e^2/h$.
This feature is the same as what  
is observed in the case of the off-diagonal distortions.

 The solid line in Fig.\  \ref{fig:conductance_de_u1} (a) 
denotes the contour for $\delta_\mathrm{e} - \delta_\mathrm{o} =\pi$,
on which the series conductance becomes zero.  
This contour runs across the region of the three-electron occupancy 
almost horizontally 
in an area with weak distortions  $\Delta \epsilon \simeq 0.0$.
It  associated with a sharp valley of the series conductance, 
which is typical of the SU(4) Kondo effect in the TTQD and 
is seen also for the off-diagonal distortions.
The SU(4) symmetry is caused by the channel degeneracy restored 
along the line at low energies,\cite{Numata2} 
and the phase shifts take the values of 
$\delta_\mathrm{e} \simeq 5\pi/4$ 
and $\delta_\mathrm{o} \simeq \pi/4$.
We can also see that 
the contour for $\delta_\mathrm{e} - \delta_\mathrm{o} =\pi$ 
is deformed significantly, in the trapezoidal $S=1$ Kondo region,  
 from the non-interacting form which is a simple straight line
shown in Fig.\ \ref{fig:conductance_deltaE_u0} (a). 
This could happen, however, continuously with increasing $U$, 
as the ground state evolves adiabatically in the case 
that the quantum dots are coupled to the leads.

The dotted lines in Fig.\  \ref{fig:conductance_de_u1} (a) 
denote the contours for 
 $\delta_\mathrm{e} - \delta_\mathrm{o}=\pi/2$ (above the solid line) 
and $3\pi/2$ (below the solid line), on which the series 
conductance takes the unitary-limit value $g_\mathrm{s} = 2e^2/h$.
It should be noted that a long and very sharp ridge 
emerges in Fig.\  \ref{fig:conductance_de_u1} (a) 
for the series conductance at $\Delta \epsilon /U \simeq -0.4$ 
and $-0.9 \lesssim \epsilon_d/U \lesssim -0.6$.
This sharp ridge runs along the lower end of the trapezoidal 
$S=1$ Kondo region, and reflects the singlet-triplet transition 
taking place in the isolated TTQD cluster for $\Gamma \to 0$.

\begin{figure}[t]
 \leavevmode
 \begin{minipage}[t]{0.49\linewidth}
  \includegraphics[width=1.0\linewidth]{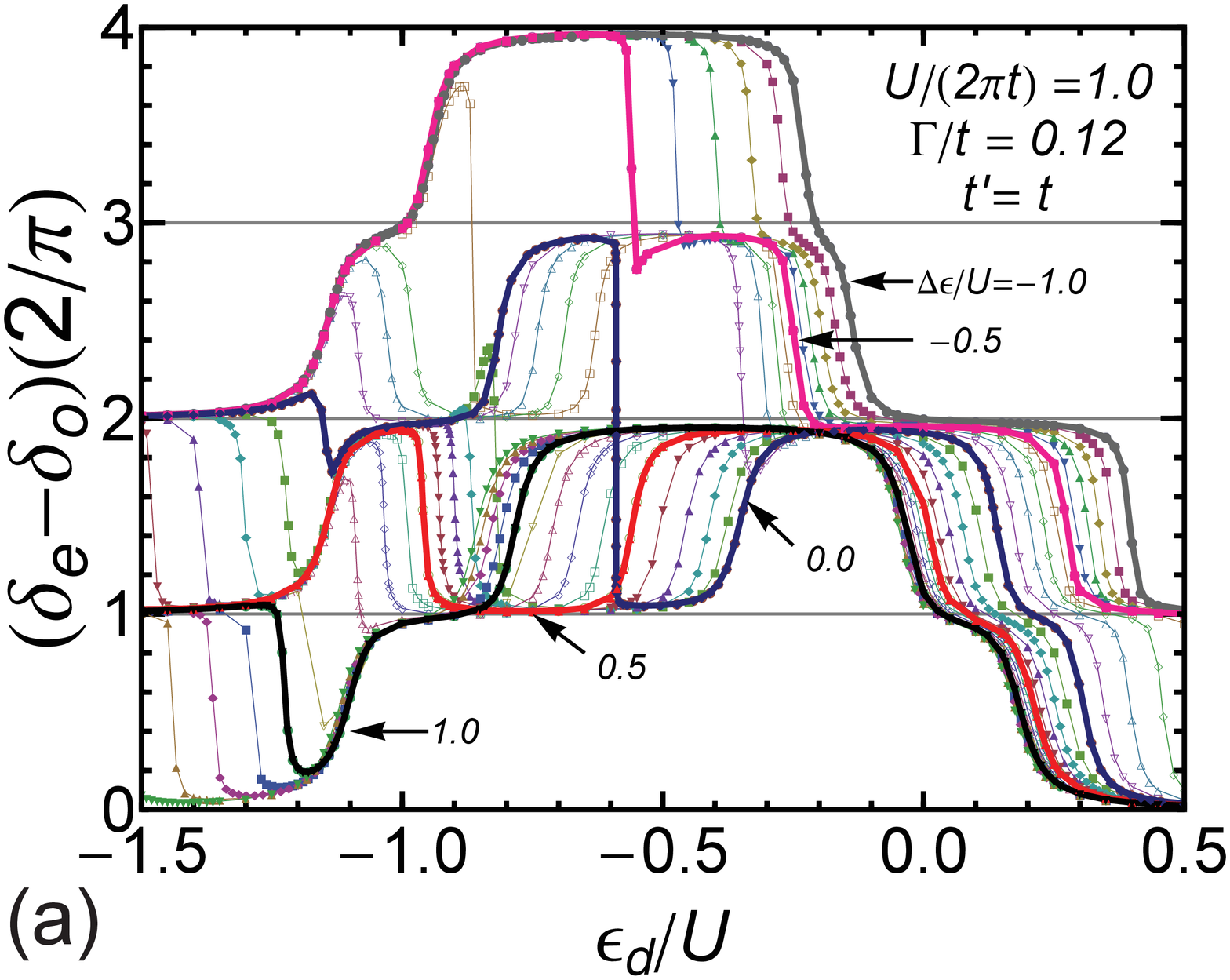}
 \end{minipage}
 \begin{minipage}[t]{0.46\linewidth}
  \includegraphics[width=1.0\linewidth, clip, trim = 0.1cm 0cm 0cm 0cm]{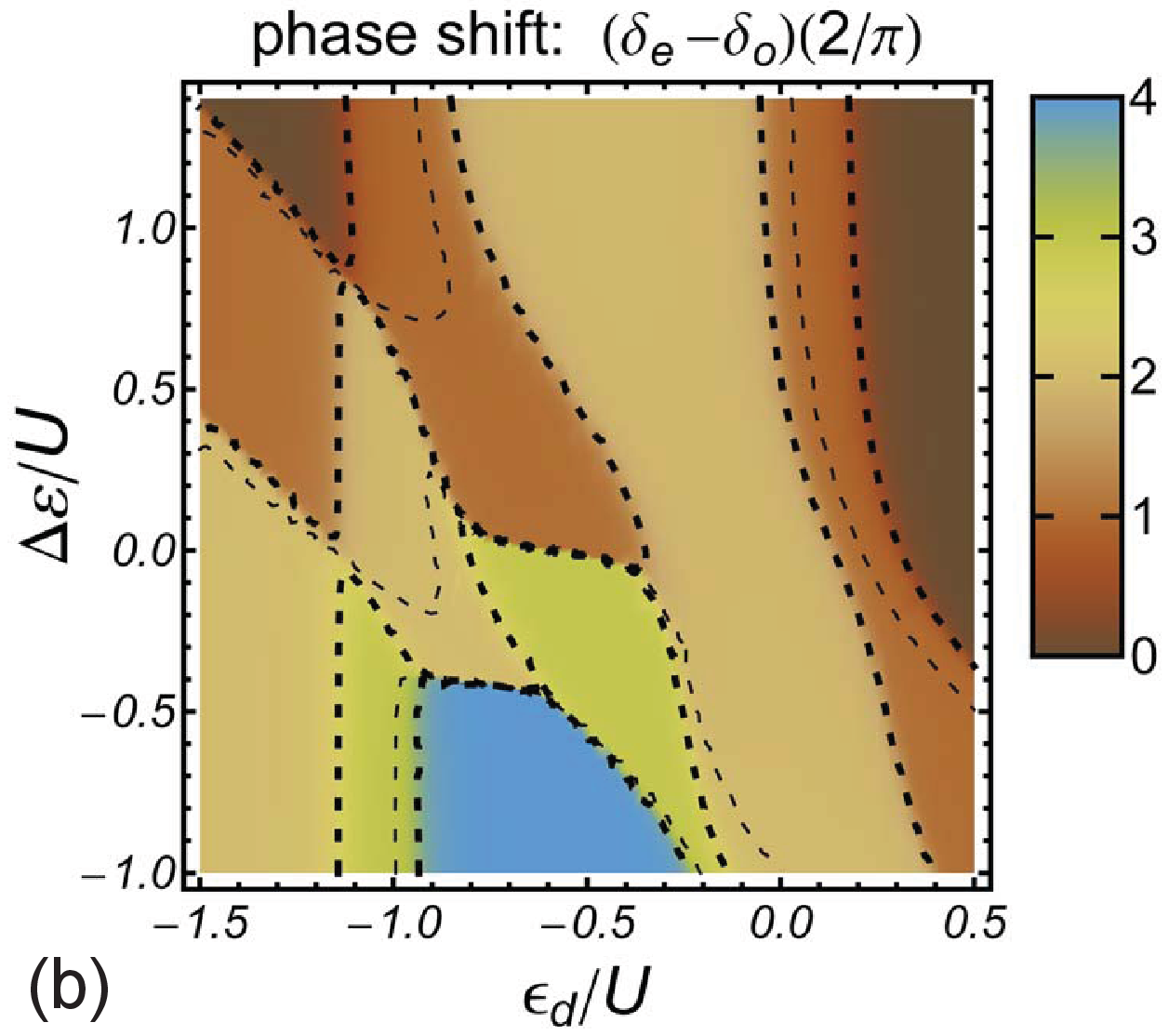}
\end{minipage}
 \caption{(Color online) 
Difference between the even and odd phase shifts 
$\Theta\equiv (\delta_\mathrm{e} -\delta_\mathrm{o})(2/\pi)$ 
for 
 $U/(2\pi t) =1.0$, $\Gamma/t =0.12$ and $t' =t$. 
The left panel (a) shows   $\Theta$  
as a function of $\epsilon_d/U$  for the values of 
$\Delta \epsilon/U=-1.0,\,-0.9,\,-0.8,\, \ldots,$ and $1.4$ 
(in steps of $0.1$ from the top to the bottom).  
The right panel (b) shows 
$\Theta$  in 
the  $\epsilon_d/U$ vs  $\Delta \epsilon/U$ plane. 
The dotted lines  in (b) are the contours for    
$(\delta_\mathrm{e} -\delta_\mathrm{o})(2/\pi) 
= 0.5,\, 1.0,\, 1.5,\, \ldots,$ and $3.5$ 
(in steps of $0.5$ from the top to the bottom).   
}
\label{fig:u1_dedo_de}
\end{figure}


\subsubsection{Phase-shift difference: 
$\,\Theta =
\frac{\displaystyle\protect\mathstrut 2}{\displaystyle\protect\mathstrut \pi}
(\delta_\mathrm{e}-\delta_\mathrm{o})\,$ 
for $\,\Delta \epsilon \neq 0$
}

Figure \ref{fig:u1_dedo_de} (a) shows the results 
of the phase-shift difference  
$\Theta \equiv (\delta_\mathrm{e}-\delta_\mathrm{o})(2/\pi)$ 
as a function of $\epsilon_d/U$ for the values of $\Delta \epsilon/U$ 
varying from $-1.0$ to $1.4$ in steps of $0.1$.
We can clearly see that there are a number of plateaus  
near integer values of 
$\Theta \simeq 1.0,\, 2.0,$ and $3.0$.
Specifically, 
the height of the plateaus approaches very close to 
exact integers in the present case 
because the coupling between the leads and the TTQD is small.  
Although we can recognize that some of them, 
for instance, the ones near $\Theta \simeq 3.0$, 
still deviate from an exact integer, 
these deviations can be controlled  
by tuning $\Gamma$ to be small.
\cite{Numata2}

In order to see the behavior of $(\delta_\mathrm{e}-\delta_\mathrm{o})(2/\pi)$
 in the parameter space,
the results are plotted also 
 in the $\epsilon_d/U$ vs  $\Delta \epsilon/U$ plane 
in  Fig.\ \ref{fig:u1_dedo_de} (b).
In this figure the dotted lines denote 
the contours for $\Theta$, particularly the thicker ones 
are the contours for the half-integer 
values:  $\Theta =0.5$, $1.5$, $2.5$, and $3.5$. 
Each of these thick dotted lines 
runs very closed to the phase boundaries for the isolated TTQD 
shown in Fig.\ \ref{fig:ground_state_isolated_u1_de} (b).
These thick contours, as a whole, cover almost all the boundaries. 
These contour lines of $\Theta$, however, evolve continuously 
from the non-interacting forms as $U$ increases. 
This is because the ground state of the whole system evolve 
adiabatically from a $U=0$ spin singlet 
to a correlated singlet described the local Fermi-liquid theory 
for finite $\Gamma$.

The sharp SU(4) Kondo valley of the series conductance 
corresponds to a steep cliff standing 
at  $-0.8 \lesssim \epsilon_d/U \lesssim -0.3$ 
for small distortions $\Delta \epsilon/U \simeq 0.0$
in Fig.\ \ref{fig:u1_dedo_de} (b). 
At the edge of this cliff,  
$(\delta_\mathrm{e}-\delta_\mathrm{o})(2/\pi)$ varies by an
amount  $2.0$ approximately.
It varies from $\Theta \simeq 1.0 $ 
 (for $\Delta \epsilon\gtrsim 0.0$) to $\Theta \simeq 2.9$ 
(for $\Delta \epsilon\lesssim  0.0$),
 taking the value of $\delta_\mathrm{e}-\delta_\mathrm{o} = \pi$ 
which corresponds to the zero point of $g_\mathrm{s}$ 
in the middle of the cliff.
The slope of this cliff determines the width of 
the SU(4) valley, as that in the case of off-diagonal distortions.

We can see another sharp cliff 
in Fig.\ \ref{fig:u1_dedo_de} (b) 
at the bottom of the $S=1$ Kondo region  
for  $\Delta \epsilon /U \simeq -0.4$ and 
$-0.9 \lesssim \epsilon_d/U \lesssim -0.6$,
where the singlet-triplet transition takes place for the isolated cluster.
Between the top and bottom of this cliff, 
the phase difference 
$\Theta$ varies rapidly from $2.0$ to $4.0$,
which causes the sharp Kondo ridge of the series conductance, 
seen in Fig.\  \ref{fig:conductance_de_u1} (a). 
Note that the slope of the cliff determines the width of the sharp peak 
of $g_\mathrm{s}$.
We can also see a narrow cliff  due to the singlet-triplet transition,
at the top of the $S=1$ Kondo region  
for $\Delta \epsilon /U \simeq 0.8$ and $\epsilon_d/U \simeq -1.2$.
At this cliff, the value of $\Theta$ changes 
from $0.0$ to $2.0$ approximately.

\section{Characteristic energy scale}
\label{sec:TK}


\begin{figure}[t]
 \leavevmode
\begin{minipage}{1.0\linewidth}
\includegraphics[width=0.75\linewidth]{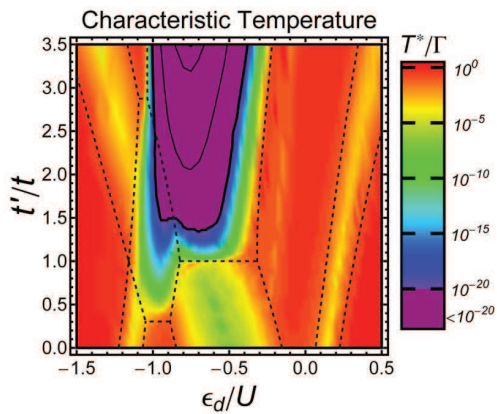}
\end{minipage}
 \caption{(Color online) 
The characteristic energy scale $T^*$ 
in the $\epsilon_d/U$ vs $t'/t$ plane 
for $\Gamma/t =0.25$,  $U/(2\pi t)=1.0$ 
and $\Delta \epsilon=0$. 
The results in the range 
of $-20<\log_{10} (T^*/\Gamma) < 0$ are 
painted in the colors varying from blue to red.
$T^*$ becomes very small in the purple region, 
and the solid lines there denote the contours for  
$\log_{10} (T^*/\Gamma) = -20,\, -40$, and  $-60$. 
The dashed lines denote 
the phase boundary in the limit of $\Gamma \to 0$, 
corresponding to 
Fig.\ \ref{fig:ground_state_isolated_u1_ts} (b).  
 }
 \label{fig:TK_ts}
\end{figure}


The ground-state properties, discussed in the above,
are determined by the behavior of 
the phase shifts for the quasi-particles. 
At finite temperature $T$, for instance, 
the structure of the plateaus and dips 
will be smeared gradually as $T$ increases.  
Nevertheless, the corrections due to finite $T$ 
can be still determined by the local Fermi-liquid theory 
for $T\lesssim T^*$, namely at temperatures 
lower than the characteristic energy scale $T^*$. 
Specifically, in the case the quantum dots have a local moment, 
$T^*$ can be regarded as the  Kondo temperature 
 such that the moment is screened at  $T \ll T^*$.
The value of $T^*$, however, depends sensitively on the parameters 
at each point of the parameter space.  
Therefore, the actual temperature, 
at which the Fermi-liquid behavior can be observed, 
is different depending on the region in the parameter space.

\begin{figure}[t]
 \leavevmode
\begin{minipage}{1.0\linewidth}
\includegraphics[width=0.75\linewidth]{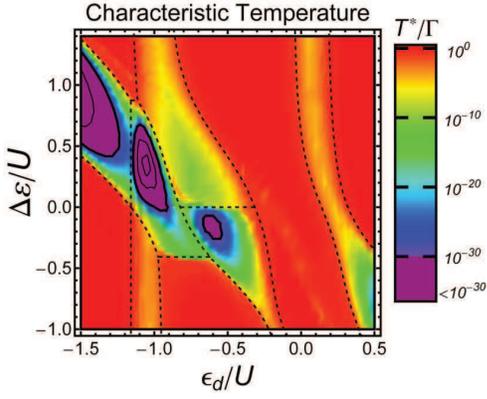}
\end{minipage}
 \caption{(Color online) 
The characteristic energy scale $T^*$ 
in the $\epsilon_d/U$ vs $\Delta \epsilon/U$ plane 
for $\Gamma/t =0.12$,  $U/(2\pi t)=1.0$, and $t'=t$. 
The results in the range 
of $-30<\log_{10} (T^*/\Gamma) < 0$ are 
painted in the colors varying from blue to red.
$T^*$ becomes very small in the purple region, 
and the solid lines there denote the contours for  
$\log_{10} (T^*/\Gamma) = -30,\, -60$, and $-90$. 
The dashed lines denote 
the phase boundary in the limit of $\Gamma \to 0$, 
corresponding to 
Fig.\ \ref{fig:ground_state_isolated_u1_de} (b).  
 }
 \label{fig:TK_de}
\end{figure}

In the NRG approach, the crossover from the high energy  
to the low energy Fermi-liquid regime can be 
seen in the trajectory of the low-energy levels of 
the discretized Hamiltonian $H_N$   
defined in Appendix \ref{sec:NRG_approach}.
The trajectory evolves 
as the number of orbitals $N$ of the conduction band increases.
In the present work we have estimated  $T^*$ 
through $N^*$ that is a particular value of $N$,  
at which the trajectory {\it almost\/} enters the low-energy regime,
as 
\begin{align}
T^* = D\,\Lambda^{-(N^*-1)/2}\;. 
\end{align}
Therefore, $T^*$ is the typical energy scale of 
the low-lying excitations 
of a finite NRG chain with $N=N^*$.
The values of $T^*$ determined in 
this way have some ambiguities, 
especially in the case where the crossover is gradual,
and our definition tends to give a smaller value 
for the characteristic energy scale. 
Nevertheless, the relative variations of $T^*$ 
in the parameter space can be extracted reasonably 
as shown in Figs.\ \ref{fig:TK_ts} and \ref{fig:TK_de}. 
 We have also confirmed that $T^*$ determined in this way 
 is consistent with the ones we had estimated 
from the entropy in the previous work
 for the equilateral triangle.\cite{Numata2}
We will see below that the variations of $T^*$ 
can be understood roughly through the distribution 
of the charge and spin in the even and odd parity orbitals, 
 described in Appendix \ref{sec:even_odd}.

 \subsection{$T^*$  vs  off-diagonal distortions ($t' \neq t$)}

 \begin{figure}[t]
 \leavevmode
 \begin{minipage}[t]{0.49\linewidth}
  \includegraphics[width=1\linewidth]{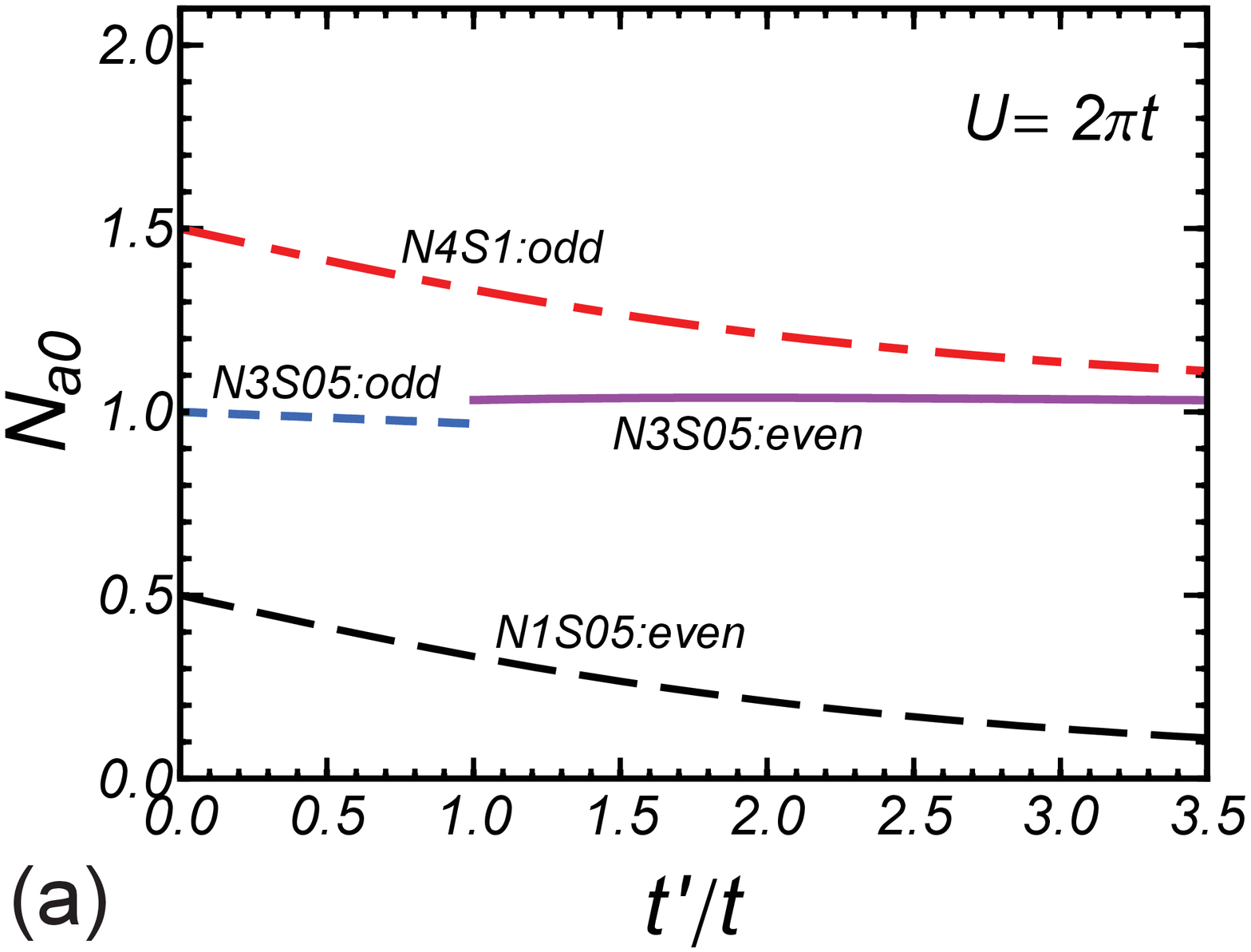}
 \end{minipage}
 \begin{minipage}[t]{0.49\linewidth}
  \includegraphics[width=1\linewidth]{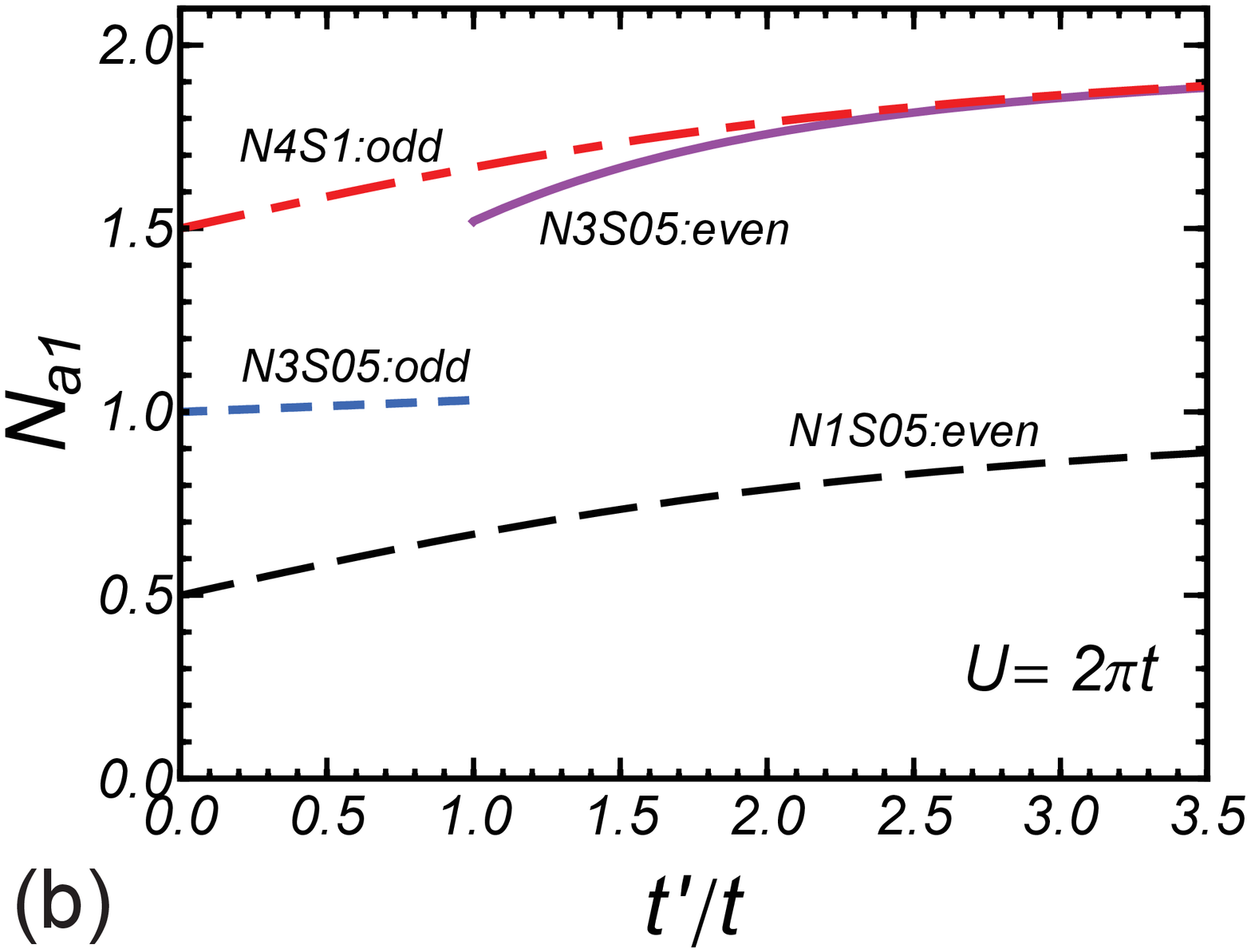}
 \end{minipage}
\caption{(Color online) 
Charge distribution in the isolated TTQD,  
 $N_{a,i} \equiv \sum_\sigma \langle n_{a,i\sigma}\rangle $   
is plotted as a function of  $t'/t$  
for (a) $N_{a0}$ the apex site, 
and for (b) the even $a_1$ orbital
described in Appendix \ref{sec:even_odd}.
The average is taken with respect to some of 
the eigenstates of $\mathcal{H}_\mathrm{dot} 
\equiv \mathcal{H}_\mathrm{dot}^0+\mathcal{H}_\mathrm{dot}^U$,
and labelled by the occupation number, spin, and parity 
(\lq\lq N3S05:even" denotes, 
\lq\lq$N_\mathrm{tot}=3$", $S=1/2$, and an even parity). 
The parameters are chosen to be 
$\Gamma=0$,  $U/(2\pi t) = 1.0$, 
and $\epsilon_\mathrm{apex} =\epsilon_d$.
Note that the occupation number takes the form 
$N_\mathrm{tot} = N_{a,0}+N_{a,1}+N_{b,1}$.
}
  \label{fig:charge_distribution_isolated_ts}
 \end{figure}

The results for $T^*$  are plotted in Fig.\  \ref{fig:TK_ts} 
using a logarithmic scale, 
as a function of $\epsilon_d/U$ and $t'/t$   
 for $\Gamma/t=0.25$ and $U/(2\pi t) =1.0$. 
The phase diagram of the isolated TTQD 
given in  Fig.\ \ref{fig:ground_state_isolated_u1_ts} (b)  
is also superposed onto   Fig.\  \ref{fig:TK_ts} 
with the dashed lines, as a guide for the eye. 
We can see that the variation of 
$T^*$ in the parameter space also relates to the plateaus 
of $\Theta$
[see Fig.\ \ref{fig:u1_dedo_ts} in Appendix \ref{sec:even_odd}].
The energy scale $T^*$ is large     
in the case that the quantum dots have no local moment,
and it  becomes smaller when the TTQD has a local moment.

We can see, nevertheless, $T^*$ is still relatively high  
in the regions of $N_\mathrm{tot} \simeq 1.0$ and $5.0$,
in which the $S=1/2$ Kondo effect  takes place.
The $S=1/2$ moment 
for  $N_\mathrm{tot} \simeq 1.0$ 
is caused by a single electron, 
which enters  an even-parity state 
and stays mainly at the $a_1$ orbital adjacent to the leads 
[see Fig.\ \ref{fig:even_odd}].
Thus the screening can be completed relatively easily in this case.
Figure \ref{fig:charge_distribution_isolated_ts} 
shows the occupation number $N_{a0}$  ($N_{a1}$) 
in the $a_0$ ($a_1$) orbital, as a function of $t'/t$,\
for the limit of  $\Gamma \to 0$:
note that $a_0$ corresponds to the apex site.
The dashed line which is labelled  \lq\lq N1S05:even" 
shows the average with respect 
to the lowest eigenvector in the subspace 
with $N_\mathrm{tot}=1$, $S=1/2$, and an even parity.
We can see that $N_{a1}$ approaches to $1.0$ as $t'/t$ increases.
In the opposite case, at $t'/t = 0.0$,  
the occupation number 
for $a_0$ and that for $a_1$ coincide  $N_{a0} = N_{a1} = 0.5$, 
but still the charge and spin fluctuations are not suppressed 
because these orbitals are still at quarter filling.

In the five-electron region for $t'/t>1.0$,
the eigenvector $|\Phi^{(5)}_\mathrm{odd}\rangle$ 
for the isolated TTQD can be expressed 
in the form of Eq.\ \eqref{eq:N5_odd_cluster}. 
The local moment in this case stays 
at the $b_1$ orbital, 
which is also adjacent to one of the leads 
[see Fig.\ \ref{fig:even_odd}],
and the conduction electrons can screen the moment 
through the usual kinetic exchange mechanism.
For another five-electron region at $0 \leq t'/t<1.0$,
the eigenvector $|\Phi^{(5)}_\mathrm{even}\rangle$ 
can be written in the form of Eq.\ \eqref{eq:N5_even_cluster}. 
The averages  $N_{a0}$ and $N_{a1}$ with respect 
to this state coincide with those 
with respect to the Nagaoka state $|\Phi^{(4)}_\mathrm{odd}\rangle$
defined in Eq.\ \eqref{eq:N4_Nagaoka_cluster},
and the results are plotted 
 in Fig.\ \ref{fig:charge_distribution_isolated_ts} 
with the dot-dash line labelled  \lq\lq N4S1:odd".
We can see that for $0 \leq t'/t < 1.0$ 
a single {\it hole\/} with a spin $1/2$ enters both 
of the even orbitals, and for $t'/t\to 0$ 
these orbitals approach to quarter filling in the hole picture.
Therefore, the screening 
is not suppressed so much also in this five-electron region.

\begin{figure}[t]
\begin{minipage}[t]{0.49\linewidth}
 \includegraphics[width=1.0\linewidth]{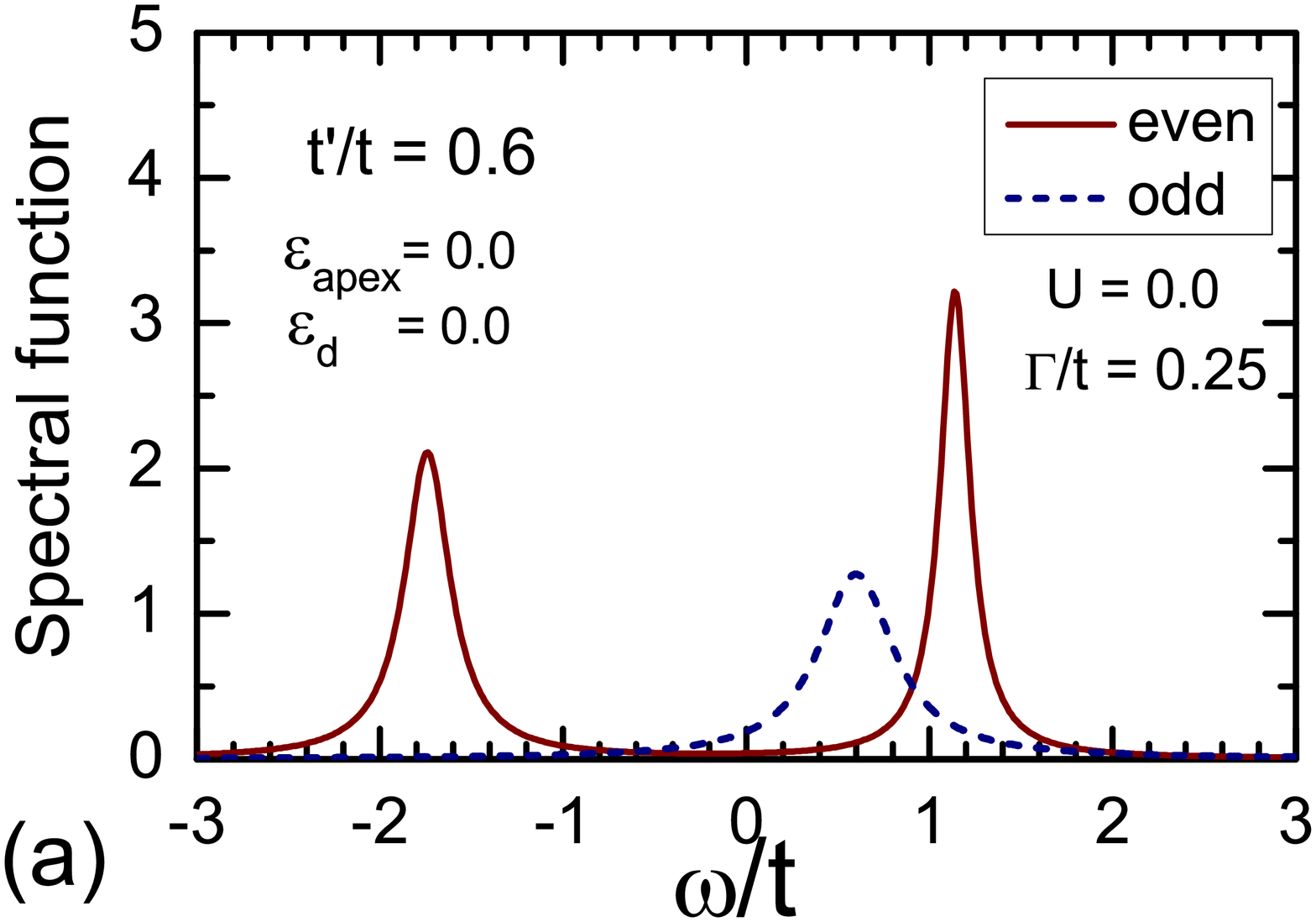}
\end{minipage}
\begin{minipage}[t]{0.49\linewidth}
 \includegraphics[width=1.0\linewidth]{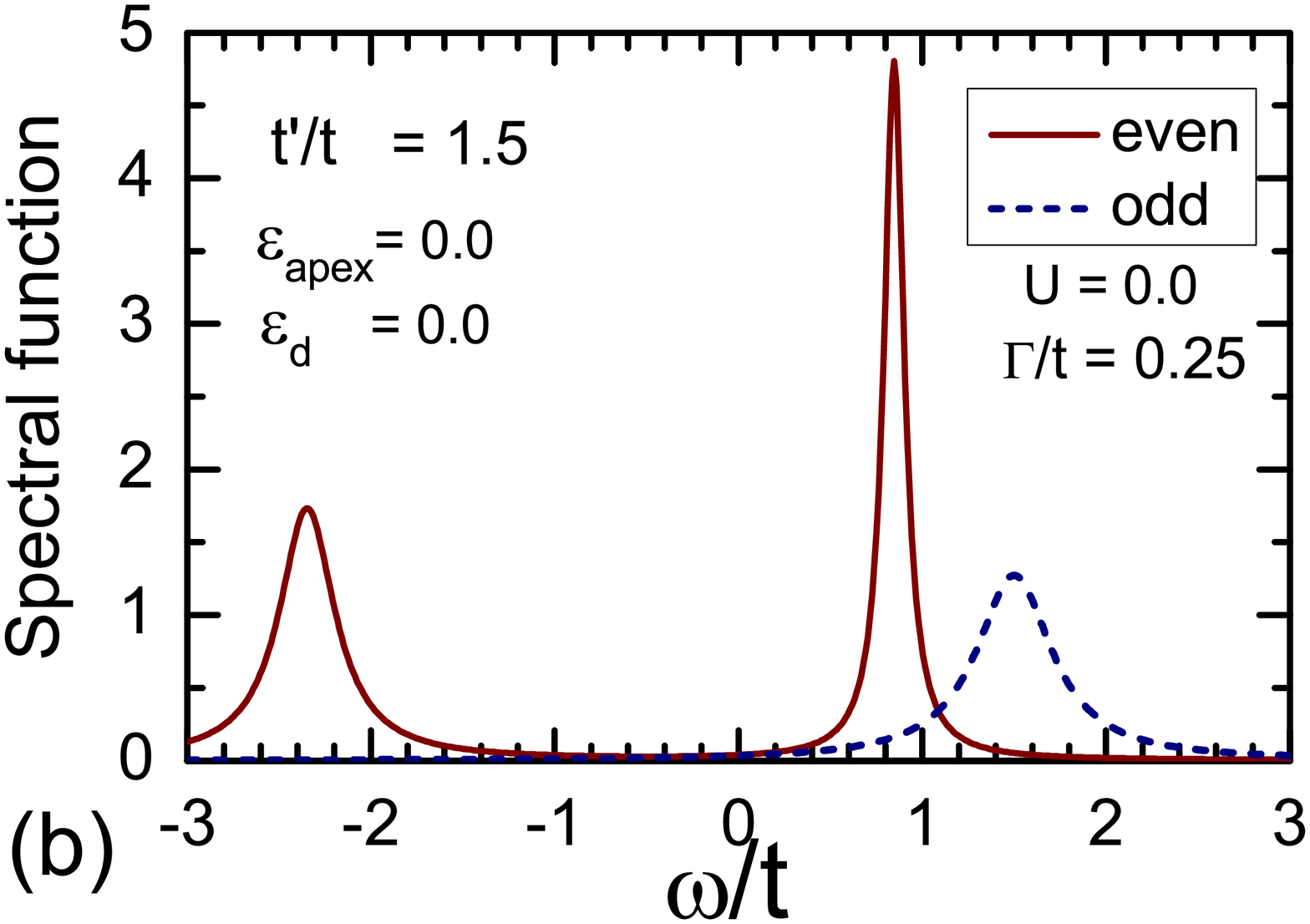}
\end{minipage}
\caption{(Color online)
Spectral functions in the noninteracting case 
for (a) $t'/t =0.6$ and (b) $t'/t =1.5$. 
The solid line is the even component 
$-\mathrm{Im}\, [G_{a0}(\omega)+ G_{a1}(\omega)]/\pi$, 
and the dotted line is the odd component 
$-\mathrm{Im}\, G_{b0}(\omega)/\pi$, 
defined in Appendix \ref{sec:even_odd}.
The other parameters are chosen to be 
 $\Delta \epsilon=0.0$,  $\epsilon_d=0.0$, 
and  $\Gamma/t = 0.25$. 
 }
\label{fig:spec_u0_ts}

\end{figure}

The screening temperature becomes small 
in the three and four electron regions, 
namely $ -1.1 \lesssim \epsilon_d \lesssim -0.3$
 in Fig.\  \ref{fig:TK_ts}. 
We can see in the three electron region, however, 
$T^*$ is still much higher for $t'/t \lesssim 1.0$ 
than for $t'/t \gtrsim 1.0$  despite 
the local moment in the TTQD is $S=1/2$ in both of the cases.  
For $t'/t \lesssim 1.0$, the ground state is an odd-parity state,
and one of the three electrons enters the $b_1$ orbital, 
and the other two electrons enter almost 
equally to the $a_0$ and $a_1$ orbitals. 
This can be confirmed through the dotted line 
labelled \lq\lq N3S05:odd"  
 in Fig.\ \ref{fig:charge_distribution_isolated_ts}. 
We can also see in Fig.\ \ref{fig:spec_u0_ts} (a) 
that the third electron enters 
the odd $b_1$ orbital for $t'/t \lesssim 1.0$ 
in the noninteracting limit.
In contrast, for $t'/t \gtrsim 1.0$ 
the ground state is an even parity state,
and the solid line labelled \lq\lq N3S05:even" 
denotes average with respect to this state.  
We see that $N_{a1}$ approaches to $2.0$ as $t'/t$ increases, 
while the occupation of the apex site 
is almost unchanged  $N_{a0} \simeq 1.0$,
and thus the occupation of the $b_1$ orbital is decreasing 
in this case. 
Therefore, the local moment  
is mainly due to the electron staying at the apex site.
Thus the screening is protracted significantly 
because the charge and spin fluctuations are suppressed 
at the nearly filled $a_1$ orbital, 
over which the conduction electrons come 
to screen the moment. 
We have also confirmed that along 
the sharp conductance valley caused by the SU(4) Kondo effect,  
at $t'/t \simeq 1.0$ and $-0.8\lesssim \epsilon_d/U \lesssim -0.3$, 
the energy scale  $T^*$ is enhanced due to the orbital degeneracy.
Note that the variations 
in the spin and charge configurations   
near the SU(4) symmetric point  
becomes wider in the TTQD than the double dot.


The properties of the local moment in the three-electron region 
also reflect the feature of the one-particle state  
which emerges as a peak of the conductances shown 
in Fig.\  \ref{fig:conductance_ttqd_u0}, 
and also the corresponding 
spectral function is shown in Fig.\ \ref{fig:spec_u0_ts}. 
Specifically for $t'/t \gtrsim 1.0$ the orbital degeneracy is lifted 
such that $E_\mathrm{o}^{(1)} > E_\mathrm{e,+}^{(1)}$.
Therefore, after two electrons occupy the lowest even-parity orbital 
with the energy $E_\mathrm{e,-}^{(1)}$,  
the third electron enters the resonance state 
corresponding to the excited even-parity orbital 
 which appears in Fig.\ \ref{fig:spec_u0_ts} (b) 
as the central peak.
This state has a dominant spectral weight 
 in the apex site, and the resonance width 
is narrower than $\Gamma$ already in the noninteracting case 
[see also Eq.\ \eqref{eq:peak_even_excited} 
in Appendix \ref{sec:even_odd}].
It should also be noted that the width of 
this resonance determines $T^*$ in the noninteracting limit.
For finite $U$, this peak may evolve into a Kondo resonance 
whose width is reduced further by the Coulomb interaction 
to the value of the order of $\,T^*$.

The wavefunction for the Nagaoka state has an odd parity,
and the charge distribution of this state 
is shown in Fig.\ \ref{fig:charge_distribution_isolated_ts}, 
with the dot-dash line labelled  \lq\lq N4S1:odd".
We can see that $N_{a0}$ and $N_{a1}$ for the Nagaoka 
state are similar to those for the three electrons 
state \lq\lq N3S05:even". 
The fraction of the local moment stays at the apex site, 
and this explains the reason why $T^*$ is small also in this case.
One extra electron enters mainly the $b_1$ orbital, 
and it provides half of 
the $S=1$ moment which can be screened at high temperature at   
the first stage of the two-stage Kondo screening.\cite{Numata2}


\subsection{$T^*$  vs diagonal distortions 
($\epsilon_\mathrm{apex} \neq \epsilon_d$) }

\begin{figure}[t]
  \leavevmode
 \begin{minipage}[t]{0.49\linewidth}
  \includegraphics[width=1\linewidth]{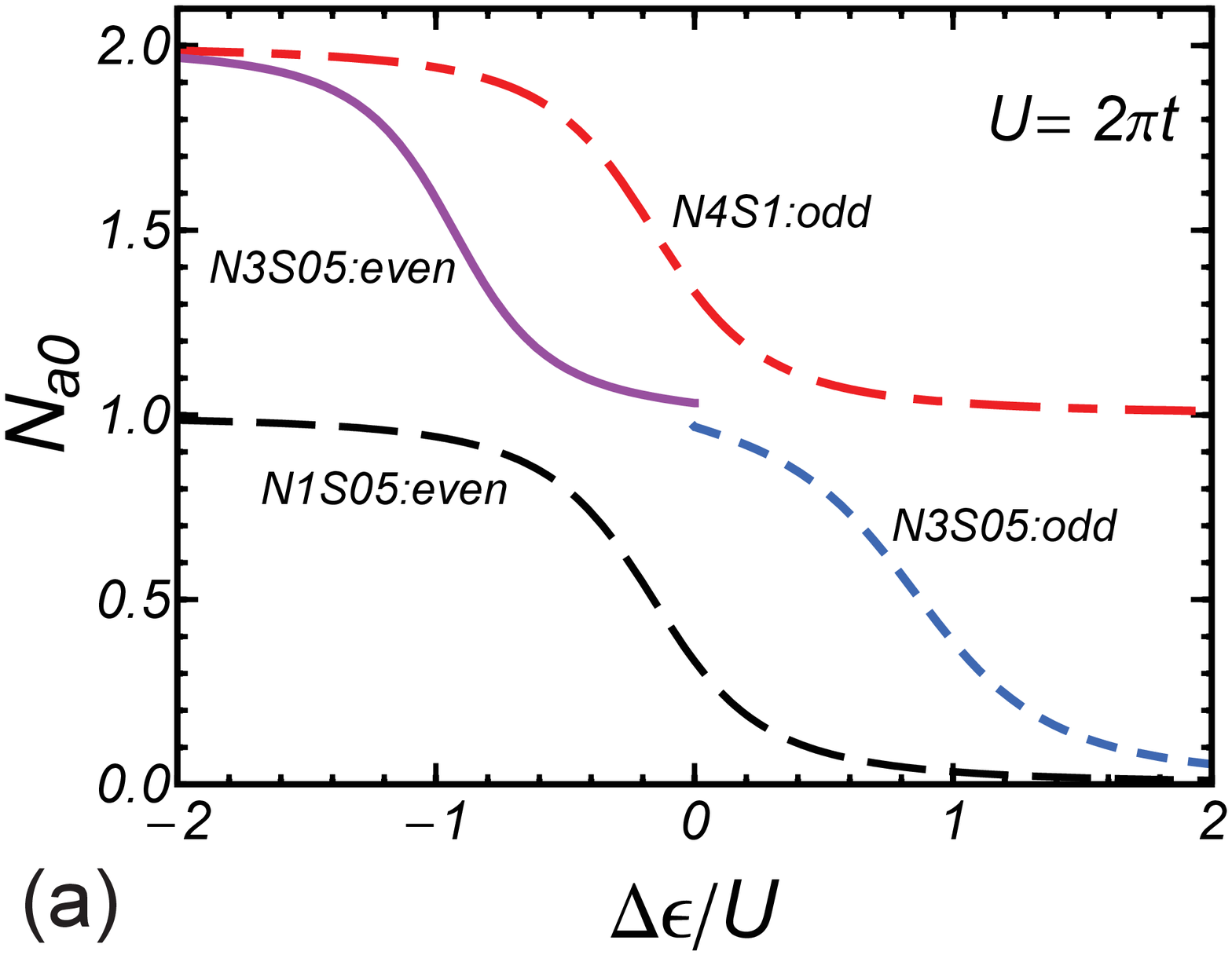}
 \end{minipage}
 \begin{minipage}[t]{0.49\linewidth}
  \includegraphics[width=1\linewidth]{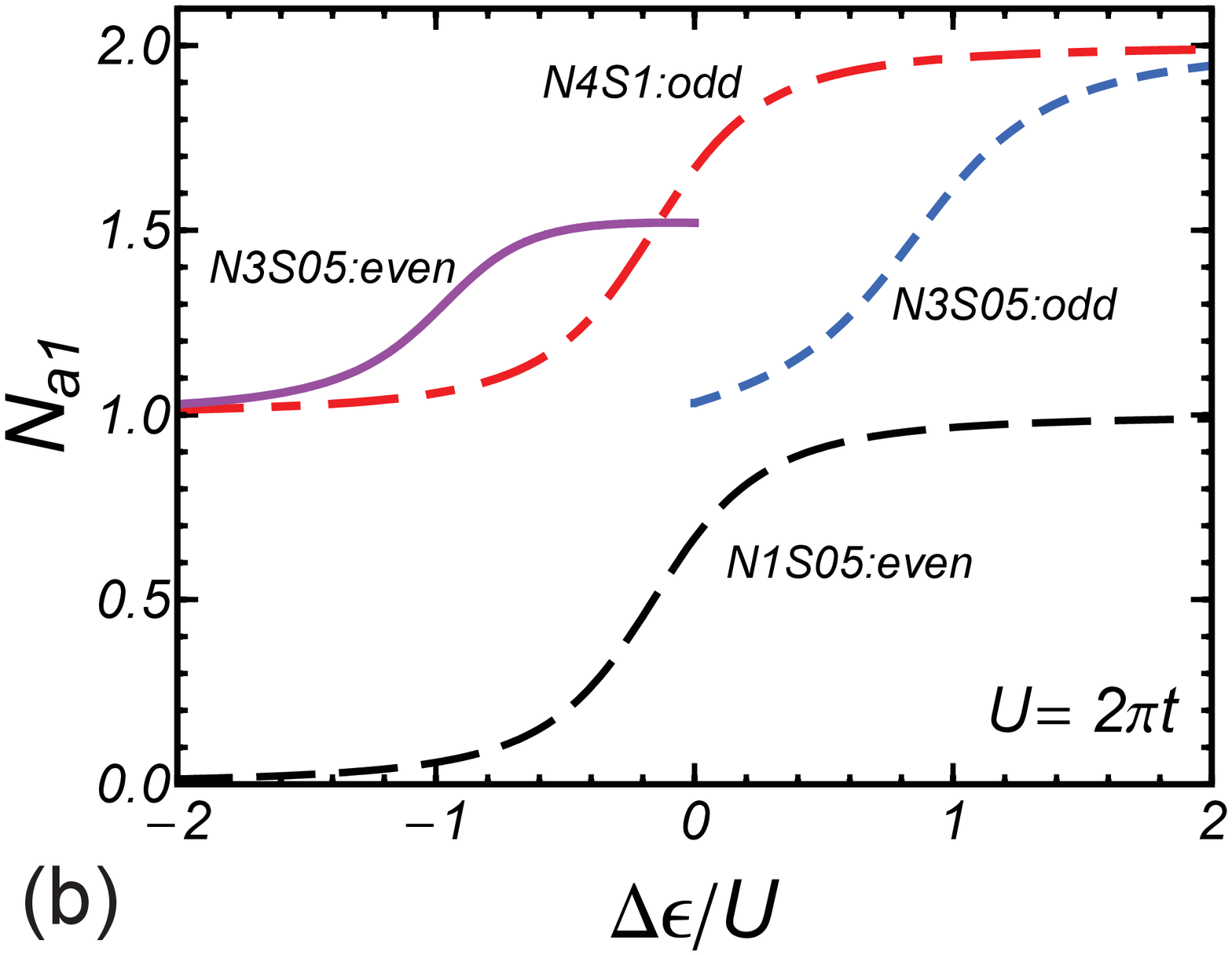}
 \end{minipage}
  \caption{(Color online) 
Charge distribution in the isolated TTQD, 
 $N_{a,i} \equiv \sum_\sigma \langle n_{a,i\sigma}\rangle $, 
is plotted as a function of  $\Delta \epsilon/U$
for (a) $N_{a0}$ the apex site, 
and for (b) the even $a_1$ orbital
described in Appendix \ref{sec:even_odd}.
The average is taken with respect to some of 
the eigenstates of $\mathcal{H}_\mathrm{dot} 
\equiv \mathcal{H}_\mathrm{dot}^0+\mathcal{H}_\mathrm{dot}^U$,
and labelled by the occupation number, spin, and parity 
(\lq\lq N4S1:odd" denotes, $N_\mathrm{tot}=4$, $S=1$, 
and an odd parity). 
The parameters are chosen to be 
$\Gamma=0$,  $U/(2\pi t) = 1.0$, 
and $t'=t$. 
Note that  
 $\Delta \epsilon \equiv \epsilon_\mathrm{apex} -\epsilon_d$,
and the occupation number takes the form 
 $N_\mathrm{tot} = N_{a,0}+N_{a,1}+N_{b,1}$,  
}
  \label{fig:charge_distribution_isolated_de}
 \end{figure}

The characteristic energy scale $T^*$ 
for the TTQD with the diagonal distortions 
is plotted in Fig.\  \ref{fig:TK_de}
using a logarithmic scale, 
as a function of $\epsilon_d/U$ and $\Delta \epsilon/U$  
for $U/(2\pi t) =1.0$. 
The phase diagram of the isolated TTQD 
given in  Fig.\ \ref{fig:ground_state_isolated_u1_de} (b) 
is also superposed onto Fig.\  \ref{fig:TK_de}.
Note that the coupling between the leads and quantum dots 
is chosen to be $\Gamma/t=0.12$, which is smaller than 
that we chose for the off-diagonal case.
Therefore, the absolute values of $T^*$  
become smaller than those in Fig.\ \ref{fig:TK_ts}.
We saw in the above that 
the Kondo screening is sensitive to the 
 electronic structure of the TTQD for the off-diagonal distortions $t' \neq t$.
In order to see the results in a similar way,
the charge distribution in the even and odd orbitals 
in the isolated TTQD for $\Gamma \to 0$ 
is also plotted in Fig.\ \ref{fig:charge_distribution_isolated_de}
as a function $\Delta \epsilon$ 
($\equiv \epsilon_\mathrm{apex} - \epsilon_d$).

We can see that $T^*$ is suppressed also in the region 
of $N_\mathrm{tot} \simeq 1.0$ 
at $\Delta \epsilon \lesssim 0.0$,
which corresponds to the area of $\epsilon_d/U \gtrsim 0.3$ 
at the right bottom of Fig.\  \ref{fig:TK_de}. 
In this parameter region, the potential profile of the onsite 
energy is such that $\epsilon_\mathrm{apex} < E_F$ and $\epsilon_d > E_F$,  
with the Fermi energy $E_F \equiv 0.0$.
Therefore, the single electron 
enters mainly the apex site, 
and  the other two dots are almost empty. 
This can also be confirmed through 
the charge distribution plotted 
with the dashed line labelled \lq\lq N1S05:even" 
in Fig.\ \ref{fig:charge_distribution_isolated_de}.
For $\Delta \epsilon \lesssim 0.0$, 
the occupation number $N_{a0}$ for the apex site 
approaches to $1.0$  while 
that for the even $a_1$ orbital, $N_{a1}$, almost vanishes. 
Therefore,  the screening of the local moment 
is achieved through a super-exchange process 
by the conduction electrons which come to the apex site 
over the potential barrier at the other two dots,
and thus $T^*$ decreases in this parameter region.

\begin{figure}[t]
\begin{minipage}[t]{0.49\linewidth}
 \includegraphics[width=1.0\linewidth]{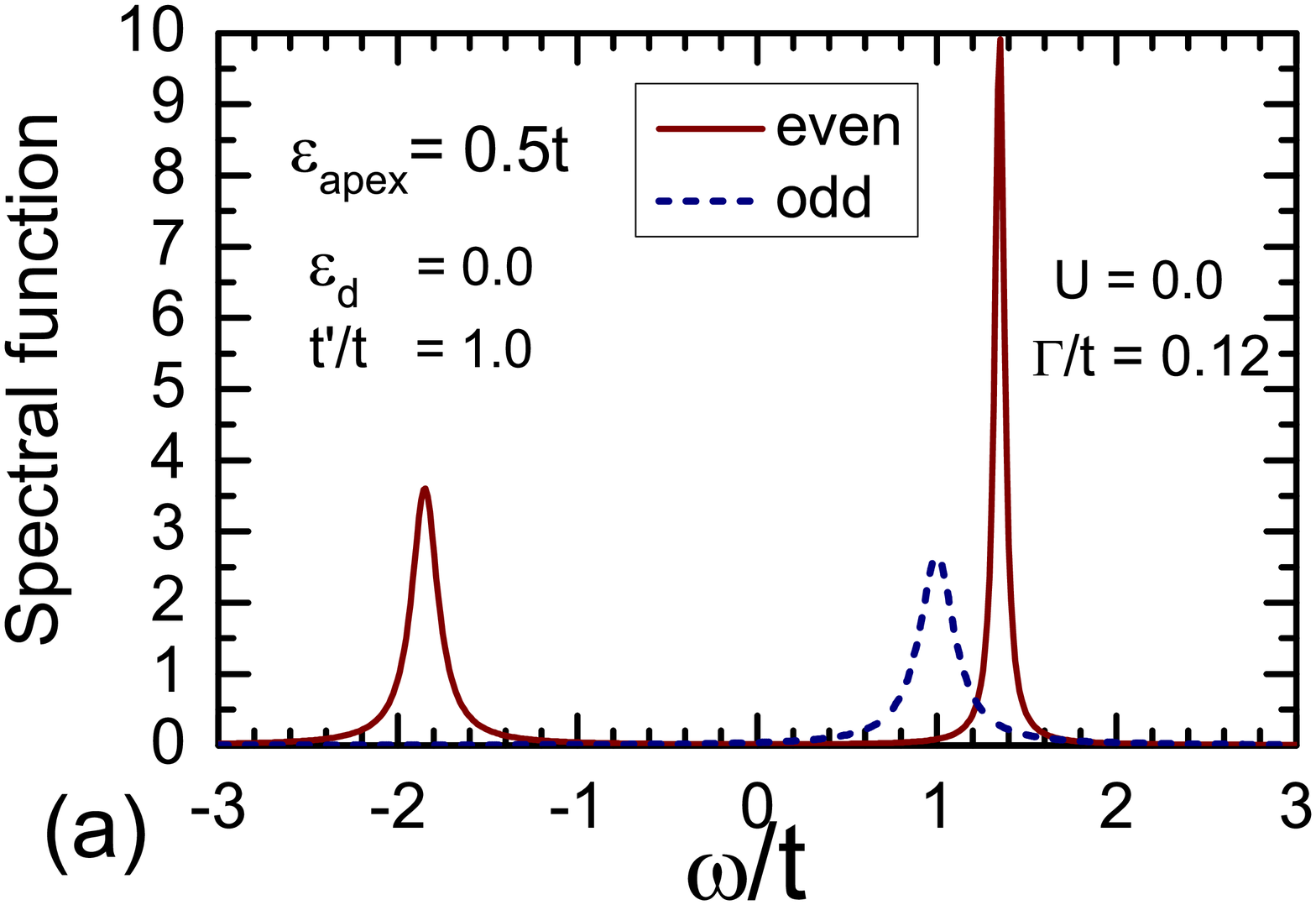}
\end{minipage}
\begin{minipage}[t]{0.49\linewidth}
 \includegraphics[width=1.0\linewidth]{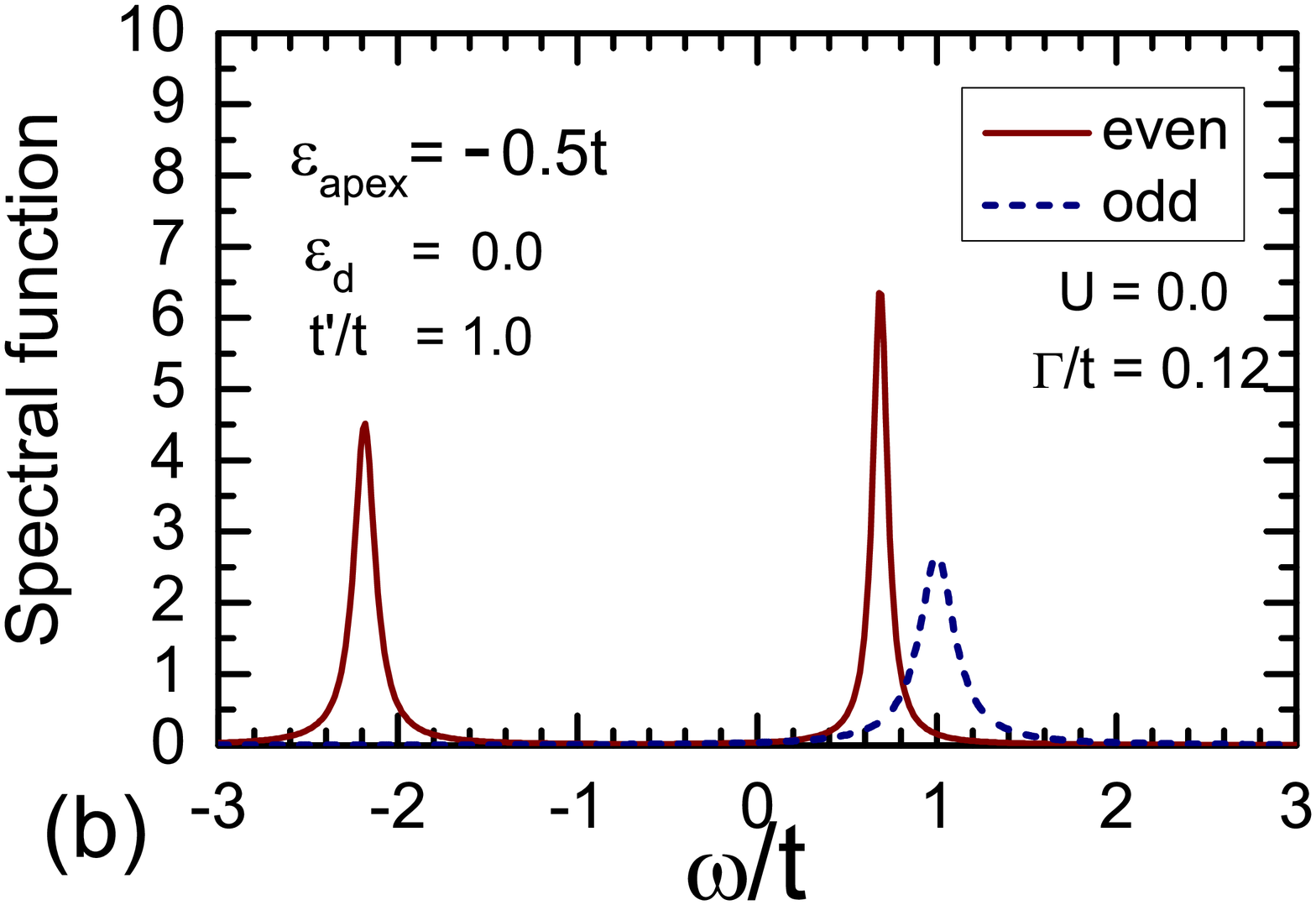}
\end{minipage}
\caption{(Color online)
Spectral functions in the noninteracting case 
for (a) $\Delta \epsilon=0.5 t$ and (b) $\Delta \epsilon=-0.5t$. 
The solid line is the even component 
$-\mathrm{Im}\, [G_{a0}(\omega)+ G_{a1}(\omega)]/\pi$, 
and the dotted line is the odd component
$-\mathrm{Im}\, G_{b0}(\omega)/\pi$, 
defined in Appendix \ref{sec:even_odd}.
The other parameters are chosen to be 
 $t'/t=1.0$, 
 $\epsilon_d=0.0$, 
and  $\Gamma/t = 0.12$. 
 }
\label{fig:spec_u0_de}

\end{figure}

Similarly, 
$T^*$ is suppressed also in the five-electron region  
for $\Delta \epsilon > 0.0$, which corresponds to 
the area for $\epsilon_d/U \lesssim -1.2$  at the  top left 
of Fig.\  \ref{fig:TK_de}. 
The ground state in this parameter region has 
 an even parity, and in the limit of $\Gamma \to 0$ 
the eigenvector is given by  $|\Phi^{(5)}_\mathrm{even}\rangle$ 
in Eq.\ \eqref{eq:N5_even_cluster}. 
The average number of electrons  $N_{a0}$ and $N_{a1}$ 
for this state coincide with those 
with respect to $|\Phi^{(4)}_\mathrm{odd}\rangle$
defined in Eq.\ \eqref{eq:N4_Nagaoka_cluster}.
The results are shown  
in Fig.\ \ref{fig:charge_distribution_isolated_de},
with the dot-dash line labelled  \lq\lq N4S1:odd".
We can see that
a single {\it hole} with a spin $1/2$ stays in the apex site
for $\Delta \epsilon \gtrsim 0.0$, 
and the other two dots are almost doubly occupied.
Therefore, $T^*$ becomes very small in this case.
This can also understood from 
the feature of the spectral function, shown in Fig.\ \ref{fig:spec_u0_de}.
The Fermi level for the five electrons in this case is situated  
in the middle of the sharp peak at $\omega \simeq 1.2 t$ 
in Fig.\ \ref{fig:spec_u0_de} (a).
The width of this resonance corresponds to $T^*$ 
in the noninteracting limit, 
and the peak will become much narrower for finite Coulomb interaction $U$.
There is another five-electron region for $\Delta <0.0$,
where the eigenvector 
is given by $|\Phi^{(5)}_\mathrm{odd}\rangle$ in 
 Eq.\ \eqref{eq:N5_odd_cluster} for the isolated TTQD.
The local moment in this case stays 
at the $b_1$ orbital, which is close to one of the leads,
and thus the screening can be achieved by the 
usual kinetic exchange mechanism of the $S=1/2$ Kondo effect.

The screening temperature becomes small also 
in the three region, and in the four electron region. 
In the three electron region for $\Delta \epsilon \gtrsim 0.0$ 
the ground state has an odd parity, 
and the charge distribution for this state is plotted 
in Fig.\ \ref{fig:charge_distribution_isolated_de},
with the dotted line labelled \lq\lq N3S05:odd".
We can see that at $\Delta \epsilon \simeq 1.0$ 
the three electrons distribute almost homogeneously  
as $N_{a0} \simeq N_{a1} \simeq 1.0$, 
and thus $N_{b1} \simeq 1.0$. 
Then, as $\Delta \epsilon$ increases,
a single electron in the apex site  moves 
 towards the even-parity $a_1$ orbital, 
and the occupation numbers approach to $N_{a0} \simeq 0.0$ and 
$N_{a1} \simeq 2.0$, keeping the occupation of the 
odd-parity orbital almost unchanged  $N_{b1} \simeq 1.0$.
Therefore, $T^*$ becomes larger as $\Delta \epsilon$ increases.

The ground state in the other three electron region,
for $\Delta \epsilon \lesssim 0.0$, 
has an even parity. The charge distribution 
with respect to this state is shown with
the solid line labelled \lq\lq N3S05:even"
in Fig.\ \ref{fig:charge_distribution_isolated_de}, 
and in this case it is such that  $N_{a0}\simeq 1.0$, $N_{a1} \simeq 1.5$, 
and $N_{b1} \simeq 0.5$ 
near $\Delta \epsilon \simeq 0.0^-$. 
Therefore, the apex site is singly occupied, and the moment protracted. 
The distribution varies as $\Delta \epsilon$ decreases, 
and the fraction of the local moment tends 
to stay close to the leads, as 
 $N_{a0}\simeq 2.0$,  $N_{a1} \simeq 0.5$, and $N_{b1} \simeq 0.5$.
We have also confirmed that  $T^*$ is enhanced 
near $\Delta \epsilon \simeq 0.0$ 
and $-0.7\lesssim \epsilon_d/U \lesssim -0.4$, 
along the sharp conductance valley caused by the SU(4) Kondo effect.  
The local moment in the three-electron region 
also reflects the properties of the one-particle state. 
Similar to the $t'/t \gtrsim 1.0$ case 
discussed in the previous subsection, the orbital degeneracy is lifted 
as $E_\mathrm{o}^{(1)} > E_\mathrm{e,+}^{(1)}$  
for $\Delta \epsilon \lesssim 0.0$.
In this case the third electron enters the resonance state 
corresponding to the excited even-parity orbital, 
appearing in the middle of Fig.\ \ref{fig:spec_u0_de} (b). 
Since the spectral weight of this state is mainly at the apex site,
the resonance width becomes narrow already in the noninteracting case 
and it evolves into a sharp Kondo resonance for finite $U$.

The $S=1$ Kondo effect takes place in the four electron region 
at $ -0.4 \lesssim \Delta \epsilon/U \lesssim 0.8$ 
in Fig.\  \ref{fig:TK_de}.
We can see that $T^*$ varies significantly inside this region 
depending on whether $\Delta \epsilon \lesssim 0.0$ or 
$\Delta \epsilon \gtrsim 0.0$, 
although the $S=1$ Nagaoka high-spin state 
which has an odd parity evolves continuously as $\Delta \epsilon$ varies. 
The eigenvector in the limit of $\Gamma \to 0$ is given by
$|\Phi^{(4)}_\mathrm{odd}\rangle$,  
and the charge distribution with respect 
to this state is shown 
in Fig.\ \ref{fig:charge_distribution_isolated_de}, 
with the  dot-dash line labelled  \lq\lq N4S1:odd".
We can see that $N_{a0}$ and $N_{a1}$ vary rapidly 
near $\Delta \epsilon \simeq 0.0$, keeping 
the filling of $b_1$ orbital unchanged $N_{b1}=1.0$. 
For $\Delta \epsilon \gtrsim 0.0$,
the local moment has a  finite component in the apex site. 
This component of the moment moves to the $a_1$ orbital near the leads  
for $\Delta \epsilon \lesssim 0.0$. 
The variation of $T^*$ inside the $S=1$ Kondo region reflects 
these changes in the charge and spin distributions.

\section{Summary}
\label{sec:summary}

We have studied the effects of distortions which break the full symmetry 
of an equilateral triangle 
of a TTQD connected to two non-interacting leads, 
 over a wide range of the gate voltage $\epsilon_d$. 
Two types of disorder have been considered, ($i$) 
an inter-dot tunneling matrix element $t'$ ($\neq t$), 
and ($ii$) a level position $\epsilon_\mathrm{apex}$ 
 ($\neq \epsilon_d$) of the dot at the apex site.
We have concentrated on the low energy behavior, restricting attention
 mainly to the regime 
with large Coulomb interaction $U$ 
and small hybridization $\Gamma$ as this leads to several different
types of the Kondo effect. 

We find that the  key variables for characterizing the low energy behavior 
are the total occupation number $N_\mathrm{tot}$ and the phase
difference $\Theta \equiv (\delta_\mathrm{e} - \delta_\mathrm{o} )(2/\pi)$. 
The two phase shifts for the renormalized quasi-particles,
$\delta_\mathrm{e}$ and $\delta_\mathrm{o}$, 
can be deduced theoretically from the low energy NRG fixed point.
The phase shifts may be deduced experimentally  
through the conductances and $N_\mathrm{tot}$.
Measurements of the AB oscillation 
in a magnetic field may also give a clue to determine 
the phase difference.

 In the parameter space for large $U$ we find plateaus with 
the integer values of $\Theta$, 
and at each plateau  the occupation number 
also approaches to an integer.
 These plateaus, therefore, can be classified with  
 the two integer  set
$(N_\mathrm{tot}, \Theta)$ 
[see Figs.\ \ref{fig:u1_dedo_ts} and \ref{fig:u1_dedo_de}]. 
The structure of these plateaus of $\Theta$ 
determines the precise feature 
of the Kondo ridges and valleys 
of the conductance 
[see Figs.\ \ref{fig:conductance_u1} 
and \ref{fig:conductance_de_u1}].

Different Kondo effects occur in different regimes.
The SU(4) Kondo effect takes place for weak distortions, 
along the contour line for $\Theta =2.0$ 
which runs in the region of $N_\mathrm{tot} \simeq 3.0$ 
in the parameter space.
This contour transverses  the middle 
of a steep cliff of $\Theta$, standing between the plateau for 
$\Theta\simeq 1.0$ and that for $\Theta\simeq 3.0$.
It can be observed as a sharp conductance valley 
between the Kondo ridges on both sides,
and the slope of the cliff determines 
the width of the conductance valley. 
The SU(4) Kondo behavior is sensitive to the perturbations 
which lower the  symmetry of the equilateral triangle. 
This is caused by the fact that the SU(4) symmetry 
relies crucially on the orbital degeneracy.
 Furthermore, the spin and charge distributions inside 
the TTQD vary  near the SU(4) symmetric point, 
 and it affects significantly the Kondo screening.

The $S=1$ Kondo effect, 
taking place at the plateau of $\Theta$ 
for $(N_\mathrm{tot}, \Theta) \simeq (4.0,\, 2.0)$,
is robust against the breaking of the symmetry of the equilateral triangle. 
This is mainly due to a size effect:
there is a finite energy separation between 
the Nagaoka high-spin state and 
the excited local singlet state in the isolated TTQD cluster.
For large distortions 
a singlet-triplet transition takes place.
It becomes a crossover between 
a Kondo and non-Kondo singlet state for finite $\Gamma$, 
and the series conductance has a peak of  
the height of $2e^2/h$ in the transient regions. 
The width of the peak is determined by 
the slope of the cliff of $\Theta$, 
which appears at the crossover region.

Apart from the phase shifts which determine the conductance
and occupation of the TTQD, another important renormalized
parameter characterizing the low energy behavior is the 
 characteristic energy scale $T^*$.  For $T \lesssim T^*$ 
the low-energy properties 
can be described by the local Fermi-liquid theory. 
In the cases where the quantum dots have a local moment, 
   $T^*$ can be regarded as the Kondo temperature.
We have estimated  $T^*$ from the region where the NRG levels
crossover to the low energy fixed point. 
The results for $T^*$ 
reflect 
the  distribution of the  charge and spin in the TTQD 
[Figs.\ \ref{fig:TK_ts} and \ref{fig:TK_de}].

Specifically,  $T^*$ tends to be small  
in the case where a partial moment remains 
in the apex site, which has no direct coupling to the leads.
The screening of such a partial moment becomes sensitive 
to the charge and spin on the other two dots 
because the conduction electrons tunneling from the leads 
have to pass through either of the two dots to get the apex site.   
In some regions of the parameter space, 
we find that the tunneling of the 
conduction electron is suppressed 
at these two dots, 
in a way analogous to a super-exchange process 
caused by a potential barrier between  
the local moment and leads.
The characteristic temperature $T^*$  can be raised, however, 
by making the coupling to the leads $\Gamma$ stronger.
Note that $T^*$ depends on $\Gamma$ not only  
through the prefactor, but also through 
the higher order contributions of the hybridization, 
which cause an exponential dependence of $T^*$ on $\Gamma$ and 
other parameters.
Specifically, $T^*$ may become large 
for the TTQD with a small charging energy $U$.
Our results provide an overview  
of how  characteristic energy scale 
varies in the different the regions in the parameter space.

A general point worthy of note is that the  two types of the distortions  
show a clear contrast in the form of  the charge distribution 
for some regions of the parameter space 
[see Figs.\   \ref{fig:charge_distribution_isolated_ts}
 and  \ref{fig:charge_distribution_isolated_de}]. 
The diagonal distortion 
 ($\epsilon_\mathrm{apex} \neq \epsilon_d$) 
affects 
directly the potential of the apex site, so that 
the charge distribution is more sensitive 
to $\Delta \epsilon$ than 
to the off-diagonal one ($t' \neq t$), 
and this difference affects 
the characteristic energy scale $T^*$ significantly 
in some regions of the parameter space.

\begin{acknowledgments}

We would like to thank S.\ Mimura for valuable discussions. 
This work is supported by JSPS Grant-in-Aid 
for Scientific Research (C)  (Grant No.\ 20540319).
One of us (ACH) acknowledges 
the support of a grant from the EPSRC  (Grant No.\ Ep/G032181/1).
Two of us (SA and ST) acknowledge JSPS Grant-in-Aid for Scientific
Research S (No.\ 19104007), MEXT Grant-in-Aid for Scientific Research on
Innovative Areas (21102003), Funding Program for World-Leading
Innovative R\&D on Science and Technology(FIRST), and DARPA QuEST grant
HR0011-09-1-0007. 
Numerical computation was partly carried out 
in Yukawa Institute Computer Facility.

\end{acknowledgments}

\appendix

 \section{Phase shifts $\delta_\mathrm{e}$ and $\delta_\mathrm{o}$}

\label{sec:app_green}

 The phase shifts for interacting electrons can be 
defined, using the Green's function 
\begin{equation} 
G_{ij}(i\omega_n) 
\, =\,  
-    \int_0^{\beta} \! d\tau \,
   \left \langle  T_{\tau} \,  
   d^{\phantom{\dagger}}_{i \sigma} (\tau) 
   \, d^{\dagger}_{j \sigma} (0) 
     \right \rangle  \, e^{i \omega_n \tau} \;.
\label{eq:G_Matsubara}
\end{equation} 
Here, $\beta= 1/T$,  
$d_{j \sigma}(\tau) = 
e^{\tau  {\cal H}} d_{j \sigma} e^{- \tau  {\cal H}}$, and 
 $\langle {\cal O} \rangle =
\mbox{Tr} \left[ \, e^{-\beta  {\cal H} }\, {\cal O}
\,\right]/\mbox{Tr} \, e^{-\beta  {\cal H} }$. 
The retarded Green's function is given by  
 $G_{ij}^{+}(\omega) \equiv  G_{ij}(\omega + i\, 0^+)$ 
via the analytic continuation, 
and the self energy $\Sigma_{ij}(z)$   
due to the interaction $\mathcal{H}_\mathrm{dot}^U$ 
can be described by the Dyson equation  
\begin{equation} 
   G_{ij}(z)    =   G^{(0)}_{ij}(z) 
     + 
 \sum_{i'=1}^{N_D}
 \sum_{j'=1}^{N_D}
 \,G^{(0)}_{ii'}(z)\,  \Sigma_{i'j'}(z)  
    \, G_{j'j}(z) \;.    
   \label{eq:Dyson}   
 \end{equation}   
 Here, the number of the dots is $N_D=3$ for the TTQD, and  
$G^{(0)}_{ij}(z)$ is the non-interacting Green's function 
 corresponding to the free Hamiltonian 
$\mathcal{H}_0 \equiv  \mathcal{H}_\mathrm{dot}^0 
+  \mathcal{H}_\mathrm{mix}  +  \mathcal{H}_\mathrm{lead}$.

At zero temperature $T=0$, the series $g_\mathrm{s}$ and 
 parallel $g_\mathrm{p}$ conductances 
are determined by the Green's functions at 
 the Fermi level $\omega=0$,\cite{aoQuasi,ONH}
\begin{align}
g_\mathrm{s}^{\phantom{0}}\, = & \  \frac{2 e^2}{h}  
    4\Gamma_R \Gamma_L \left| G_{N_D 1}^{+}(0)\right|^2  
         \;,  
\label{eq:cond_s} 
\\
g_\mathrm{p}^{\phantom{0}} 
= & \ 
{2 e^2 \over h} 
 \, 
\Bigl[\,
-\Gamma_L\, \mathrm{Im}\, G_{11}^+(0)
-\Gamma_R\, \mathrm{Im}\, G_{N_DN_D}^+(0)
\,\Bigr] \;.
\label{eq:def_g_p_Green}
\end{align}
Note that the contributions from the vertex correction do not appear here 
 due to the property that the imaginary part of the self-energy vanishes 
 $\mathrm{Im}\,\Sigma_{ij}^{\pm}(0)=0$ at $T=0$ and $\omega=0$.\cite{aoFermi}
Furthermore, 
for the symmetric coupling 
 $\Gamma_L = \Gamma_R$ ($\equiv \Gamma$), 
the Green's functions can be expressed in the forms, 
\begin{align}
G_{11}^+(0)  = G_{N_DN_D}^+(0) 
    \,=& \  \frac{1}{2 \Gamma} 
    \left[\, 
    \frac{1}{\kappa_\mathrm{e} + {i}}
    \,+\, \frac{1}{\kappa_\mathrm{o} + i}
    \,\right] \;, 
\label{eq:G_11_sym}
\\
G_{N_D 1}^+(0)  
   \,=& \  \frac{1}{2 \Gamma} 
   \left[\, 
   \frac{1}{\kappa_\mathrm{e} + i}
   \,-\, \frac{1}{\kappa_\mathrm{o} + i}
   \,\right] \;. 
\label{eq:G_N1_sym}
\end{align}
Here,  $\kappa_\mathrm{e}   =   - \cot \delta_\mathrm{e}$ 
and $\kappa_\mathrm{o} = - \cot \delta_\mathrm{o}$ 
include all the many-body corrections, through 
the real part of the self-energy $\mathrm{Re}\,\Sigma_{ij}^+(0)$.\cite{ONH}  
Equations \eqref{eq:gs}--\eqref{eq:gp} follow 
from Eqs.\ \eqref{eq:cond_s}--\eqref{eq:G_N1_sym}.


\begin{figure}[t]

\begin{minipage}{0.5\linewidth}
 \includegraphics[width=1.0\linewidth]{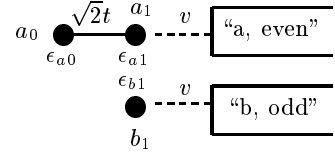}

\end{minipage}

\caption{Even and odd orbitals:
the onsite potential of each orbital is given by 
$\,\epsilon_{a0}=\epsilon_\mathrm{apex}$,
$\,\epsilon_{a1}=\epsilon_d -t'$,
and $\,\epsilon_{b1}=\epsilon_d +t'$. 
}
\label{fig:even_odd}
\end{figure}

\section{Even and odd orbitals}
\label{sec:even_odd}

The eigenstates of the Hamiltonian $\mathcal{H}$ defined  
in Eq.\ \eqref{eq:H} 
can be classified according to the parity 
in the case that the system has an inversion symmetry,
using the even-odd basis defined by 
$a^{}_{0 \sigma} \equiv d^{}_{2 \sigma}$, 
\begin{align}
 a^{}_{1 \sigma} \equiv 
\frac{d^{}_{1 \sigma} + \,d^{}_{3 \sigma}}{\sqrt{2}}
\;, \qquad \quad 
 b^{}_{1 \sigma} \equiv 
\frac{d^{}_{1 \sigma} - \,d^{}_{3 \sigma}}{\sqrt{2}}
\;.
\label{eq:even_odd_start_appendix}
\end{align}
The labels $0$ and $1$ for the even-odd basis 
are assigned in the way that is shown in Fig.~\ref{fig:even_odd}.
The odd parity  $b_1$ orbital 
corresponds to the eigenstate for $E_\mathrm{o}^{(1)}$, defined  
in Eq.\ \eqref{eq:U0_eigen_cluster2}, 
for the noninteracting TTQD cluster. Similarly 
the eigenstate for $E_\mathrm{e,\pm}^{(1)}$ 
is given by a linear combination of the even $a_0$ and $a_1$ orbitals.
\begin{align}
|\Phi^{(1)}_\mathrm{e,+}\rangle 
\,=& \  
\left(
u_+ \,a^{\dagger}_{0\sigma} \,+\,
u_-\,a^{\dagger}_{1\sigma} 
\right)
|0\rangle , 
\\
|\Phi^{(1)}_\mathrm{e,-}\rangle 
\,=& \ 
\left(
-u_- \,a^{\dagger}_{0\sigma} \,+\,
u_+ \,a^{\dagger}_{1\sigma}
\right)
|0\rangle \;.
\end{align}
Here,  $|0\rangle$ is a vacuum. 
The coefficients are normalized such that $u_+^2+u_-^2 =1$,   
\begin{align}
u_{\pm}^2 = & \  \frac{1}{2} \left( 1 \pm 
\frac{\Delta \epsilon +t'}{\sqrt{(\Delta \epsilon +t')^2 +8t^2}}\right) 
\label{eq:spectral_weight_a0}
\end{align}
and $u_+^2$ ($u_-^2$)
corresponds to the spectral weight for the $a_0$ ($a_1$) component 
in the excited state $|\Phi^{(1)}_\mathrm{e,+}\rangle$. 
When the TTQD is coupled to the leads, 
these states become the resonance levels, 
which can be described by the Green's functions for the 
$a_0$, $a_1$, and $b_1$ orbitals 
\begin{align}
G_{a0}^{(0)}(\omega)=&\  
\frac{1}{\omega-\epsilon_\mathrm{apex} -
\frac{2t^2}{\omega -\epsilon_d +t'+ i\Gamma}
}
\;,\\ 
G_{a1}^{(0)} (\omega)=&\   \frac{1}{\omega -\epsilon_d +t' + i\Gamma
-\frac{2t^2}{\omega-\epsilon_\mathrm{apex}}} \;,\\
G_{b1}^{(0)} (\omega)=&\   \frac{1}{\omega -\epsilon_d-t'+ i\Gamma}
\;.
\end{align}
Specifically for $\Delta \epsilon +t'>0$ the coefficients 
take the value of $u_+^2 > 0.5$ and $u_-^2 < 0.5$.
In this case the even excited state $|\Phi^{(1)}_\mathrm{e,+}\rangle$ 
becomes a sharp resonance peak, 
the spectral weight of which is mainly on the apex site, 
and the spectral function 
near $\omega \simeq E_\mathrm{e,+}^{(1)}$ takes the form
\begin{align}
-\mathrm{Im}\, G_{a0}^{(0)}(\omega) \  \simeq & 
\ \    
u_+^2\,
\frac{u_-^2\Gamma}
{\left(\omega-E_\mathrm{e,+}^{(1)}\right)^2 +  \left(u_-^2 \Gamma\right)^2}
\;, 
\label{eq:peak_even_excited}
\\ 
-\mathrm{Im}\, G_{a1}^{(0)}(\omega) \  \simeq & \  \   
u_-^2\,
\frac{u_-^2\Gamma}
{\left(\omega-E_\mathrm{e,+}^{(1)}\right)^2 +  \left(u_-^2 \Gamma\right)^2}
\label{eq:peak_even_lower}
\;.
\end{align}
The Green's functions near the lower  
level $\omega \simeq E_\mathrm{e,-}^{(1)}$ 
can also be written in similar forms, 
just by replacing  $+$ ($-$) in the suffix by $-$ ($+$) in  
 Eqs.\ \eqref{eq:peak_even_excited} and \eqref{eq:peak_even_lower}.

The interaction Hamiltonian defined in Eq.\ \eqref{eq:HC^U} 
can be expressed, in terms of these even-odd orbitals,
in the form 
\begin{align}
\mathcal{H}_\mathrm{dot}^U  
=& \ U n_{a,0\uparrow}\,n_{a,0\downarrow}
+ \frac{U}{2} \left(
n_{a,1\uparrow}
\,n_{a,1\downarrow}
+n_{b,1\uparrow}\,n_{b,1\downarrow} \right)
\nonumber \\ 
& 
+ \frac{U}{2} \left(
\frac{1}{2} n_{a,1}\,n_{b,1}
-2 \vec{\bm{S}}_{a,1}\cdot\vec{\bm{S}}_{b,1}
\right)
\nonumber \\ 
& 
+ \frac{U}{2} \left(
 a^{\dagger}_{1 \uparrow} a^{\dagger}_{1 \downarrow} 
b^{}_{1 \downarrow} b^{}_{1 \uparrow}
+ 
 b^{\dagger}_{1 \uparrow} b^{\dagger}_{1 \downarrow} 
a^{}_{1 \downarrow} a^{}_{1 \uparrow}
\right).
\label{eq:even_odd}
\end{align}
Here, 
$\vec{\bm{S}}_{a,i}= \sum_{\sigma \sigma'}
a^{\dagger}_{i \sigma} \vec{\bm{\sigma}}_{\sigma\sigma'} 
a^{}_{i \sigma'}/2 $,   $\,\,\vec{\bm{\sigma}}$ the Pauli matrices,
$n_{a,i\sigma}= 
a^{\dagger}_{i \sigma}a^{}_{i \sigma} $, and 
$n_{a,i}= \sum_{\sigma}n_{a,i \sigma} $. 
The operators $n_{b,1\sigma}$ and $\vec{\bm{S}}_{b,1}$ 
for the odd-parity orbital $b_{1\sigma}$ are defined in the same way.

The Hilbert space for the isolated TTQD cluster, 
described by 
$\mathcal{H}_\mathrm{dot}^{}  =
\mathcal{H}_\mathrm{dot}^0 +
\mathcal{H}_\mathrm{dot}^U$, can be constructed 
from the three orbitals. 
For instance, 
the Nagaoka state for $\mathcal{H}_\mathrm{dot}^{}$ 
has an odd parity.
It has one electron in the $b_1$ orbital, and 
the eigenvector takes the form  
\begin{align} 
& 
|\Phi^{(4)}_\mathrm{odd}\rangle 
=  
\  
\alpha_0\, 
|\mathrm{I} \rangle  
+ \alpha_1\,
|\mathrm{II} \rangle  
\;, \\
& 
\ \ 
|\mathrm{I} \rangle  =   
  b^{\dagger}_{1\uparrow}
 a^{\dagger}_{1\uparrow}
 a^{\dagger}_{0\uparrow}
 a^{\dagger}_{0\downarrow} 
|0\rangle , 
 \quad   
|\mathrm{II} \rangle  
 =  
 b^{\dagger}_{1\uparrow}
 a^{\dagger}_{1\uparrow}
 a^{\dagger}_{1\downarrow} 
 a^{\dagger}_{0\uparrow}
|0\rangle 
\;.
\label{eq:N4_Nagaoka_cluster}
\end{align} 
Here, 
$\alpha_0$ and $\alpha_1$ are the coefficients.
The eigenvectors for five electrons can be expressed in the form
\begin{align} 
& |\Phi^{(5)}_\mathrm{even}\rangle 
=\,   
  b^{\dagger}_{1\downarrow}
|\Phi^{(4)}_\mathrm{odd}\rangle 
\;, 
\label{eq:N5_even_cluster}
\\  
&|\Phi^{(5)}_\mathrm{odd}\rangle 
 = \,   
 b^{\dagger}_{1\uparrow}
 a^{\dagger}_{1\uparrow}
 a^{\dagger}_{1\downarrow} 
 a^{\dagger}_{0\uparrow}
 a^{\dagger}_{0\downarrow}
|0\rangle  . 
\label{eq:N5_odd_cluster}
\end{align} 
Therefore, 
the odd-parity $b_1$ orbital is fully occupied 
for $|\Phi^{(5)}_\mathrm{even}\rangle$,
while the even $a_0$ and $a_1$ orbitals 
are fully occupied 
for $|\Phi^{(5)}_\mathrm{odd}\rangle$.
The distribution of the charge and spin 
in the three orbitals affects significantly 
on the way  the screening by the conduction electrons 
is carried out for finite $\Gamma$, 
 when the leads are connected to the TTQD.

\section{NRG approach}
\label{sec:NRG_approach}

We provide an explicit form of the discretized Hamiltonian 
 $H_N$ of the NRG in this appendix.
The non-interacting leads are transformed 
into the tight-biding chains in the NRG approach,  
through the logarithmic discretization 
with the parameter $\Lambda$. 
Then, a sequence of the Hamiltonian $H_N$ with a finite size 
is introduced in the form\cite{KWW,KWW2}
\begin{align}
&H_N \ =  \Lambda^{(N-1)/2} 
\left( 
\mathcal{H}_\mathrm{dot}^0 +
\mathcal{H}_\mathrm{dot}^U 
 +  H_\mathrm{mix}^{\phantom{0}}  +   H_\mathrm{lead}^{(N)}
 \right) ,
\label{eq:H_N} 
\\
&H_\mathrm{mix}^{\phantom{0}}  \, = \, \bar{v} \, 
       \sum_{\sigma}
\left(\,
f^{\dagger}_{0,L\sigma} d^{\phantom{\dagger}}_{ 1,\sigma}
\,+\, 
d^{\dagger}_{ 1,\sigma}  f^{\phantom{\dagger}}_{0,L\sigma} 
     \right)
     \nonumber \\
 & \qquad \quad  
+  \bar{v}\, 
       \sum_{\sigma} 
       \left(\,
f^{\dagger}_{0,R \sigma} d^{\phantom{\dagger}}_{N_C, \sigma} 
   \,+\, 
d^{\dagger}_{N_C, \sigma} f^{\phantom{\dagger}}_{0,R \sigma} 
         \,  \right)  \;,
\label{eq:H_mix_NRG}
\\
& H_\mathrm{lead}^{(N)} \,=\,
D\,{1+1/\Lambda \over 2} \,
\sum_{\nu=L,R}
\sum_{\sigma}
\sum_{n=0}^{N-1} 
\, \xi_n\, \Lambda^{-n/2}
\nonumber \\
& \qquad \qquad \times  
\left(\,
  f^{\dagger}_{n+1,\nu\sigma}\,f^{\phantom{\dagger}}_{n,\nu\sigma}
  +  
 f^{\dagger}_{n,\nu\sigma}\, f^{\phantom{\dagger}}_{n+1,\nu\sigma}
 \,\right) \;.
\label{eq:H_lead_NRG}
\end{align}
Here,  $D$ is the half-width of the conduction band, and 
the other parameters are defined by
\begin{align}
  \bar{v}
&\,=\, \sqrt{ \frac{2D\,\Gamma A_{\Lambda}}{\pi} }
\;,
\qquad
A_{\Lambda}  \,=\,  \frac{1}{2}\, 
 {1+1/\Lambda \over 1-1/\Lambda }
\,\log \Lambda
\;, 
\label{eq:A_lambda}
\\
\xi_n &\,=\,    
{ 1-1/\Lambda^{n+1}  
\over  \sqrt{1-1/\Lambda^{2n+1}}  \sqrt{1-1/\Lambda^{2n+3}} 
} 
\;.
\label{eq:xi_n}
\end{align}


\end{document}